# Dynamic anticipation by Cdk2/Cyclin A-bound p27 mediates signal integration in cell cycle regulation

Maksym Tsytlonok[1]*, Hugo Sanabria[2,3]*, Yuefeng Wang[4,5]*, Suren Felekyan[3], Katherina Hemmen[3], Mi-Kyung Yun[4], Brett Waddell[4,6], Cheon-Gil Park[4], Sivaraja Vaithiyalingam[4,6], Luigi Iconaru[4], Stephen W. White[4], Peter Tompa[1,7]**, Claus A. M. Seidel[3]**, and Richard Kriwacki[4,8]**

[1]VIB Center for Structural Biology, Vrije Universiteit Brussel, Brussels, Belgium.

[2]Department of Physics and Astronomy, Clemson University, Clemson, SC, 29634, USA.

[3]Lehrstuhl für Molekulare Physikalische Chemie, Heinrich-Heine-Universität, Düsseldorf, 40225, Germany.

[4]Department of Structural Biology, St. Jude Children's Research Hospital, 262 Danny Thomas Place, Memphis, TN 38105, USA.

[5]Current address: Department of Radiation Oncology, West Cancer Center, University of Tennessee Health Sciences Center, Memphis, TN 38134, USA

[6]Molecular Interaction Analysis Shared Resource, St. Jude Children's Research Hospital, 262 Danny Thomas Place, Memphis, TN 38103, USA.

[7]Institute of Enzymology, Research Centre for Natural Sciences of the Hungarian Academy of Sciences, Budapest, Hungary.

[8]Department of Microbiology, Immunology and Biochemistry, University of Tennessee Health Sciences Center, Memphis, TN 38163, USA.

*, ** these authors contributed equally to the work

**Correspondence: tompa@enzim.hu (PT), cseidel@hhu.de (CAMS) and richard.kriwacki@stjude.org (RK); Lead contact: richard.kriwacki@stjude.org (RK).

***Abstract***

p27$^{Kip1}$ (p27) is an intrinsically disordered protein (IDP) that folds upon binding to cyclin-dependent kinase (Cdk)/cyclin complexes (*e.g.,* Cdk2/cyclin A), inhibiting their catalytic activity and causing cell cycle arrest. However, cell division progresses when stably Cdk2/cyclin A-bound p27 is phosphorylated on one or two structurally occluded tyrosine residues [tyrosines 88 (Y88) and 74 (Y74)] and a distal threonine residue [threonine 187 (T187)]. These events trigger ubiquitination and degradation of p27, fully activating Cdk2/cyclin A to drive cell division. Using an integrated approach comprising structural, biochemical, biophysical and single-molecule fluorescence methods, we show that Cdk2/cyclin A-bound p27 samples lowly-populated conformations that dynamically anticipate the sequential steps of this signaling cascade. "Dynamic anticipation" provides access to the non-receptor tyrosine kinases, BCR-ABL and Src, which sequentially phosphorylate Y88 and Y74 and promote intra-assembly phosphorylation (of p27) on distal T187. Tyrosine phosphorylation also allosterically relieves p27-dependent inhibition of substrate binding to Cdk2/cyclin A, a phenomenon we term "cross-complex allostery". Even when tightly bound to Cdk2/cyclin A, intrinsic flexibility enables p27 to integrate and process signaling inputs, and generate outputs including altered Cdk2 activity, p27 stability, and, ultimately, cell cycle progression. Intrinsic dynamics within multi-component assemblies may be a general mechanism of signaling by regulatory IDPs, which can be subverted in human disease, as exemplified by hyper-active BCR-ABL and Src in certain cancers.

**Keywords:** intrinsically disordered proteins; signaling; intrinsic flexibility; cell cycle regulation; single-molecule fluorescence; allostery; dynamic anticipation; protein NMR spectroscopy.

**Introduction**

Many eukaryotic proteins lack stable 3D structures and exhibit conformational heterogeneity in isolation but fold into discrete conformations upon binding to other molecules (e.g. proteins, nucleic acids, small molecule ligands, metal ions, etc.) (Wright and Dyson, 2009). These so-called intrinsically disordered proteins (IDPs) often function to regulate complex biological processes (van der Lee et al., 2014), with their bound conformations exerting a basal level of regulatory control over their targets. However, this basal level of control can often be modulated, or switched, through post-translational modifications to achieve complex regulatory behavior (Tompa, 2014; Van Roey et al., 2013), as exemplified by the cell-cycle regulator, $p27^{Kip1}$ (p27), which folds upon binding to cyclin-dependent kinase (Cdk)/cyclin complexes that control cell division (Galea et al., 2008). The basal function of p27 is the inhibition of the kinase activity of Cdk/cyclin complexes, which can be relieved through phosphorylation of its tyrosine residue(s) by non-receptor tyrosine kinases (NRTKs)(Chu et al., 2007; Grimmler et al., 2007). Relief of Cdk inhibition triggers a sequence of additional post-translational modifications, including phosphorylation of T187 to activate a phosphodegron within the same p27 molecule, followed by its ubiquitination and degradation, triggering full activation of Cdk/cyclin complexes and progression to S phase during cell division.

The existence of regulatory modification sites within bound and folded regions of p27 (and IDPs in general) raises a key question regarding how these sites become accessible for enzymatic modification in signal transduction. It has been previously suggested that regions of bound disordered proteins experience dynamic fluctuations (Grimmler et al., 2007; Tompa and Fuxreiter, 2008). Therefore, we hypothesized that regions within IDPs subject to function-altering post-translational modifications have evolved to dynamically sample different conformational states that provide accessibility to modifying enzymes. The ensuing enzymatic modifications remodel the conformational landscape of the IDP, enabling further changes and controlled integration of incoming signals. Herein, we tested this hypothesis by studying p27 bound to Cdk2/cyclin A by NMR spectroscopy, single-molecule multiparameter fluorescence detection (smMFD) and other biophysical techniques, which revealed dynamic fluctuations ("dynamic structural anticipation") that allow sequential phosphorylation of two of its tyrosines, Y88 (Grimmler et al., 2007) and Y88 and Y74 (Chu et al., 2007) by the NRTKs, Breakpoint cluster region-Abelson murine leukemia viral oncogene homolog 1 (BCR-ABL) and Src, respectively. We suggest that remodeling of dynamic structural ensembles of bound states by post-translational modifications might be a general mechanism of signal integration by regulatory IDPs.

## Results

### p27 phosphorylation restores Cdk2 activity

p27 is tethered to the Cdk2/cyclin A assembly via two discontinuous subdomains, D1 and D2, within its kinase inhibitory domain (KID); D1 binds to a hydrophobic surface patch on cyclin A and D2 binds to the active site of Cdk2 (for subdomain nomenclature, see Figure 1A, for the structure of the assembly, see Figure 1B). Phosphorylation of Y88 by BCR-ABL was previously shown to partially relieve p27-dependent inhibition of Cdk2/cyclin A, triggering p27 ubiquitination and degradation, and full activation of Cdk2 to drive progression into S phase of the cell division cycle(Chu et al., 2007; Grimmler et al., 2007). Tyrosine 88, located within D2 subdomain, is inserted into the ATP-binding pocket of Cdk2 (Figure 1B) (Galea et al., 2008; Grimmler et al., 2007). Interestingly, p27 exhibits a second tyrosine within its KID, Y74 (Figure 1), and both Y88 and Y74 were previously shown to be phosphorylated by Src in a large fraction of hyper-proliferative breast cancer cell lines (Chu et al., 2007). Dual phosphorylation of Y88 and Y74 was shown to be associated with heightened Cdk2 activity (Chu et al., 2007; Grimmler et al., 2007) but the molecular mechanism of this effect was not investigated. In Cdk2 activity assays with p27-KID, in which Y88 (pY88-p27-KID) or both Y74 and Y88 (pY74/pY88-p27-KID) are phosphorylated, we found that Y88 phosphorylation restored ~20 % of full Cdk2 activity at saturating concentrations of the inhibitor, as previously observed (Grimmler et al., 2007), whereas dual Y phosphorylation restored ~50 % of kinase activity (Figure 2A, Suppl. Figure 1A). Similar Cdk2 reactivation was observed with Y88- and Y74/Y88-phosphorylated full-length p27 (Suppl. Figure 1A, B).

### Phosphorylation of tyrosine residues in p27 is sequential

Next, we asked if these phosphorylation reactions occurred in a specific order, because both Y88 and Y74 are buried against the surface of Cdk2 in the Cdk2/cyclin A/p27-KID structure (Figure 1B, PDB 1JSU; 4 % and 0 % solvent accessible surface area, respectively). We have previously shown that Y88 of p27 was accessible for phosphorylation by ABL kinase in cells and by ABL kinase domain (ABL-KD) *in vitro* (when bound to Cdk2/cyclin A) (Grimmler et al., 2007). To address the issue of accessibility of Y74, we performed phosphorylation assays using Src kinase domain (Src-KD) and pY88-p27-KID, as well as p27-KID constructs with Y88 or both Y74 and Y88 mutated to phenylalanine (Y88F-p27-KID and Y74F/Y88F-p27-KID, respectively). In the absence of Cdk2/cyclin A, Y74 and Y88 were phosphorylated equally by Src-KD (Figure 2B). In the presence of Cdk2/cyclin A, Y74 was phosphorylated within p27-KID and pY88-p27-KID but not within Y88F-p27-KID (Figure 2B). These results indicated that prior phosphorylation of Y88 is required for Y74 to become accessible for phosphorylation by Src-KD, suggesting that the two tyrosine residues are sequentially phosphorylated. How priming phosphorylation of Y88, with p27 tightly bound to Cdk2/cyclin A occurs, however, is enigmatic.

**Tyrosine phosphorylation of p27 exerts rheostat-like control of Cdk2 activity and T187 phosphorylation**

Reactivation of Cdk2 through phosphorylation of Y88 stimulates Cdk2-dependent phosphorylation of T187 within the flexible C-terminus of p27 through a pseudo uni-molecular mechanism within the Cdk2/cyclin A/pY88-p27 ternary assembly (Das et al., 2016; Grimmler et al., 2007). Because Src-dependent dual Y phosphorylation of p27 is associated with reduced p27 protein levels in breast cancer cell lines (Chu et al., 2007), and because Cdk2-dependent phosphorylation of T187 is associated with p27 ubiquitination and degradation (Montagnoli et al., 1999), we next asked if dual Y phosphorylation of p27 within Cdk2/cyclin A complexes would further enhance intra-assembly phosphorylation of T187. To this end, we compared the effect of mono Y and dual Y phosphorylation of p27 on T187 phosphorylation *in vitro*. As shown previously (Grimmler et al., 2007), a low level of T187 phosphorylation can be observed in 1:1:1 Cdk2/cyclin A/p27 assemblies (Figure 2C, Suppl. Figure 2A); this arises from bi-molecular reactions (termed the "inter-molecular" mechanism) between a small amount of free (and active) Cdk2/cyclin A and T187 within either free or Cdk2/cyclin A-bound p27. T187 phosphorylation is enhanced by prior phosphate incorporation into pY88-p27 (Figure 2C), and dual Y phosphorylation causes a further 2-fold increase of phosphate incorporation. The data exhibited the linear concentration dependence indicative of an intra-complex mechanism (Grimmler et al., 2007) (Figure 2C). Thus, mono and dual Y phosphorylation of p27 exert rheostat-like control of Cdk2 activity and phosphorylation of p27 on T187.

**Reactivated Cdk2: phosphorylation of intra-assembly *versus* inter-molecular substrates**

These observations show that Y phosphorylation of p27 tethered to Cdk2/cyclin A facilitates intra-assembly phosphorylation of T187 (Figure 2C), whereas results with histone H1 showed facilitation of phosphorylation of a non-specific substrate (Figure 2A). Next, we asked whether reactivation enhances phosphorylation of physiological substrates of Cdk2, such as retinoblastoma protein (Rb) and p107 (Adams, 2001), which are recognized by activated Cdk2/cyclin A via the same hydrophobic surface patch on cyclin A that binds the D1 subdomain of p27 KID. Simultaneous monitoring of phosphorylation of T187 within p27 and of Rb or p107 showed that phosphorylation of Y residues within p27 enhanced Cdk2-dependent phosphorylation of both Rb and p107 (especially upon dual Y74/Y88 phosphorylation of p27), although to a lesser extent than that of the intra-assembly substrate, T187 of p27 (Figure 2D, Suppl. Figure 2B,C). These results showed that activation of Cdk2 through Y phosphorylation of p27 primarily promoted intra-assembly phosphorylation of T187 but non-co-assembled substrates also gain access to the active site of Cdk2 and/or the substrate binding pocket on cyclin A.

**Structural mechanism of Cdk2 reactivation by tyrosine phosphorylation of p27**

The data discussed above suggest that p27 bound to Cdk2/cyclin A can sense and integrate activation signals from different NRTKs. However, due to the tight binding of p27 to Cdk2/cyclin A, the structural mechanism that enables these phosphorylation events is unclear. We hypothesized that it could be linked with the distributed interaction interface between p27-KID and Cdk2/cyclin A, which may enable phosphorylation to selectively interfere with small portions of this interface causing transient local (segmental) release of the bound inhibitor.

In accord, 2D NMR analysis previously showed that phosphorylation of Y88 displaces the C-terminal half of the D2 subdomain containing Y88 from the ATP binding pocket of the kinase (Grimmler et al., 2007), whereas dual Y74/Y88 phosphorylation displaces the entire D2 subdomain (Figure 3A). These results indicate that Y phosphorylation activates Cdk2 by making the active site sterically accessible. Next, we showed that the biochemical effect of this phosphorylation-dependent displacement could be mimicked by deletion of residues 80-94 from p27-KID (p27-KID-$\Delta$C), which advanced our understanding of the extent (Suppl. Figure 3A) and structural basis (Suppl. Figure 3B, Suppl. Table 1) of partial Cdk2 activation. Despite accessibility of the active site to ATP, only 20% of full kinase activity was regained because p27-KID-$\Delta$C altered the structure of Cdk2 near the active site. In particular, residues 74-79 of p27-KID-$\Delta$C formed an inter-molecular $\beta$-strand with $\beta$-strand 2 ($\beta$2) of Cdk2, displacing the $\beta$1 strand that otherwise forms the G-loop that binds the phosphates of ATP (Figure 3B, Suppl. Figure 3B). This intermolecular $\beta$-strand can also be observed with p27-KID (Russo et al., 1996), where it reinforces full inhibition mediated by Y88. In the case of pY88-p27-KID and p27-KID-$\Delta$C, in which the ATP binding pocket of Cdk2 is accessible, this feature explains the observed partial catalytic activity of the enzyme.

**Intrinsic dynamics in Cdk2-bound p27 enable tyrosine phosphorylation: insights from smMFD measurements**

To address the key question of how Y88, and subsequently Y74, within the tight p27-Cdk2/cyclin A assembly becomes accessible for phosphorylation by Src-KD, we introduced fluorescent dyes at several positions within p27, using native and non-native cysteine residues (*cf.* Figure 1B), and monitored their local and global structural and dynamic features using single-molecule multiparameter fluorescence detection (smMFD) of freely diffusing protein molecules. To map local flexibility, single cysteine residues at positions 29, 40, 54, 75, and 93 were labeled with BODIPY using a short linker (termed p27-C29, p27-C40, p27-C54, p27-C75, and p27-C93, respectively) and studied by single-molecule fluorescence anisotropy (smFA) experiments. To additionally map global structural dynamics, three p27 constructs were prepared with pairs of Cys residues (at positions 29-54, 54-93, and 75-110) and labelled with Alexa Fluor 488 and Alexa Fluor 647 (termed p27-C29-54, p27-C54-93, and p27-C75-110, respectively) for use in single-molecule Förster resonance energy transfer (smFRET) experiments.

The question of the enigmatic priming phosphorylation of Y88 was resolved by analyzing the local flexibility of BODIPY-labeled pY88-p27-C93 by smFA experiments. The two-dimensional histogram of fluorescence anisotropy $r_D$ *vs.* fluorescence-weighted average lifetime ($\langle\tau_{D(0)}\rangle_f$) showed broad distribution of anisotropy values of single molecule bursts (Figure 4A, for all smFA fit parameters, see Suppl. Tables 2-4). We applied probability distribution analysis (PDA) to properly account for the shot noise in the histograms (Kalinin et al., 2007). PDA disclosed two underlying long lived (> 1 ms) states – one of high ($r_D^H$, light blue line, more rigid state) and another of low ($r_D^L$, dark blue line, more flexible state) anisotropy. The single-molecule data revealed an equilibrium, where, even in the absence of Y88 phosphorylation, this region samples a minor, flexible state (33 %), which increases in population (to 66 %) significantly upon Y88 phosphorylation (Figure 4A). The existence of this lowly populated, solvent-exposed conformer of region 83-89 of p27-KID provides an explanation for the accessibility of Y88 for phosphorylation by BCR-ABL and Src in cells and the corresponding kinase domains *in vitro* (Chu et al., 2007; Grimmler et al., 2007).

smFA data for p27 labeled with BODIPY on positions C54 or C75 (Figure 4B) revealed further conformational features of D2 and also of the LH subdomain of p27-KID. In both regions, a major, highly populated state exhibited a high $r_D$ value ($r_D > 0.2$), supporting the view that positions 54 and 75 are mostly rigid both before and after Y88 phosphorylation. The minor state with a low anisotropy value ($r_D \sim 0.1$; Figure 4B, Suppl. Figure 4) increased in population upon dual Y74/Y88 phosphorylation of p27, markedly for position 54 and slightly for position 75, consistent with NMR results showing that the entire D2 subdomain was released upon dual Y phosphorylation (Figure 3A). Also in line with NMR observations are the smFA results for other regions of p27 KID as reported by C40 and C29 within the D1 subdomain. C29, which can be resolved with NMR, exhibits a highly populated rigid (high $r_D$) state, whereas for position C40 (unresolved by NMR) a flexible (low $r_D$) state is the major population. For C40, $r_D$ populations were nearly unchanged upon Y88 and Y88-Y74 phosphorylation, whereas for C29, there is some increase in the flexible state already upon Y88 phosphorylation (Figure 4B, Suppl. Figure 4).

To elucidate the structural context of the dynamic motions observed by smFA, we performed smFRET experiments and analyzed them by smMFD-plots and correlation analysis (Figure 4C, Suppl. Figure 5, all determined FRET observables are compiled in Suppl. Tables 5-7). First, we displayed the data by plotting a two-dimensional (2D) frequency smMFD-histogram for the two FRET indicators averaged donor-acceptor distance $\langle R_{DA}\rangle_E$ (derived from the intensity ratio of donor over acceptor signal) *vs.* the average donor fluorescence lifetime $\langle \tau_{D(A)}\rangle_f$, which will be shortened by FRET. This representation allows us to directly assess the structural heterogeneity and dynamics of the sample on different time scales(Sisamakis et al., 2010). The smFRET data for p27-C54-93 bound to Cdk2/cyclin A (Figure 4C) showed a broadly distributed population peaking at $\langle R_{DA}\rangle_E$ values ~ 45 Å and $\langle \tau_{D(A)}\rangle_f$ ~ 2 ns. The maximum population in the 2D histogram is shifted to the right of the static FRET line (Figure 4C, green curve; Suppl. Table 5), which is a hallmark for fast dynamic mixing of at least two different limiting states (Sisamakis et al., 2010). Moreover, the dynamic FRET populations are very broad which indicates differently averaged populations in slow exchange. We applied PDA analysis to appropriately account for the broadening of the histograms due to shot noise and distinct acceptor brightness values (Kalinin et al., 2007; Kalinin et al., 2010). This way we could reveal at least two averaged conformational states: a highly populated, major state (83 %) with $\langle R_{DA}\rangle_E$ = 45 Å (Figure 4C, dark blue), and a somewhat more extended minor state (17 %) with $\langle R_{DA}\rangle_E$ = 52 Å (Figure 4C, light blue). We assign the major state to a p27 conformation, in which Y88 is bound within the ATP pocket of Cdk2 [as observed in the Cdk2/cyclin A/p27-KID structure (Russo et al., 1996)], and the minor state to a conformation in which Y88 was released from this pocket, giving a longer inter-dye distance. This interpretation is confirmed by comparing the measured inter-dye $\langle R_{DA}\rangle_E$ values with expected distances calculated from the X-ray structure of p27-KID bound to Cdk2/cyclin A (1JSU) (Russo et al., 1996) (Suppl. Table 8), which was extended by molecular dynamics simulations (Galea et al., 2008) for the unresolved (> 90) residues of p27 (Figure 1B). To describe the dye behavior, we performed coarse grained accessible volume simulations (Suppl. Table 8) (Sindbert et al., 2011), and compared the calculated interdye distances to the highly populated states of p27-C29-54 (and also of p27-C54-93, and p27-C75-110, see later). In all cases we found that the smMFD-derived and simulated distances were in agreement (within 10 %); hence, this state indeed represents p27 bound to Cdk2/cyclin A.

We next observed that the minor state population increased to 46 % upon Y88 phosphorylation. Although ejected due to Y88 phosphorylation based on NMR data (Figure 3A), 54 % of the molecules still exhibited the shorter $\langle R_{DA}\rangle_E$ values (Figure 4C,D), suggesting that the pY88 region transiently interacts with the ATP pocket. Based on the smFA results, which showed no change after

Y88 phosphorylation for position C54 (Figure 4B), we conclude that the observed structural and local flexibility changes detected by smFRET with pY88-p27-C54-93 occur near Y88, rather than near position 54. Interestingly, despite a significant increase of the low-anisotropy state of C54 upon Y74 phosphorylation in smFA (Figure 4B), smMFD studies with the p27-C54-93 construct showed no major changes in the conformational features or populations of the two states following dual Y74/Y88 phosphorylation in comparison with the mono Y88-phosphorylated form (Figure 4D; Suppl. Figure 5B). In addition, smFA data for BODIPY-labeled pY74/pY88-p27-C75 indicated only a small change in the populations of the high and low anisotropy states in comparison with the corresponding mono Y phosphorylated construct (Figure 4B). In contrast, smFA indicated liberation of position C54 from its more rigid, bound conformation (Figure 4B). Because NMR results showed the release of the entire D2 subdomain from Cdk2 upon dual Y phosphorylation (Figure 3A), and smFRET results suggest that the region between C54 and C93 maintains conformational topology similar to that of the bound state, we interpret these results by the existence of a partially populated secondary structure in this region (Sivakolundu et al., 2005a) and/or its structural compaction in the free state, which restricts fluctuations of the residue pair, C54 and C93, and also residue 75. Such compaction driven by hydrophobic residues within D2 subdomain has been observed by NMR (Iconaru et al., 2015), and is also supported by smFRET data with p27-C75-110. Here, the fraction of molecules with short $\langle R_{DA}\rangle_E$ increased strongly after mono pY88 phosphorylation. As smFA experiments conducted on p27-C75 did not show a change upon Y88 phosphorylation (Figure 4B), the observed change in distance distribution stems primarily from the liberation of the D2 subdomain, consistent with the NMR data (Figure 3A).

**Direct observation of intrinsic p27 dynamics across broad timescales: dynamic anticipation**

Irrespective of the actual structural state of liberated D2 subdomain, the coexistence of two anisotropy- and FRET-populations and their position in the 2D diagrams (Figure 4A, C; Suppl. Figures 4-5) already in the non-phosphorylated complex indicates that p27-KID is in a slow dynamic exchange (on timescales identical to or slower than the diffusion time of ~ 1.7 ms in the confocal volume, Suppl. Table S9) between a tightly Cdk2/cyclin A-bound state (*cf.* $K_D$-values in Table 1) and a minor, loosely bound state. To study also the additional fast dynamic processes affecting p27-KID on its transition from the high FRET (HF) to the low FRET (LF) state and within each sub-state, we computed the species-autocorrelation (sACF) and -cross correlation (sCCF) functions of the sub-states depicted in Figure 4E for the FRET pair p27-C54-C93 (SI Section 1.4) (Felekyan et al., 2012a). The recovered dynamic structural fluctuations with relaxation times between 50 ns and 1 ms report on chain (60 ns) and local (~ 1.5 µs, 20 µs and 250 µs, for details of the fFCS parameters, cf. Suppl. Table 9) dynamics. The fact that these dynamics are always present, even in the absence of phosphorylation, demonstrates the inherent dynamic anticipation of the complex, which is manifested in the liberation of local regions of p27-KID from the Cdk2/cyclin A complex, making Y88 and Y74 accessible for phosphorylation. While for all FRET-pairs the relaxation times change only marginally upon mono or dual phosphorylation (*cf.* Suppl. Figure S6 and Table S9), p27-KID liberation results in a marked increase in the relative fraction of the component showing fast chain dynamics in the sCCF and HF-HF sACF and of slow dynamics in the LF-LF sACF. Altogether, smFA and smFRET PDA (Figures 4A-D) and species correlation functions (Figure 4E) consistently indicate a gradual shift of the binding mode from the “bound” state with a higher average FRET efficiency to a “more liberated” state with a lower average FRET efficiency due to progressive phosphorylation of Y88 and Y74.

In summary, based upon integration of biochemical, NMR and diverse smMFD data, p27 bound to Cdk2/cyclin A is best described as a highly dynamic assembly with multiple conformers in a multi-level energy landscape, whose structural properties and binding interfaces are modulated by the degree of phosphorylation. Prior to phosphorylation, Y88 samples lowly populated solvent exposed conformations that enable its phosphorylation by NRTKs. Once Y88 is phosphorylated, the region 83-89 is ejected from the ATP binding pocket of Cdk2 (Grimmler et al., 2007), which – as shown by smMFD data – can still transiently interact with the catalytic pocket of Cdk2. This initial Y phosphorylation partially activates Cdk2 (Grimmler et al., 2007), and also exposes Y74 for phosphorylation, which causes displacement of subdomain D2 from Cdk2 and further kinase activation. Interestingly, the displaced D2 subdomain appears to maintain conformational features similar to the Cdk2-bound state, consistent with past observations with isolated p27-KID (Iconaru et al., 2015; Sivakolundu et al., 2005a). A critical feature of the Y phosphorylation-dependent mechanism that modulates p27-dependent regulation of Cdk2 is that both pY88- and pY88/pY74-phosphorylated p27 remain tethered to the Cdk2/cyclin A assembly through binding of the D1 subdomain to cyclin A, enabling intra-assembly phosphorylation of T187 within the flexible p27 C-terminus. These results solve the enigma regarding how Y88, which directly participates in Cdk2 inhibition, is made accessible for phosphorylation by NRTKs; this residue dynamically anticipates the Y phosphorylated state, enabling the critical first step of the multi-step phosphorylation cascade that controls Cdk2 activity and cell division.

### Thermodynamic and kinetic signatures of the tyrosine phosphorylation-dependent p27 rheostat

To add further dimensions to this model, we performed surface plasmon resonance (SPR) and isothermal titration calorimetry (ITC) experiments and monitored how Y phosphorylation affects interactions between p27 and Cdk2/cyclin A. Phosphorylation of Y residues did not affect the kinetics of p27 association with Cdk2/cyclin A but did slightly but progressively increase the rate of dissociation (Suppl. Figure 7, Suppl. Table 10); similar results were obtained with p27 and p27-KID constructs. These changes were associated with a phosphorylation-dependent increase in apparent $K_D$ values but affinity for Cdk2/cyclin A and cyclin A alone remained high (apparent $K_D$ values of <2.9 nM and <52 nM, respectively), consistent with results from NMR and smMFD. Next, ITC was used to determine the number of residues of Y-phosphorylated p27 that fold upon binding to Cdk2/cyclin A, as demonstrated previously for unmodified p27 (Lacy et al., 2004a). The extent to which the folding of p27 upon binding is reduced is a surrogate for Y phosphorylation-dependent displacement of p27 from Cdk2/cyclin A. For these experiments, we used constructs in which substitution of Y88 or both Y88 and Y74 with glutamic acids mimicked the biochemical effects of phosphorylation. In agreement with the progressive partial release of p27-KID from Cdk2/cyclin A upon phosphorylation, ITC shows different extents of folding upon binding to Cdk2/cyclin A for p27 and the mono and dual Y phosphomimetic forms, Y88E-p27 and Y74E/Y88E-p27 (Table 1, Suppl. Figure 8). With p27 and Y88E-p27, 99 and 82 residues folded, respectively, upon binding to Cdk2/cyclin A (values of $\Re$, Table 1). This result suggests that a 17 residue-long region of subdomain D2 is displaced from Cdk2 in Y88E-p27, consistent with the biochemical Cdk2 inhibition data, and past SPR and NMR data for p27-KID and pY88-p27-KID (Grimmler et al., 2007) on the extent of folding for p27-KID binding to Cdk2/cyclin A (Lacy et al., 2004a). The $K_D$ values for these interactions were identical within error (4.9 ± 1.5 nM and 3.4 ± 1.1 nM, respectively), which suggests a significant decrease in the entropic component and entropy-enthalpy compensation for binding of Y88E-p27 in comparison to p27 ($-T\Delta S$, Table 1). These observations provide strong support for increased dynamics in the bound state upon Y->E mutation. As demonstrated by smFRET experiments with the mono- and dual Y phosphomimetic forms (Suppl. Figure 9), they also show that C-terminal residues outside of the N-terminal KID experienced folding when the KID (comprised of 62 residues) folded upon binding to Cdk2/cyclin A. An even smaller number of residues folded when dual phosphorylation-mimicking Y74E/Y88E-p27 bound to Cdk2/cyclin A (48 residues; Table 1) and the $K_D$ value increased to 14.0 ± 1.6 nM, consistent with the displacement of the entire D2 region from Cdk2 (Lacy et al., 2004a). These conclusions were further supported by NMR data for the two phosphomimetic (Y->E) p27-KID mutants bound to Cdk2/cyclin A, which showed that the Y88 region and the D2 and LH subdomains, respectively, were displaced from Cdk2 by the Y74E and Y74E/Y88E mutations (Suppl. Figure 10). This displacement of the entire D2 subdomain from Cdk2 upon dual Y phosphorylation of p27 explains substantial reactivation of the kinase (50 % of full Cdk2 activity with pY74/pY88-p27-KID and 60 % with Y74E/Y88E-p27, Suppl. Figure 10). Thus, the SPR and ITC data, with supporting biochemical and NMR data, support the model discussed above in which Y phosphorylation displaces portions of inhibitory subdomain D2 of p27 from Cdk2, restoring partial kinase activity, while interactions between subdomain D1 and cyclin A are maintained.

**Tyrosine phosphorylation of p27 alters interactions with cyclin A through cross-complex allostery**

Our collective results thus far reveal features of the conformational landscape of Cdk2/cyclin A-bound p27 that involve interactions of its D2 subdomain with the kinase subunit of the assembly to regulate kinase activity. However, p27 also inhibits substrate phosphorylation by Cdk2/cyclin A through the binding of its D1 subdomain to a conserved surface on cyclin A, inhibiting substrate recruitment (Russo et al., 1996; Schulman et al., 1998) (*cf.* Figure 1). Therefore, we considered the possibility that tyrosine phosphorylation within subdomain D2 could in some way alter interactions between subdomain D1 of p27 and cyclin A and thus affect this second inhibitory mechanism. To test this hypothesis, we utilized p27 constructs labeled on C29 alone (p27-29C) or paired with C54 (p27-C29-54), in smMFD experiments. smFRET experiments showed a major (84%) state with a longer $\langle R_{\mathrm{DA}}\rangle_E$ = 52.3 Å and a minor (16%) state with a shorter $\langle R_{\mathrm{DA}}\rangle_E$ = 43.1 Å distance for p27-C29-54 bound to Cdk2/cyclin A (Figure 4D, Suppl. Figure 5). After mono and dual phosphorylation, the fraction of the shorter $\langle R_{\mathrm{DA}}\rangle_E$ state increased gradually, most probably due to partial release and compaction of the 29-54 region. This is consistent with the dynamic release seen in smFA with BODIPY-labeled p27-C29 bound to Cdk2/cyclin A (Figure 4B, Suppl. Figure 4). In all cases, the major state exhibited an $r_D$ value ≥ 0.25, consistently with C29 binding to cyclin A, whereas the minor state has an $r_D$ value ≤ 0.08, indicative of increased local mobility relative to the major state. Because BODIPY-labeled p27-C40 remained relatively constant upon mono and dual Y phosphorylation, the changes primarily occur at C29 and the fluctuations of subdomain D1 did not propagate into subdomain LH (Figure 4B, Suppl. Figure 4B). We suggest that the increased population of the minor state in the smFRET data for the C29-54 dye pair with phosphorylation of both Y74 and Y88 is due to allosteric effects that increase the dynamic fluctuations in the region around position 29. Hence, we term this long-range effect between D1 and D2 *cross-complex allostery.*

This long-range effect may account, in part, for p27 Y phosphorylation-dependent enhanced phosphorylation of the Cdk2 substrates, Rb and p107, and may be consistent with the non-linear increase of their phosphorylation upon phosphate incorporation into p27 (Figure 2D), as opposed to that of the non-specific substrate H1 (Figure 2A).

## Discussion

**Intrinsic dynamics within a multi-component enzyme assembly is essential for the regulatory function of an IDP**

Our study has addressed a key mechanistic question regarding the roles of IDPs as regulatory switches in signaling pathways; specifically, we addressed how sites within a bound IDP (p27 in complex with Cdk2/cyclin A) can become accessible to, and integrate, regulatory modifications despite apparent steric inaccessibility within a rigid 3D structure (Russo et al., 1996). p27 experiences phosphorylation on two structurally inaccessible tyrosines, which signal p27 degradation and promote cell division (Chu et al., 2007; Grimmler et al., 2007). This is achieved by the intrinsic dynamics of p27, which, within the p27/Cdk2/cyclin A assembly is far from being static. Here, we have shown that the two events are sequential: the phosphorylation of Y88 is required for the subsequent phosphorylation of Y74, resulting in the D2 region to be fully displaced from Cdk2.

By integrating biochemical (enzyme activity and substrate phosphorylation) and biophysical (NMR, X-ray, smMFD, ITC, and SPR) methods, we have discovered that the priming phosphorylation of Y88 is possible because this region structurally fluctuates between a major bound and a minor partially released state compatible with NRTK-mediated phosphorylation (Figure 5). Upon phosphorylation, the solvent-exposed conformation of Y88 becomes more populated, and sterically enables phosphorylation of the minor, partially released state of Y74, which then displaces the entire D2 subdomain and facilitates intra-complex phosphorylation of T187. The activated phosphodegron mediates p27 poly-ubiquitination and degradation and, ultimately, cell cycle progression. In all, tightly bound p27 within its complex with Cdk2/cyclin A *dynamically anticipates* the conformational changes caused by sequential Y phosphorylation (Figure 5).

**Rheostat-like control of a critical cell cycle kinase by an IDP**

These structural fluctuations and remodeling of the conformational ensemble enable Cdk2/cyclin A-bound p27 to actively integrate upstream NRTK signals with rheostat-like precision. This rheostat has four settings: with unphosphorylated p27, Cdk2 is “off”; with pY88-p27, Cdk2 is “20 % on”; and with pY74/pY88-p27, Cdk2 is “50 % on” (Figure 2A), to be turned “100 % on” by T187 phosphorylation and subsequent elimination of p27 by the ubiquitin-proteasome system. The phosphorylation-dependent functional properties of p27 are reminiscent of observations with the folded protein, dihydrofolate reductase (DHFR), that was shown using NMR spectroscopy to dynamically anticipate a sequence of structural changes that accompanied cofactor binding, substrate binding and product release (Boehr et al., 2006). Tuning structural fluctuations by post-translational modification of p27 on Y88 and Y74 through *cross-complex allostery* may also trigger activation of cyclin A in the Cdk2/cyclin A complex, providing access to its substrate-binding pocket to specific substrates (Figure 5). This may be reflected in the non-linear response of substrate modification to p27 phosphorylation, and may be essential for phosphorylation of external substrates, such as Rb and p107. Overall, our data are consistent with a model of sequential remodeling of the conformational energy landscape of bound p27 (Figure 5), which might represent a general mechanistic theme of how IDPs sense, integrate and propagate signals within signaling pathways.

**Author contributions**

M.T. prepared samples for smFL experiments, participated in smFL measurements and analyzed smFL data; H.S. performed smFL experiments, analyzed data and prepared figures; Y.W. performed biochemical assays and NMR experiments, and prepared associated figures, and prepared samples for SPR experiments; S.F. and K.H. analyzed smFL data and prepared associated figures; M-KY and S.W.W. performed X-ray crystallography experiments and structure determination; B.W. performed and analyzed data from SPR experiments and prepared associated tables and figures; C-GP performed molecular biology and protein biochemistry experiments to prepare protein samples for smFL, SPR, ITC and X-ray diffraction experiments; S.V. performed analyzed data from ITC experiments and prepared associated tables and figures; L.I. performed NMR experiments and prepared associated figures; P.T. supervised M.T. and performed smFL experimental design, analyzed data and wrote the paper; C.A.M.S supervised H.S. and performed smFL experimental design, analyzed data and wrote the paper; and R.W.K. supervised the St. Jude co-authors, conceived the project, performed experimental design, analyzed data and wrote the paper.

**Acknowledgements**

This work was supported by the Odysseus grant G.0029.12 from Research Foundation Flanders (FWO) for P.T. M.T. was supported by a Marie-Curie postdoctoral fellowship. H.S. acknowledges a starting fund from Clemson University. C.A.M.S. was supported by the ERC Advanced Grant hybridFRET (671208). R.W.K. was supported by US National Institutes of Health (NIH) grants R01CA082491 and R01GM083159, a US National Cancer Institute Cancer Center Support Grant P30CA21765 (at St. Jude Children's Research Hospital), and ALSAC.

**References**

Adams, P.D. (2001). Regulation of the retinoblastoma tumor suppressor protein by cyclin/cdks. Biochim Biophys Acta *1471*, M123-133.

Adams, P.D., Grosse-Kunstleve, R.W., Hung, L.W., Ioerger, T.R., McCoy, A.J., Moriarty, N.W., Read, R.J., Sacchettini, J.C., Sauter, N.K., and Terwilliger, T.C. (2002). PHENIX: building new software for automated crystallographic structure determination. Acta Crystallogr D Biol Crystallogr *58*, 1948-1954.

Boehr, D.D., McElheny, D., Dyson, H.J., and Wright, P.E. (2006). The dynamic energy landscape of dihydrofolate reductase catalysis. Science *313*, 1638-1642.

Chu, I., Sun, J., Arnaout, A., Kahn, H., Hanna, W., Narod, S., Sun, P., Tan, C.K., Hengst, L., and Slingerland, J. (2007). p27 phosphorylation by Src regulates inhibition of cyclin E-Cdk2. Cell *128*, 281-294.

Das, R.K., Huang, Y., Phillips, A.H., Kriwacki, R.W., and Pappu, R.V. (2016). Cryptic sequence features within the disordered protein p27Kip1 regulate cell cycle signaling. Proc Natl Acad Sci U S A.

Emsley, P., and Cowtan, K. (2004). Coot: model-building tools for molecular graphics. Acta Crystallogr D Biol Crystallogr *60*, 2126-2132.

Felekyan, S., Kalinin, S., Sanabria, H., Valeri, A., and Seidel, C.A. (2012a). Filtered FCS: species auto- and cross-correlation functions highlight binding and dynamics in biomolecules. Chemphyschem *13*, 1036-1053.

Felekyan, S., Kalinin, S., Sanabria, H., Valeri, A., and Seidel, C.A.M. (2012b). Filtered FCS: species auto- and cross-correlation functions highlight binding and dynamics in biomolecules. ChemPhysChem *13*, 1036-1053.

Galea, C.A., Nourse, A., Wang, Y., Sivakolundu, S.G., Heller, W.T., and Kriwacki, R.W. (2008). Role of intrinsic flexibility in signal transduction mediated by the cell cycle regulator, p27 Kip1. J Mol Biol *376*, 827-838.

Grimmler, M., Wang, Y., Mund, T., Cilensek, Z., Keidel, E.M., Waddell, M.B., Jakel, H., Kullmann, M., Kriwacki, R.W., and Hengst, L. (2007). Cdk-inhibitory activity and stability of p27Kip1 are directly regulated by oncogenic tyrosine kinases. Cell *128*, 269-280.

Iconaru, L.I., Ban, D., Bharatham, K., Ramanathan, A., Zhang, W., Shelat, A.A., Zuo, J., and Kriwacki, R.W. (2015). Discovery of Small Molecules that Inhibit the Disordered Protein, p27(Kip1). Scientific reports *5*, 15686.

Kalinin, S., Felekyan, S., Antonik, M., and Seidel, C.A. (2007). Probability distribution analysis of single-molecule fluorescence anisotropy and resonance energy transfer. J Phys Chem B *111*, 10253-10262.

Kalinin, S., Valeri, A., Antonik, M., Felekyan, S., and Seidel, C.A.M. (2010). Detection of structural dynamics by FRET: a photon distribution and fluorescence lifetime analysis of systems with multiple states. J Phys Chem B *114*, 7983-7995.

Lacy, E.R., Filippov, I., Lewis, W.S., Otieno, S., Xiao, L., Weiss, S., Hengst, L., and Kriwacki, R.W. (2004a). p27 binds cyclin-CDK complexes through a sequential mechanism involving binding-induced protein folding. Nat. Struct. Mol. Biol. *11*, 358-364.

Lacy, E.R., Filippov, I., Lewis, W.S., Otieno, S., Xiao, L., Weiss, S., Hengst, L., and Kriwacki, R.W. (2004b). p27 binds cyclin-CDK complexes through a sequential mechanism involving binding-induced protein folding. Nat Struct Mol Biol *11*, 358-364.

Montagnoli, A., Fiore, F., Eytan, E., Carrano, A.C., Draetta, G.F., Hershko, A., and Pagano, M. (1999). Ubiquitination of p27 is regulated by Cdk-dependent phosphorylation and trimeric complex formation. Genes Dev *13*, 1181-1189.

Murshudov, G.N., Vagin, A.A., and Dodson, E.J. (1997). Refinement of macromolecular structures by the maximum-likelihood method. Acta Crystallogr D Biol Crystallogr *53*, 240-255.

Myszka, D.G. (1999). Improving biosensor analysis. J Mol Recognit *12*, 279-284.

Neidhardt, F.C., Bloch, P.L., and Smith, D.F. (1974). Culture medium for enterobacteria. J. Bact. *119*, 736-747.

Otwinowski, Z., and Minor, W. (1997). Processing of X-ray Diffraction Data Collected in Oscillation Mode. Meth. Enzymol. *276*, 307-326.

Russo, A.A., Jeffrey, P.D., Patten, A.K., Massague, J., and Pavletich, N.P. (1996). Crystal structure of the p27Kip1 cyclin-dependent-kinase inhibitor bound to the cyclin A-Cdk2 complex. Nature *382*, 325-331.

Schulman, B.A., Lindstrom, D.L., and Harlow, E. (1998). Substrate recruitment to cyclin-dependent kinase 2 by a multipurpose docking site on cyclin A. Proc Natl Acad Sci U S A *95*, 10453-10458.

Seeliger, M.A., Young, M., Henderson, M.N., Pellicena, P., King, D.S., Falick, A.M., and Kuriyan, J. (2005). High yield bacterial expression of active c-Abl and c-Src tyrosine kinases. Protein Sci *14*, 3135-3139.

Sindbert, S., Kalinin, S., Nguyen, H., Kienzler, A., Clima, L., Bannwarth, W., Appel, B., Muller, S., and Seidel, C.A. (2011). Accurate distance determination of nucleic acids via Forster resonance energy transfer: implications of dye linker length and rigidity. J Am Chem Soc *133*, 2463-2480.

Sisamakis, E., Valeri, A., Kalinin, S., Rothwell, P.J., and Seidel, C.A.M. (2010). Accurate single-molecule FRET studies using multiparameter fluorescence detection. Methods in Enzymology *475*, 456-514.

Sivakolundu, S.G., Bashford, D., and Kriwacki, R.W. (2005a). Disordered p27Kip1 exhibits intrinsic structure resembling the Cdk2/cyclin A-bound conformation. J Mol Biol *353*, 1118-1128.

Sivakolundu, S.G., Bashford, D., and Kriwacki, R.W. (2005b). Disordered p27Kip1 exhibits intrinsic structure resembling the Cdk2/cyclin A-bound conformation. Journal of Molecular Biology *353*, 1118-1128.

Spolar, R.S., and Record, M.T., Jr. (1994). Coupling of local folding to site-specific binding of proteins to DNA. Science *263*, 777-784.

Tompa, P. (2014). Multisteric regulation by structural disorder in modular signaling proteins: an extension of the concept of allostery. Chem Rev *114*, 6715-6732.

Tompa, P., and Fuxreiter, M. (2008). Fuzzy complexes: polymorphism and structural disorder in protein-protein interactions. Trends Biochem Sci *33*, 2-8.

Vagin, A., and Teplyakov, A. (2010). Molecular replacement with MOLREP. Acta Crystallogr D Biol Crystallogr *66*, 22-25.

van der Lee, R., Buljan, M., Lang, B., Weatheritt, R.J., Daughdrill, G.W., Dunker, A.K., Fuxreiter, M., Gough, J., Gsponer, J., Jones, D.T.*, et al.* (2014). Classification of intrinsically disordered regions and proteins. Chem Rev *114*, 6589-6631.

Van Roey, K., Dinkel, H., Weatheritt, R.J., Gibson, T.J., and Davey, N.E. (2013). The switches.ELM resource: a compendium of conditional regulatory interaction interfaces. Sci Signal *6*, rs7.

Wright, P.E., and Dyson, H.J. (2009). Linking folding and binding. Curr Opin Struct Biol *19*, 31-38.

**Figure Legends**

**Figure 1. Domain structure of p27.** (A) Full-length p27 contains the kinase inhibitory domain (KID), which is subdivided into domain 1 (D1), linker helix (LH) and domain 2 (D2). Tyrosine phosphorylation sites are labeled as Y74 and Y88 and the threonine phosphorylation site is labeled as T187. The labeling positions of cysteine residues are indicated as C29, C40, C54, C75, C93 and C110. (B) The structure of p27 in complex with Cdk2/cyclin A based on pdb:1JSU (Russo et al., 1996) and molecular dynamics simulations (Sivakolundu et al., 2005b). Labeled sites are represented as circles, green only for smFA experiments and green/red circles where used in both smFA and smFRET experiments.

**Figure 2. Incremental phosphorylation of tyrosine residues in p27 exerts rheostat-like control over Cdk2/cyclin A and promotes phosphorylation of p27 on T187.** (A) p27-KID completely inhibits the kinase activity of Cdk2/cyclin A toward histone H1 (squares; IC50 value, 2.8 ± 0.4 nM) while pY88-p27-KID (circles; IC50 value, 12 ± 1 nM) and pY74/pY88-p27-KID (circles; IC50 value, 29 ± 7 nM) are associated with 20% and 43% residual Cdk2 activity at saturating concentrations. (B) Y88 and Y74 are sequentially phosphorylated by Src-KD. Various p27-KID constructs were used as sustrates for tyrosine phosporylation by Src-KD in the absence (left lanes) or presence (right lanes) of Cdk2/cyclin A. The substrate, pY88-p27-KID, was prepared by prior treatment with ABL-KD. (C) Phosphorylation of Y88 (purple) and Y74 and Y88 (yellow) enables intra-complex phosphorylation of p27 on T187 within ternary complexes with Cdk2/cyclin A. T187 within unphosphorylated p27 is phosphorylated by Cdk2/cyclin A to a small extent due to intermolecular reactions (Grimmler et al., 2007). (D) The phosphorylation by Cdk2 of T187 is enhanced to a greater extent by phosphorylation of Y88 or Y74 and Y88 within p27 (that is bound to Cdk2/cyclin A) than phosphorylation of the inter-molecular Cdk2 substrates, Rb and p107. Fold enhancements of phosphorylation of T187 [in the presence of Rb (top) and p107 (bottom)], Rb and p107 are normalized to the extent of phosphorylation observed with unphosphorylated (on Y74 and Y88) p27.

**Figure 3. Tyrosine phosphorylation-dependent displacement of portions of the D2 region of p27 from Cdk2 mediates kinase reactivation.** (A) NMR analysis of the influence of tyrosine phosphorylation on interactions between p27-KID and Cdk2/cyclin A. Chemical shift differences for residues in non-phosphorylated (top) and tyrosine phosphorylated (pY88-p27-KID, middle; and pY74/pY88-p27-KID, bottom) p27-KID bound to Cdk2/cyclin A. Residues near Y88 and within the entire D2 subdomain adopt free state-like conformations in pY88-p27-KID and pY74/pY88-p27-KID, respectively. $\Delta\delta$ values were calculated using the equation: $\Delta\delta = [(\Delta\delta\ ^{1}H_N)^2 + 0.0289 \times (\Delta\delta\ ^{15}N_H)^2]^{1/2}$.

(B) Views of the region of Cdk2 (cyan) bound by p27 (orange) in various structures of Cdk2/cyclin A: Cdk2/cyclin A/ATP (PDB 1JST, left; ATP is shown in ball and stick format), Cdk2/cyclin A/p27-KID-ΔC (determined in this study, middle; a model of ATP is shown in semi-transparent ball and stick format), and Cdk2/cyclin A/p27-KID (PDB 1JSU; the position of ATP in the absence of p27 is shown in semi-transparent ball and stick format). Both p27-KID-ΔC and p27-KID displace β-strand 1 (β1) from the N-terminal β-sheet of Cdk2 but, due to deletion of residues displaced by phosphorylation of Y88 in p27-KID-ΔC, the active site is accessible to ATP (modeled based upon the Cdk2/cyclin A/ATP structure). However, displacement of the β1 strand of Cdk2 (including residues of the G-loop) in the Cdk2/cyclin A/p27-KID-ΔC structure gives rise to an incompletely formed ATP binding pocket, consistent with the limited biochemical activity of Cdk2 within this complex.

**Figure 4. Local dynamic fluctuations of tyrosine residues within p27 bound to Cdk2/cyclin A.**

(A) Time-window analysis (Δt =2 ms) for BODIPY-labeled p27-C93 (blue) and pY88-p27-C93 (purple) bound to Cdk2/cyclin A is shown as a two-dimensional histogram of scatter-corrected fluorescence anisotropy ($r_D$) *vs.* the fluorescence-weighted average BODIPY fluorescence lifetime ($\langle \tau_{D(0)} \rangle_f$). One-dimensional histograms are the projected distributions over a single variable. Probability distribution analysis (PDA) is used to fit the $r_D$ distributions with two shot-noise limited states (high $r_D$ and low $r_D$, light and dark colors, respectively). These values are used to calculate the proper rotational correlation times for each state using Perrin's equation (Suppl. Information, eq. (9); Suppl. Table 1-3; Figure 4). (B) Global fits using PDA for three time-windows (Δt =1 ms, 2 ms and 3 ms) are presented for each single Cys variant (C29, C40, C54, C75 and C93). The corresponding fractions and anisotropy values for the low and high $r_D$ are presented in light and dark colors, respectively. (C) Time-window analysis (Δt =3 ms) for dual labeled p27-C54-93 (blue) and pY88-p27-C54-93 (purple) bound to Cdk2/cyclin A is shown as two-dimensional histogram of the FRET averaged donor-acceptor distance ($\langle R_{DA} \rangle_E$) *vs.* the fluorescence-weighted average fluorescence donor lifetime in presence of acceptor ($\langle \tau_{D(A)} \rangle_f$). One-dimensional histograms are the projected distributions over a single variable. Grey shaded area indicates area of Donor-only labeled molecules. PDA is used to fit $\langle R_{DA} \rangle_E$ distributions with two shot-noise limited inter-dye distances (low $\langle R_{DA} \rangle_E$ and high $\langle R_{DA} \rangle_E$, light and dark colors, respectively). Horizontal lines are added for visual identification of both mean FRET states (Suppl. Table 5-6, Figure 5). Static FRET lines calculated based on the fluorescence dye properties and PDA are shown in green (Suppl. Table 4): No P: $\langle R_{DA} \rangle_E$ = $(1/(3.8306/(0.0079*\langle \tau_{D(A)} \rangle_f^2+1.0179*\langle \tau_{D(A)} \rangle_f +-0.1618)-1))^{(1/6)}*53.0$; pY88: $\langle R_{DA} \rangle_E$ = $(1/(3.9997/(0.0202*\langle \tau_{D(A)} \rangle_f^2+0.9655*\langle \tau_{D(A)} \rangle_f +-0.1527)-1))^{(1/6)}*53.0$. (D) Global fits using PDA for three time-windows (Δt =1 ms, 2 ms and 3 ms) are presented for each of the FRET-samples (C29-54, C54-93, and C75-110). The corresponding fractions and inter-dye distances for low $\langle R_{DA} \rangle_E$ and high $\langle R_{DA} \rangle_E$ are presented in light and dark colors, respectively. Error bars on $\langle R_{DA} \rangle_E$ represent the half-width of the distribution of each state when fitting with PDA. (E) The sACFs and sCCFs of filtered fluorescence correlation analysis for p27-C54-93 in complex with Cdk2/cyclin A without phosphorylation and with phosphorylated Y88 map the complex multi-level dynamics of p27 reflecting the change of chain mobility (fit results see Suppl. Table 9).

**Figure 5. Structural dynamics within Cdk2/cyclinA/p27 complex.** p27 bound to Cdk2/cyclinA can be best described as a highly dynamic ensemble with multiple conformational states (approximated here by 4) in a multi-level energy landscape, whose structural properties and binding interfaces are tuned by the degree of phosphorylation. Prior to phosphorylation, Y88 already samples solvent exposed conformations that enable its phosphorylation by NRTKs (by Abl). This initial Y phosphorylation (of Y88, marked in purple) partially activates Cdk2, exposes Y74 for phosphorylation (by Src, marked in yellow), which causes displacement of subdomain D2 from Cdk2 (while preserving its local compact conformation), further kinase activation and intramolecular phosphorylation of T187, also facilitated by partial release of the substrate-binding region on cyclin A. Flexibility is color coded from rigid (blue) to flexible (red).

**Table 1. Equilibrium dissociation constant ($K_D$) values and thermodynamic parameters for the binding of p27, Y88E-p27 and Y74E/Y88E-p27 to Cdk2/cyclin A determined at 25 °C using isothermal titration calorimetry (ITC).**

The ΔCp values were determined from ΔH values measured at 5 °C, 10 °C, 15 °C, 20 °C, and 25 °C and are reported in Suppl. Figure 8.

| p27 species binding to Cdk2/cyclin A | $K_D$ (nM) | ΔG (kcal $mol^{-1}$) | ΔH (kcal $mol^{-1}$) | -TΔS (kcal $mol^{-1}$) | ΔCp (cal $mol^{-1}$ $K^{-1}$) | $\mathfrak{R}$ (# of residues that fold) |
|---|---|---|---|---|---|---|
| p27 | 4.9 ± 1.5 | -11.4 ± 0.2 | -50.6 ± 1.2 | +39.2 ± 1.2 | -1267 ± 108 | 99 |
| Y88E-p27 | 3.4 ± 1.1 | -11.6 ± 0.2 | -38.5 ± 1.1 | +26.9 ± 1.0 | -1119 ± 36 | 82 |
| Y74E/Y88E-p27 | 14.0 ± 1.6 | -10.8 ± 0.1 | -34.4 ± 0.5 | +23.6 ± 0.4 | -630 ± 29 | 48 |

**Supplemental Information**

**Table of Contents**

**Online content**

Methods, along with any additonal Extended Data display items and Source Data, are available in the online version of the paper. References unique to this section only appear in the online paper.

**Online Methods**

**Protein expression and purification**

cDNA for residues 22–104 (p27-KID), and 1–198 (full length) of human p27 were sub-cloned into pET28a (Novagen). A p27 construct in which Tyrosine 89 was mutated to F (Y89F) was used in our studies because cellular ABL phosphorylates only Y88 but the ABL kinase domain (ABL-KD) phosphorylates both Y88 and Y89 *in vitro* (Grimmler et al., 2007). Non-physiological phosphorylation of Y89 is eliminated with the p27-KID-Y89F construct, hereafter referred to as p27-KID. p27-KID-ΔC was prepared by deletion of residues 80-94 from p27-KID in the pET28a expression vector. To mimic tyrosine phosphorylation, we prepared constructs with tyrosine 88 (Y88) or both Y88 and tyrosine 74 (Y74) mutated to glutamate (Y88E or Y88E/Y74E) using site directed mutagenesis. pET28a-based expression vectors for p27 constructs with one or two non-native Cys residues for smMFD experiments were prepared using site-directed mutagenesis starting with p27-Y89F. First, four wild-type cysteine (Cys) residues were mutated to Ala or Ser, as follows: Cys 29 to Ala, Cys 49 to Ala, Cys 99 to Ser, and Cys 148 to Ser (termed Cys-less p27). One construct was prepared with only the latter three mutations, leaving a single native Cys residue at position (p27-C29). Then, constructs with one or two non-native Cys residues were prepared in the Cys-less p27 mutant background, as follows: Glu 40 to Cys, p27-C40; Glu 54 to Cys, p27-C54; Arg 93 to Cys, p27-C93; Cys 29 and Glu 54 to Cys, p27-C29-54; Glu 54 to Cys and Arg 93 to Cys, p27-C54-93; and Glu 75 to Cys and Ser 110 to Cys, p27-C75-110. The various mono- and dual-Cys mutant p27 constructs were shown to exhibit Cdk2/cyclin A inhibitory activity indistinguishable from that of p27-Y89F (data not shown). The various p27 constructs, human Cdk2, T160-phosphorylated Cdk2, and truncated human cyclin A (residues 173–432 of human cyclin A) were expressed in *E. coli* and purified as described (Lacy et al., 2004b). The SH3 and kinase domains of murine ABL (termed ABL-KD), and the C-terminal domain of chicken Src (residues 251-533; termed Src-KD), were amplified by PCR, sub-cloned into pET28a, expressed, and purified from *E. coli* using established procedures (Seeliger et al., 2005). The C-terminus of human Rb (residues 713-928) was sub-cloned into pET28a with an N-terminal, non-cleavable fusion tag (NusA), expressed and purified from *E. coli* using $Ni^{2+}$-affinity chromatography (termed Rb). The C-terminus of human p107 (residues 385-949) was expressed with glutatione S-transferase (GST) and $(His)_6$ fusion tags in *E. coli* and purified with a GST affinity column (termed p107). p27 constructs phosphorylated on Y88 were prepared through incubation with ABL-KD at 30 °C for 2 hours followed by $Fe^{3+}$-affinity chromatography. p27 constructs doubly phosphorylated on Y74 and Y88 were prepared through incubation with ABL-KD followed by incubation with Src-KD. Isotope-labeled samples of p27-KID and p27 for NMR experiments were prepared in *E. coli* as previously described (Grimmler et al., 2007) using MOPS-based minimal media (Neidhardt et al., 1974) and $^{15}N$-ammonium chloride, $^{2}H/^{13}C$-glucose, and $^{2}H_2O$.

**_In Vitro_ Cdk2 kinase activity assays and kinetic studies of T187 phosphorylation**

The inhibitory profiles of p27 and p27-KID constructs in which either Y88 or Y74 and Y88 were phosphorylated or mutated to glutamate were determined by measuring *in vitro* Cdk2/cyclin A kinase activity toward histone H1 while titrating the different p27 species (Grimmler et al., 2007). $IC_{50}$ values were determined using nonlinear regression analysis using Graphpad Prism software. Kinetic analysis of the T187 phosphorylation reaction was performed as described previously (Grimmler et al., 2007) with five different concentrations (0.25, 0.5, 1.0, 2.0, and 4.0 μM) with ternary complex of p27, pY88-p27 or pY74/pY88-p27 with cyclin A/Cdk2. These reagents were equilibrated with γ-[$^{32}P$]-ATP for four time intervals (20, 40, 60, and 120 min) and were followed by analysis using SDS-PAGE and phosphoimager analysis (GE Healthcare, Piscataway, NJ) of $^{32}PO_4$-T187 within the p27 species. To compare the influence of p27 tyrosine phosphorylation on Cdk2-dependent phosphorylation of the intra-complex substrate, T187, and two inter-molecular substrates, Rb and p107, we performed the T187 phosphorylation reactions (using 1.0 and 2.0 μM of the ternary complexes) described above in the presence of equimolar amounts of either Rb or p107 for 40 minutes. The phosphorylation on T187 and either Rb or p107 was quantitated by analysis using SDS-PAGE and phosphoimager analysis. In addition, to determine the order of phosphorylation of Y74 and Y88 of p27, we performed phosphorylation reactions using 3 μM Src-KD, 5 μM p27 species (p27, Y88F-p27 or pY88-p27), 6 μM Cdk2/cyclin A and γ-[$^{32}P$]-ATP. The phosphorylation of tyrosine residues of p27 was determined using SDS-PAGE and phosphoimager analysis.

**X-ray crystallography**

p27-KID-ΔC and the Cdk2/cyclin A binary complex were mixed at a 1.1:1.0 mole ratio and concentrated to ~15 mg/ml in 20 mM HEPES, pH 7.5, 5 mM DTT, and 300 mM NaCl by ultrafiltration. The ternary Cdk2/cyclin A/p27-KID-ΔC complex was isolated using gel filtration chromatography (Superdex 200, GE Healthcare, Piscataway, NJ) using the same buffer followed by concentration by ultrafiltration to 15 mg/ml. Crystals of Cdk2/cyclin A/p27-KID-ΔC were grown by the hanging drop vapor diffusion method using the following reservoir solution: 0.1 M HEPES, pH 7.2, 1.4 M lithium sulfate, and 4 mM DTT. The hanging drops contained equal volumes of the reservoir and the protein solution (15 mg/ml in gel filtration buffer) and were equilibrated against the reservoir solution at 18 °C. Crystals were cryoprotected with 24% glycerol and flash frozen in liquid nitrogen. Diffraction data were collected at the Southeast Regional Collaborative Access Team (SER-CAT) 22-BM beamline at the Advanced Photon Source and processed using HKL2000 (Otwinowski and Minor, 1997). The structure was determined by molecular replacement using MOLREP (Vagin and Teplyakov, 2010) with the Cdk2/cyclin A/p27-KID structure (PDB ID 1JSU) as a search model. The structure was refined and optimized using REFMAC (Murshudov et al., 1997), PHENIX (Adams et al., 2002) and COOT (Emsley and Cowtan, 2004). Data collection and refinement statistics are summarized in Suppl. Table 1. The atomic coordinates for the Cdk2/cyclin A/p27-KID-ΔC complex have been deposited and validated by the Protein Data Bank (PDB 6ATH) for release upon publication.

**NMR spectroscopy**

$^{2}H/^{15}N$-labeled p27-KID constructs (p27-KID, pY88-p27-KID and pY74/pY88-p27) and their complexes with Cdk2/cyclin A were analyzed using NMR spectroscopy at 800 MHz using methods described previously (Grimmler et al, 2007). Briefly, the samples were dissolved in 20 mM potassium phosphate, pH 6.5, 50 mM arginine, 8% (v/v) $^{2}H_2O$, 5 mM DTT-$D_6$ and 0.02% (w/v) sodium azide. The isolated p27-KID constructs were analyzed using 2D $^{1}H$-$^{15}N$ HSQC and their complexes with Cdk2/cyclin A using 2D $^{1}H$-$^{15}N$ TROSY at 35 °C using a Bruker Avance 800 MHz spectrometer equipped with cryogenically-cooled TCI probe. The spectra were interpreted based upon previously established resonance assignments (Grimmler et al., 2007). The NMR experiments with the $^{2}H/^{15}N$-labeled Y88E and Y74E/Y88E variants of p27-KID were performed similarly and resonances assigned by inspection. Assignment ambiguities were resolved through analysis of 3D HNCA, HN(CO)CA, HNCACB, HN(CO)CACB spectra of the isolated $^{13}C/^{15}N$-labeled Y to E mutant p27-KID constructs.

**Surface plasmon resonance**

Binding studies were performed at 25 °C using a BIACORE 3000 optical biosensor (GE Healthcare). Unphosphorylated and tyrosine-phosphorylated p27-KID constructs were covalently attached to a carboxymethyl dextran-coated gold surface (CM-4 Chip; GE Healthcare). The carboxymethyl groups of dextran were activated with *N*-ethyl-*N*´-(3-dimethylaminopropyl) carbodiimide (EDC) and *N*-hydroxysuccinimide (NHS), and p27-KID constructs were attached at pH 4.0 in 10 mM sodium acetate. Any remaining reactive sites were blocked by reaction with ethanolamine.

The kinetics of association and dissociation were monitored at a flow rate of 50 µl/min. Cdk2/cyclin A, cyclin A, and Cdk2 were prepared in 10 mM Tris (pH 8.0), 300 mM NaCl, 1 mM EGTA, 5 mM DTT, 0.1 mg/mL bovine serum albumin, and 0.005% Tween20. Binding was measured for concentration ranges of 391 pM – 25 nM for Cdk2/cyclin A, 6.3 – 400 nM for cyclin A, and 31 nM – 2 µM for Cdk2. To account for injection artifacts, a series of sensorgrams was recorded throughout the experiment after injecting only buffer (blank injections). The chip surface was regenerated with a 30 s injection of 6 M guanidine-HCl. Data reported are the difference in SPR signal between the flow cells containing the p27-KID constructs and a reference cell lacking these constructs. Additional instrumental contributions to the signal were removed by subtraction of the average signal of the blank injections from the reference-subtracted signal (Myszka, 1999). Triplicate injections were made at each concentration, and the data were analyzed globally by simultaneously fitting association and dissociation phases at all concentrations using the program Scrubber2 (Version 2.0b, BioLogic Software). The kinetic rate constants were determined by fitting the data to a 1:1 (Langmuir) interaction model. Equilibrium dissociation constants ($K_D$) were calculated as the quotient $k_d/k_a$.

**Isothermal titration calorimetry**

Thermodynamic parameters for the interaction of p27, Y88E-p27 and Y74E/Y88E-p27 with Cdk2/cyclin A were measured using a MicroCal auto-iTC 200 (Malvern Instruments) isothermal titration calorimeter. Protein samples were exchanged into 20 mM HEPES (pH 7.5), 300 mM NaCl and 2 mM Tris (2-carboxyethyl) phosphine prior to the experiments. Titrations were performed by first injecting 0.5 µl of 10 µM p27 or p27 Y to E variants into a solution of 1 µM Cdk2/cyclin A binary complex, followed by additional 3.1 µl or 2.5 µl injections. Experiments were carried out from 5 °C to 25 °C in intervals of 5 °C. Results were analyzed using Origin software (OriginLab). Equilibrium dissociation constant values ($K_D$) and thermodynamic parameters were determined fitting the data to a single-site binding model using a nonlinear least-squares fitting algorithm. The reported values are averages and standard deviations of the mean for three replicates. The values of the heat capacity change ($\Delta Cp$) associated with the different p27 species binding to Cdk2/cyclin A were determined as the slope of the temperature dependence of the enthalpy change of binding ($\Delta H$) values and the numbers of residues that folded upon binding ($\mathfrak{R}$) were determined using the formalism of Spolar and Record (Spolar and Record, 1994), as previously described (Lacy et al., 2004b).

**Production of dye-labeled p27 for single-molecule fluorescence studies**

We prepared p27 constructs in which most or all native cysteine (Cys) residues (at positions 29, 49, 99, and 148) were replaced by serine or alanine, and non-native Cys residues were introduced to allow specific labeling with fluorescent dyes. Before labeling the proteins, all buffers were sterile filtered and degassed. p27 was concentrated to 50-70 µM in buffer A (20 mM Tris-HCl pH 8, 10 mM NaCl) with 10 mM DTT. 2.5 ml of concentrated protein were loaded onto PD10 column and the protein was eluted with freshly degassed 3.5 ml of buffer A without DTT. The eluted p27 was first labeled with the acceptor Alexa Fluor 647 maleimide fluorophore (Invitrogen) at a 1:1 ratio, followed by labeling with the donor Alexa Fluor 488 maleimide fluorophore (Invitrogen) at 1:2 ratio. The dual-labeled p27 was separated from the homo-labeled and unlabeled species using ion-exchange chromatography. p27 Y88E and Y88E/Y74E mutants were prepared in the same way. The presence of energy transfer was analyzed upon excitation at 485 nm of the donor and acceptor dyes at 519 nm and 666 nm, respectively, using Perkin Elmar LS55 Luminescence Spectrometer. Single-cysteine variants for anisotropy measurements were labeled with 2x-access of BODIPY-FL or Alexa Fluor 488 (Invitrogen) and purified as above. Labeled p27 was then analyzed by gel-filtration chromatography at 1 µM protein concentration; the protein eluted at a volume expected for the monomer without any evidence of oligomerization or degradation. Labeled p27 was phosphorylated with ABL-KD and Src-KD in the same way as non-labelled protein. To ensure the maximal efficiency of phosphorylation, phosphorylated p27 was separated from non-phosphorylated p27 using Pro-Q® Diamond Phosphoprotein Enrichment Kit (Invitrogen) and the results were analyzed by Western-blot using Phospho-Tyrosine Mouse mAb (Bioke). Labeled p27 showed inhibition in Cdk2 activity assays.

**Single-molecule Multiparameter Fluorescence Detection (smMFD)**

smMFD for confocal high-precision Förster Resonance Energy Transfer (hpFRET) studies of single-molecules was done using a 485 nm diode laser (LDH-D-C 485 PicoQuant, Germany, operating at 64 MHz, power at objective 110 µW) exciting freely diffusing labeled molecules that passed through a detection volume of the 60X, 1.2 NA collar (0.17) corrected Olympus objective. The emitted fluorescence signal was collected through the same objective and spatially filtered using a 100 µm pinhole, to define an effective confocal detection volume. We used a new detection and data registration scheme to measure dead time free species cross correlation functions. For that, the signal was divided into parallel and perpendicular components at two different colors ("green" and "red") through band pass filters, HQ 520/35 and HQ 720/150, for green and red respectively, and split further with 50/50 beam splitters. In total eight photon-detectors are used, four for green (τ-SPAD, PicoQuant, Germany) and four for red channels (APD SPCM-AQR-14, Perkin Elmer, Germany). A time-correlated single-photon counting (TCSPC) module 5 (HydraHarp 400, PicoQuant, Germany) with Time-Tagged Time-Resolved (TTTR) mode and 8 synchronized input channels was used for data registration.

For smMFD (smFA and smFRET) measurements, samples were diluted (buffer used was 20 mM HEPES, pH 7.5, 300 mM NaCl, 40 µM TROLOX) to pM concentration assuring ~1 burst per second. In long measurements, to avoid drying out of the immersion water, an oil immersion liquid with refraction index of water was used (Immersol, Carl Zeiss Inc., Germany). NUNC chambers (Lab-Tek,

Thermo Scientific, Germany) were used with 500 µL sample volume. Standard controls consisted of measuring water to determine the instrument response function (IRF), buffer for background subtraction, and the nM concentration of green and red standard dyes (Rh110 and Rh101) in water solutions for calibration of green and red channels, respectively. To calibrate the detection efficiencies we used a mixture solution of dual-labeled DNA oligonucleotides with known distance separation between donor and acceptor dyes.

**Fluorescence analysis of smFA and smFRET experiments**

smFA and smFRET experiments were analyzed using smMFD (Sisamakis et al., 2010). We used Probability Distribution Analysis (PDA) (Kalinin et al., 2007; Kalinin et al., 2010) to determine the anisotropy and distance distributions and their corresponding uncertainties. Details on the analysis are given in Supplemental Methods. Filtered Fluorescence Correlation Spectroscopy (fFCS) (Felekyan et al., 2012b) was used to identify the species-specific interconversion rates in smFRET experiments. Details are given in Supplemental Methods and Tables. The software used to perform the analysis, written in house, can be downloaded from http://www.mpc.hhu.de/software.html.

Figure 1

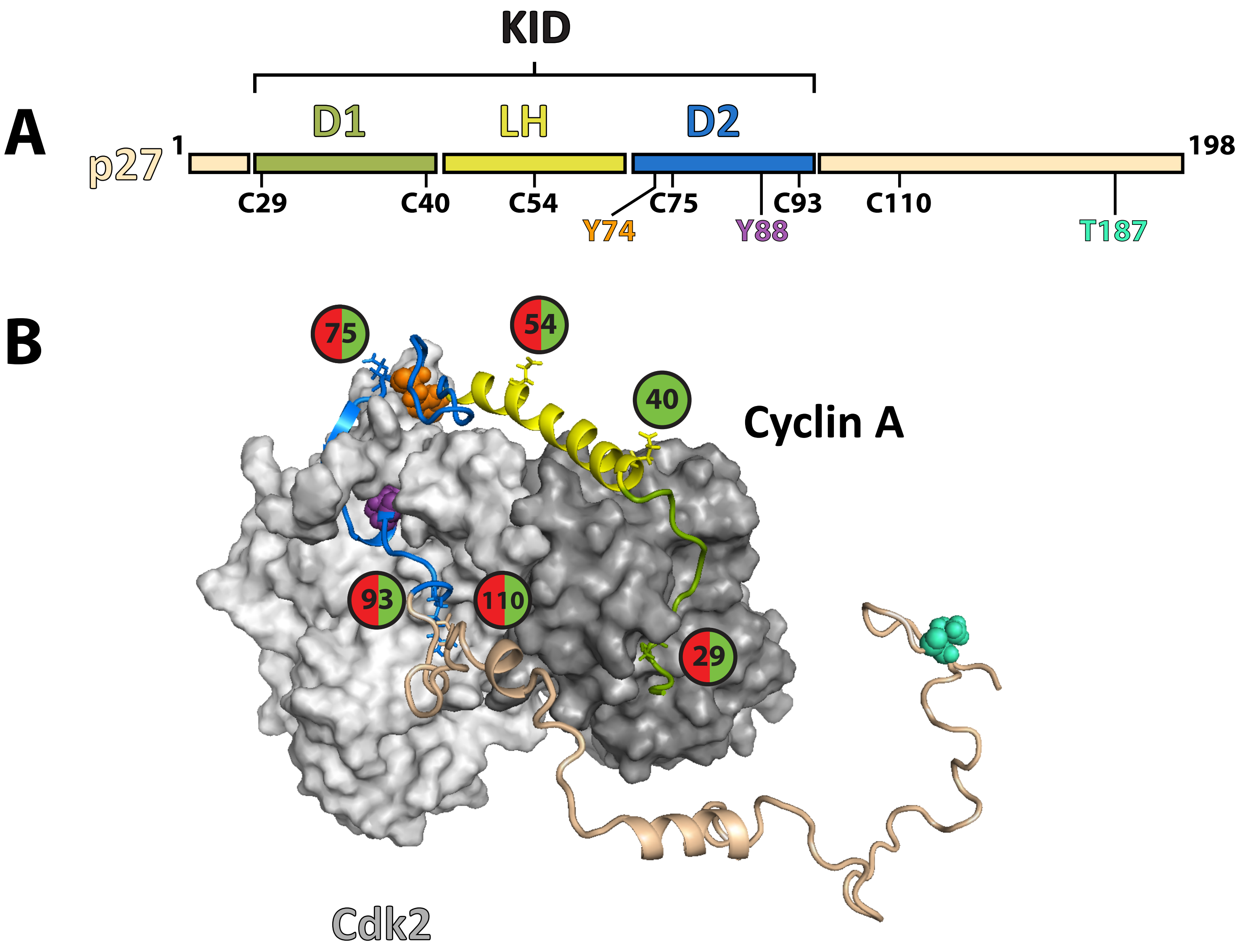

# Figure 2

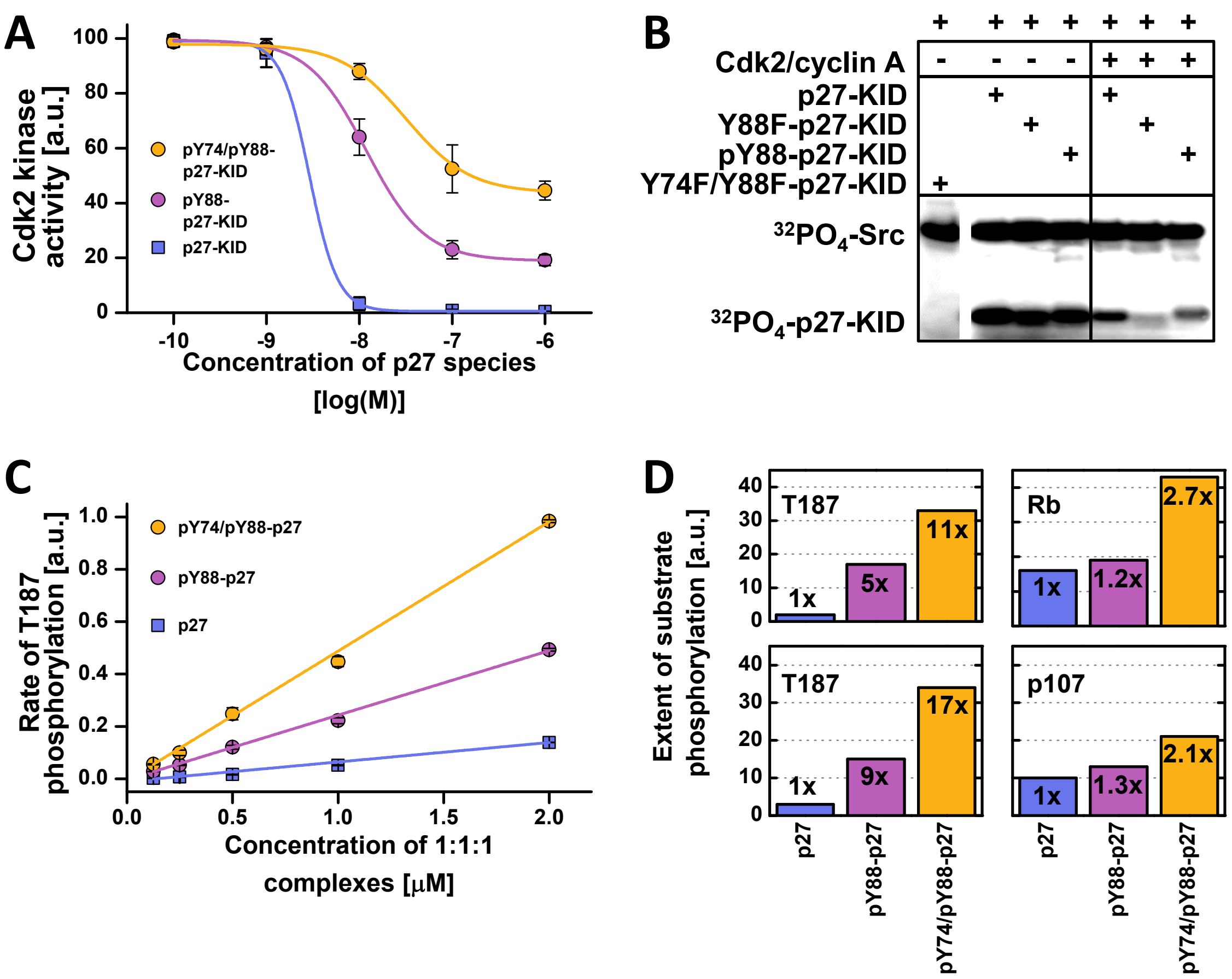

# Figure 3

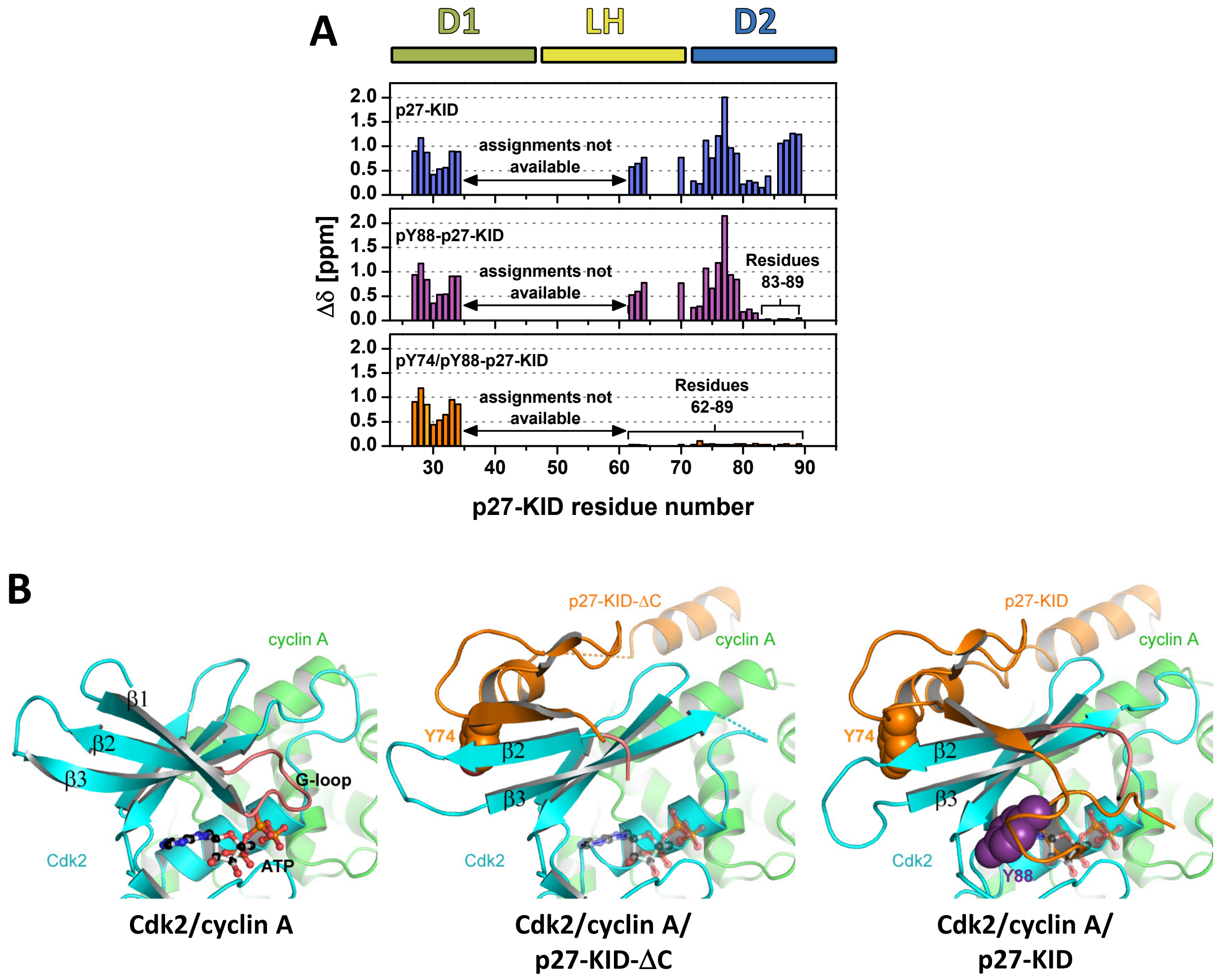

# Figure 4

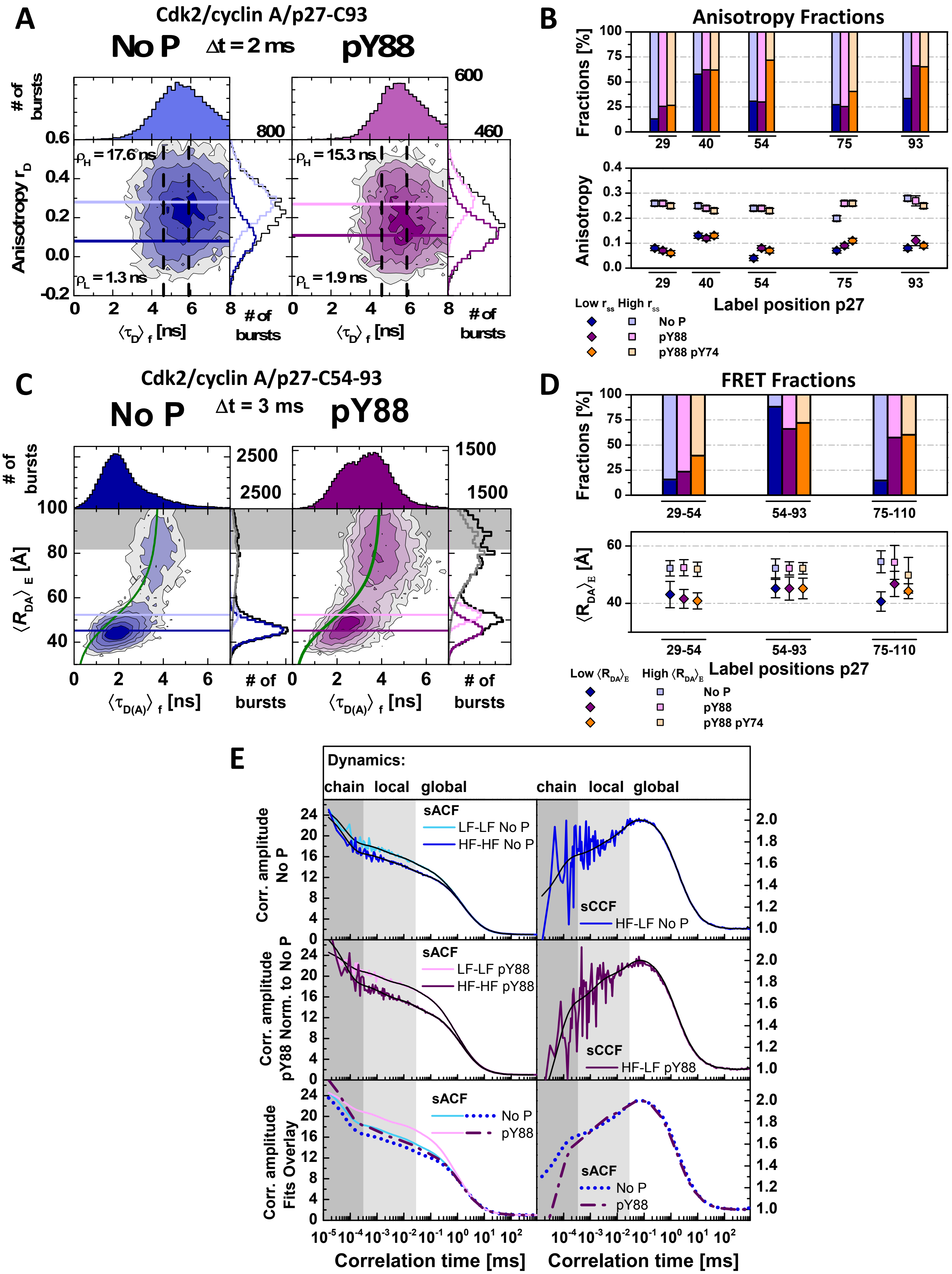
A
Cdk2/cyclin A/p27-C93
No P
Δt = 2 ms
pY88
# of bursts
Anisotropy r_D
ρ_H = 17.6 ns
ρ_L = 1.3 ns
ρ_H = 15.3 ns
ρ_L = 1.9 ns
⟨τ_D⟩_f [ns]
# of bursts
B
Anisotropy Fractions
Fractions [%]
Anisotropy
Label position p27
Low r_ss High r_ss
No P
pY88
pY88 pY74
C
Cdk2/cyclin A/p27-C54-93
No P
Δt = 3 ms
pY88
⟨R_DA⟩_E [Å]
⟨τ_D(A)⟩_f [ns]
D
FRET Fractions
29-54
54-93
75-110
Label positions p27
Low ⟨R_DA⟩_E High ⟨R_DA⟩_E
E
Dynamics:
chain local global
sACF
LF-LF No P
HF-HF No P
sCCF
HF-LF No P
LF-LF pY88
HF-HF pY88
HF-LF pY88
Corr. amplitude No P
Corr. amplitude pY88 Norm. to No P
Corr. amplitude Fits Overlay
Correlation time [ms]

# Figure 5

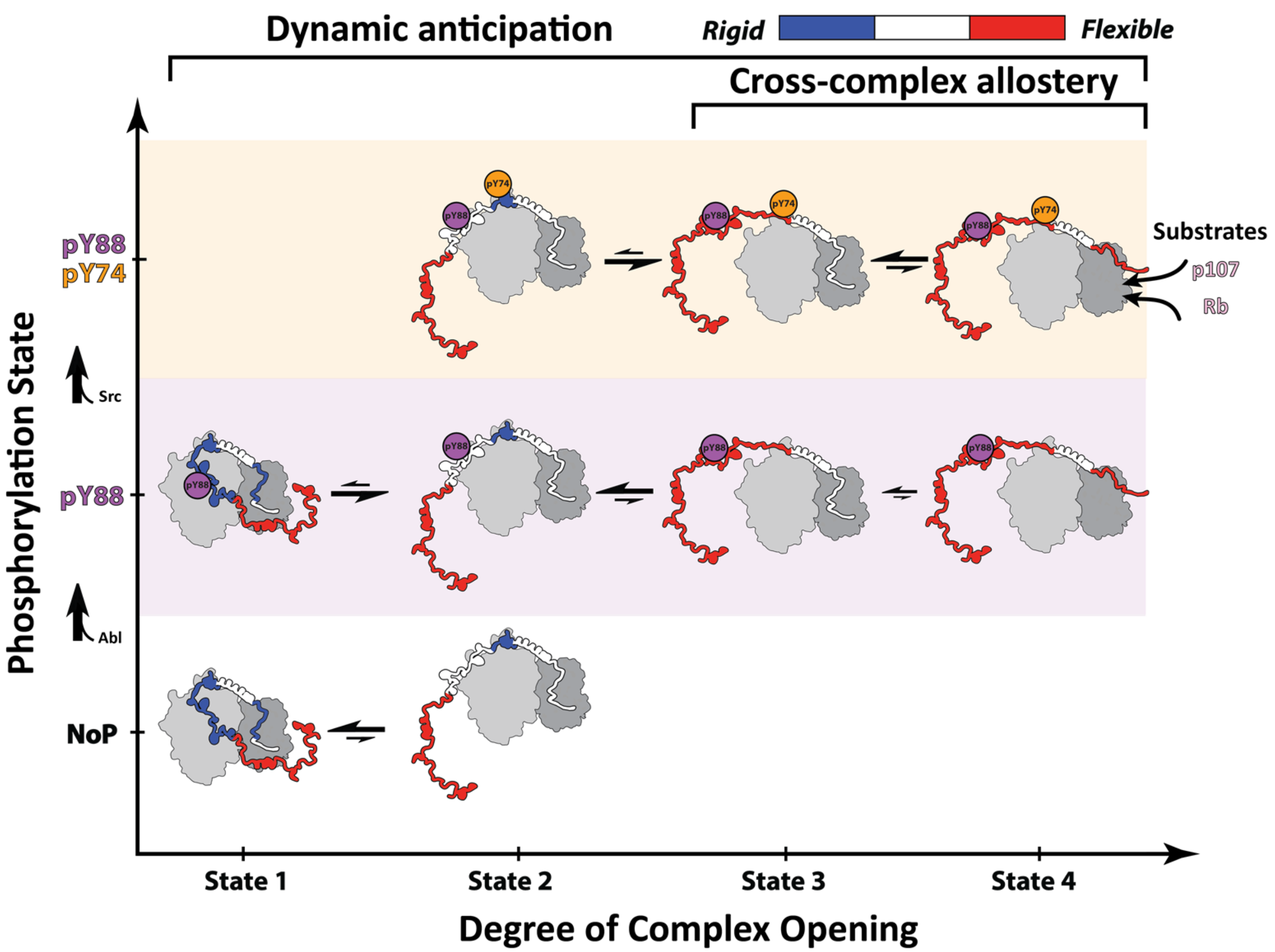

# Dynamic anticipation by Cdk2/Cyclin A-bound p27 mediates signal integration in cell cycle regulation

Maksym Tsytlonok[1]*, Hugo Sanabria[2,3]*, Yuefeng Wang[4,5]*, Suren Felekyan[3], Katherina Hemmen[3], Mi-Kyung Yun[4], Brett Waddell[4,6], Cheon-Gil Park[4], Sivaraja Vaithiyalingam[4,6], Luigi Iconaru[4], Stephen W. White[4], Peter Tompa[1,7]**, Claus A. M. Seidel[3]**, and Richard Kriwacki[4,8]**

[1]VIB Center for Structural Biology, Vrije Universiteit Brussel, Brussels, Belgium.

[2]Department of Physics and Astronomy, Clemson University, Clemson, SC, 29634, USA.

[3]Lehrstuhl für Molekulare Physikalische Chemie, Heinrich-Heine-Universität, Düsseldorf, 40225, Germany.

[4]Department of Structural Biology, St. Jude Children's Research Hospital, 262 Danny Thomas Place, Memphis, TN 38105, USA.

[5]Current address: Department of Radiation Oncology, West Cancer Center, University of Tennessee Health Sciences Center, Memphis, TN 38134, USA

[6]Molecular Interaction Analysis Shared Resource, St. Jude Children's Research Hospital, 262 Danny Thomas Place, Memphis, TN 38103, USA.

[7]Institute of Enzymology, Research Centre for Natural Sciences of the Hungarian Academy of Sciences, Budapest, Hungary.

[8]Department of Microbiology, Immunology and Biochemistry, University of Tennessee Health Sciences Center, Memphis, TN 38163, USA.

*, ** these authors contributed equally to the work

**Correspondence: tompa@enzim.hu (PT), cseidel@hhu.de (CAMS) and richard.kriwacki@stjude.org (RK)

**Supplemental Information**

**Table of Contents**

## S1. Supplemental Methods

### S1.1. Burst analysis and parametric lines for FRET and anisotropy multidimensional histograms

To identify single-molecule events we use "burstwise" or "time-window" selection with 2σ criteria out of the mean background count rate. Cutoff times may vary from sample to sample depending on the background signal. Time-resolved fluorescence histograms of each single-molecule event, with a minimum of 60 photons, is processed and fitted using a maximum likelihood algorithm (Maus et al., 2001) in custom developed programs coded in LabVIEW (National Instruments Co.). Fluorescent bursts are plotted in 2D histograms (Origin 8.6, OriginLab Co).

For single-molecule Förster Resonance Energy Transfer (smFRET) experiments, bursts are shown using a parametric relationship between the ratio of the donor fluorescence over the acceptor fluorescence ($F_D/F_A$) and the fluorescence-weighted donor lifetime obtained in burst analysis $\langle\tau_{D(A)}\rangle_f$. $F_D/F_A$ depends on specific experimental parameters such as count rate per color channel ($\langle S_G\rangle$ and $\langle S_R\rangle$), the fluorescence quantum yields of the dyes ($\Phi_{FD(0)}$ and $\Phi_{FA}$ for donor and acceptor respectively), background ($\langle B_G\rangle$ and $\langle B_R\rangle$ for green and red channels), detection efficiencies ($g_G$ and $g_R$ for green and red respectively) and crosstalk ($\alpha$) following these relationships

$$F_D = \frac{S_G - \langle B_G\rangle}{g_G}, \tag{1}$$

$$F_A = \frac{S_R - \alpha F_G - \langle B_R\rangle}{g_R}. \tag{2}$$

In the $F_D/F_A$ vs. $\langle\tau_{D(A)}\rangle_f$ 2D representations it is useful to represent a static FRET line that represent the parametric relationship between ($F_D/F_A$) and $\langle\tau_{D(A)}\rangle_f$ which include the dynamics of the fluorophore's linker. Consequently, the linker flexibility generates a distribution of distances instead of a single distance. Mathematically, the FRET lines for $F_D/F_A$ vs. $\langle\tau_{D(A)}\rangle_f$ and $R_{DA}$ vs. $\langle\tau_{D(A)}\rangle_f$, corrected for linker mobility, are

$$\left(\frac{F_D}{F_A}\right)_{\text{static,L}} = \frac{\Phi_{FD(0)}}{\Phi_{FA}} \cdot \left( \frac{\tau_{D(0)}}{\sum_{i=0}^{3} A_{i,L} \left(\langle\tau_{D(A)}\rangle_{x,L}\right)^i} - 1 \right)^{-1}. \tag{3a}$$

$$R_{DA\text{static,L}} = R_0 \cdot \left( \frac{1}{\frac{\tau_{D(0)}}{\sum_{i=0}^{2} B_{i,L} \left(\langle\tau_{D(A)}\rangle_{x,L}\right)^i} - 1} \right)^{-1/6}. \tag{3b}$$

The $L$ sub index notation is to identify and specify the linker effects and $\tau_{D(0)}$ is the donor fluorescence lifetime in the absence of acceptor. The "$A_{i,L}$" and "$B_{i,L}$" coefficients are empirically determined by a polynomial approximation of the following parametric relationship between the species average lifetime $\langle\tau_{D(A)}\rangle_{x,L}$ and fluorescence weighted average lifetime $\langle\tau_{D(A)}\rangle_{f,L}$ for a range for $\langle R_{DA}\rangle$ =[1 Å to 5 $R_0$] Å using the following relationships

$$\langle \tau_{D(A)} \rangle_{x,L} = \sum_{i=0}^{3} A_{i,L} \left( \langle \tau_{D(A)} \rangle_{f,L} \right)^{i}$$
$$\langle \tau_{D(A)} \rangle_{f} = \langle \tau_{D(A)} \rangle_{f,L} = \frac{\int \tau_{D(A)}{}^{2} p(R_{DA}) dR_{DA}}{\langle \tau_{D(A)} \rangle_{x,L}} \tag{4}$$
$$\langle \tau_{D(A)} \rangle_{x,L} = \int \tau_{D(A)} p(R_{DA}) dR_{DA}$$

Here, the distribution of distances is assumed to follow a Gaussian probability function with a mean FRET distance $\langle R_{DA} \rangle$ and standard deviation $\sigma_{DA}$,

$$p(R_{DA}) = \frac{1}{\sqrt{2\pi}\sigma_{DA}} \exp\left( -\frac{\left(R_{DA} - \langle R_{DA} \rangle\right)^2}{2\sigma_{DA}^2} \right). \tag{5}$$

Keep in mind that in Eq. 4 there is a $R_{DA}$ for each $\tau_{D(A)}$ following the Förster relationship

$$\tau_{D(A)} = \tau_{D(0)} \cdot \left( 1 + \left( \frac{R_0}{R_{DA}} \right)^6 \right)^{-1}, \tag{6}$$

where $R_0$ is the Förster distance. For our particular set of dyes $R_0$ = 52 Å and we assume isotropic reorientation of dyes ($\kappa^2$ = 2/3) in the determination of $R_0$. The static FRET line corrected for linker dynamics is only valid for cases where there is no dynamic interexchange between states.

In addition to the static FRET line, we use the dynamic FRET line to show the interexchange between states. In this case, a mixed fluorescence species arises from the interconversion between two conformational states, where each state follows the Gaussian distribution stated above. For the simplest case the dynamic FRET line can be presented as(Kalinin et al., 2010)

$$\left( \frac{F_D}{F_A} \right)_{\mathrm{dyn,L}} = \frac{\Phi_{FD(0)}}{\Phi_{FA} \tau_{D(0)}} \cdot \frac{\langle \tau_1 \rangle_f \cdot \langle \tau_2 \rangle_f}{\left( \langle \tau_1 \rangle_f + \langle \tau_2 \rangle_f - \sum_{i=0}^{3} C_{i,L} \left( \langle \tau_{D(A)} \rangle_f \right)^{i} \right) - \frac{\langle \tau_1 \rangle_f \cdot \langle \tau_2 \rangle_f}{\tau_{D(0)}}}, \tag{7}$$

where $\langle \tau_{D(A)} \rangle_{f,L}$ is the mixed fluorescence lifetime, and $\langle \tau_1 \rangle_f$ and $\langle \tau_2 \rangle_f$ are two donor fluorescence lifetimes in presence of acceptor corresponding to the states that give rise to the dynamic exchange. The "$C_{i,L}$" coefficients are determined for each FRET pair and differ from the "$A_{i,L}$" coefficients in the static FRET lines. The $L$ sub index notation is to identify and specify the linker effects.

To represent single-molecule Fluorescence Anisotropy (smFA) we chose the scatter-corrected fluorescence anisotropy per burst ($r_D$), which is calculated as (Schaffer et al., 1999)

$$r_D = \frac{G_r \left( S_{\parallel} - \langle B_{\parallel} \rangle \right) - \left( S_{\perp} - \langle B_{\perp} \rangle \right)}{G_r \left( S_{\parallel} - \langle B_{\parallel} \rangle \right)(1 - 3l_2) + \left( S_{\perp} - \langle B_{\perp} \rangle \right)(2 - 3l_1)}, \tag{8}$$

where $G_r$ is the ratio of the detection efficiencies ( $g_{\perp}/g_{||}$) on the perpendicular ($\perp$) and parallel (||) channels, sometimes referred as $G$-factor, $l_1$ = 0.01758 and $l_2$ = 0.0526 are correction factors for the depolarization created by the microscope objective (Koshioka et al., 1995; Schaffer et al., 1999), and the signal in the parallel and perpendicular detector are $S_{||}$ and $S_{\perp}$ respectively.

In smFA experiments, the parametric histogram used is $r_D$ vs. $\langle \tau_D \rangle$ which relates the steady state anisotropy and the average fluorescence lifetime per burst. In ensemble conditions there is a similar relation called Perrins' Eq.

$$r_D = r_0 \left(1 + \frac{\langle \tau_D \rangle}{\rho}\right)^{-1}, \qquad (9)$$

where $r_0$ is the fundamental anisotropy of the fluorophore and is set to be 0.38. $\rho$ is the rotational correlation time and $\langle \tau_D \rangle$ is the average fluorescence lifetime per burst.

**S1.2. Probability Distribution Analysis (PDA) of single-molecule Fluorescence Anisotropy (smFA) and Förster Resonance Energy Transfer (smFRET) experiments**

To model the shape of the anisotropy and $F_D/F_A$ distributions, we use probability distribution analysis or PDA. Anisotropy-PDA is an extension of the FRET-PDA theory which was derived first. The theory behind these can be found in (Antonik et al., 2006; Kalinin et al., 2007). In short, the measured fluorescence signal *S,* consisting of fluorescence (*F*) and background (*B*) photons are expressed in photon count numbers per time window (Δt) of a fixed length. In Multiparameter Fluorescence Detection the signal is split into two spectral windows termed "green" and "red" each with two polarization components (Parallel "||" and Perpendicular "⊥"). The probability of observing a certain combination of photon counts in two detection channels 1 and 2 (e.g., "1=green" and "2=red" or "1=||" and "2=⊥") and measured by two or more single photon counting detectors., $P(S_1, S_2)$, is given by a product of independent probabilities

$$P(S_1, S_2) = \sum_{F_1+B_1=S_1;\, F_2+B_2=S_2} P(F) P(F_1, F_2 \mid F) P(B_1) P(B_2). \qquad (10)$$

$P(F)$ describes the fluorescence intensity distribution, i.e., the probability of observing exactly $F$ fluorescence photons per time window (Δt). $P(B_1)$ and $P(B_2)$ represent the background intensity distributions. $P(F_1, F_2 \mid F)$ is the conditional probability of observing a particular combination of $F_1$ and $F_2$, provided the total number of fluorescence photons is $F$.This can be expressed as

$$P(F_1, F_2 \mid F) = \frac{F!}{F_1! F_2!} p_1^{F_1} p_2^{F_2} = \frac{F!}{F_1!(F - F_1)!} p_1^{F_1} (1 - p_1)^{F - F_1}. \qquad (11)$$

$p_1$ stands for the probability of a detected photon to be registered by the first detector (e.g., green in a FRET experiment or parallel in an anisotropy experiment). For the case of smFA experiments $p_1 = p_{||}$ and it is written as

$$p_{\parallel} = \frac{1 + \langle r \rangle (2 - 3l_1)}{1 + \langle r \rangle (2 - 3l_1) + G_r - G_r \langle r \rangle (1 - 3l_2)}; \quad p_{\perp} = 1 - p_{\parallel} \qquad (12)$$

where $G_r$, $l_1$ and $l_2$ were previously defined. Consequently, $p_2 = p_{\perp}$.

For smFRET $p_1$ is unambiguously related to the FRET efficiency $E$ according to

$$p_{1|} = \left(1 + \alpha + \frac{E \Phi_{FA}}{(1 - E) G \Phi_{FD(0)}}\right) \quad p_2 = 1 - p_{1|}. \qquad (13)$$

Here, $G$ stands for the ratio of the detection efficiencies in the spectral windows ($G = g_G/g_R$) and the quantum yields ($\Phi_{FD(0)}$ and $\Phi_{FA}$) were previously defined.

The distribution $P(F)$ in Eq. (10) is not directly measurable, instead the total signal intensity distribution $P(S)$ is measured, which is given by

$$P(S) = P(F) \otimes P(B), \tag{14}$$

where $P(B)$ is the distribution probability of background counts. Details on the deconvolution procedure are described elsewhere (Kalinin et al., 2007). Finally, Eq (10) can be extended for multiple species with the brightness correction used in this work (Kalinin et al., 2008). Each species distributions has a half width ($hw_{DA}$) which depends mostly on shot noise and photophysical properties of the acceptor fluorophore.

**S1.3. Statistical uncertainties in PDA**

Confidence intervals estimation for multiple fit parameters is performed as follows. All free fit parameters are varied simultaneously in a random manner. The $\chi_r^2$ value is calculated at 100000 random points yielding 100-1000 points with $\chi_r^2$ values below $\chi_{r,\max}^2$

$$\chi_{r,\max}^2 = \chi_{r,\min}^2 + \sqrt{\frac{2}{N}} \tag{15}$$

where $N$ is the number of bins and $\chi_{r,\min}^2$ is the reduced chi-squared value of the best fit. The threshold of $\chi_{r,\max}^2$ is assigned as 1σ confidence interval. One could calculate thresholds. Alternative methods for threshold determination are available (Soong, 2004); however, in practice $\chi_{r,\min}^2$ is often affected by experimental imperfections and can be considerably larger than one. For this reason, we prefer this test to measure the robustness of the fits providing numerical uncertainties of the free parameters.

**S1.4. Filtered Fluorescence Correlation Spectroscopy**

To separate species, we use filtered FCS (fFCS) (Böhmer et al., 2002; Felekyan et al., 2012b). fFCS differs from standard FCS (Elson and Magde, 1974) and FRET-FCS (Felekyan et al., 2013) by interrogating the "species" (conformational states) fluctuations instead of photon count rates (Felekyan et al., 2013). We define the species auto- or cross-correlation function as

$$G^{(i,m)}(t_c) = \frac{\left\langle F^{(i)}(t) \cdot F^{(m)}(t+t_c) \right\rangle}{\left\langle F^{(i)}(t) \right\rangle \cdot \left\langle F^{(m)}(t+t_c) \right\rangle} = \frac{\left\langle \left( \sum_{j=1}^{d \cdot L} f_j^{(i)} \cdot S_j(t) \right) \cdot \left( \sum_{j=1}^{d \cdot L} f_j^{(m)} \cdot S_j(t+t_c) \right) \right\rangle}{\left\langle \sum_{j=1}^{d \cdot L} f_j^{(i)} \cdot S_j(t) \right\rangle \cdot \left\langle \sum_{j=1}^{d \cdot L} f_j^{(m)} \cdot S_j(t+t_c) \right\rangle}, \tag{16}$$

where $(i)$ and $(m)$ are two selected "species" in a mixture. When $i=m$ we say it is the species auto-correlation function (sACF), and when $i \neq m$ it is the species cross-correlation function (sCCF). The difference from standard FCS is that in fFCS we introduce a set of filters, $f_j^{(i)}$, that depend on the arrival time of each photon after each excitation pulse. The signal $S_j(t)$, obtained via pulsed excitation is recorded at each $j$ = 1 ... $L$ TCSPC-channel. The signal and filters per detector, $d$, are stacked in a single array with dimensions $d \cdot L$ for global minimization as previously shown (Felekyan et al., 2012b). Filters are defined in such a way that the relative "error" difference between the photon count per species ($w^{(i)}$) and the weighted histogram $f_j^{(i)} \cdot H_j$ is minimized as defined in Eq. (17).

$$\left\langle \left( \sum_{j=1}^{d\cdot L} f_j^{(i)} \cdot H_j - w^{(i)} \right)^2 \right\rangle \to \min . \tag{17}$$

where brackets represent time averaging.

The requirement is that the decay histogram $H_j$ can be expressed as a linear combination of the conditional probability distributions $p_j^{(i)}$, such as $H_j = \sum_{i=1}^{n} w^{(i)} p_j^{(i)}$, with $\sum_{j=1}^{d\cdot L} p_j^{(i)} = 1$. Hence, the sCCF provides maximal contrast for intercrossing dynamics (Felekyan et al., 2012b). One major advantage of sCCF is that if photophysical properties are decoupled from species selection the intercrossing dynamics (Felekyan et al., 2013) is recovered with great fidelity.

To properly fit the species auto- and cross-correlation function we used a set of equations previously presented (Felekyan et al., 2012b)

$$\begin{aligned}
G_{i,i}(t_c) &= 1 + \frac{1}{N_{Br}} \cdot G_{diff}^{(i)}(t_c) \cdot \left[ 1 - T^{(i)} + T^{(i)} \cdot \mathbf{exp}\left(-t_c/t_T^i\right) + \sum_{R=1}^{4} AC_{i,i}^{(R)} \cdot \left( \mathbf{exp}\left(-\frac{t_c}{t_R}\right) - 1 \right) \right] \cdot G_B^{(i)}(t_c) \\
G_{m,m}(t_c) &= 1 + \frac{1}{N_{Br}} \cdot G_{diff}^{(m)}(t_c) \cdot \left[ 1 - T^{(m)} + T^{(m)} \cdot \mathbf{exp}\left(-t_c/t_T^m\right) + \sum_{R=1}^{4} AC_{m,m}^{(R)} \cdot \left( \mathbf{exp}\left(-\frac{t_c}{t_R}\right) - 1 \right) \right] \cdot G_B^{(m)}(t_c) \\
G_{i,m}(t_c) &= 1 + \frac{1}{N_{CC}} \cdot G_{diff}^{(i,m)}(t_c) \cdot \left[ 1 - CC_{i,m} \cdot \sum_{R=1}^{4} X_{i,m}^{(R)} \cdot \mathbf{exp}\left(-\frac{t_c}{t_R}\right) \right] \cdot \left( 1 - B_{i,m} \cdot \mathbf{exp}\left(-\frac{t_c}{t_B}\right) \right)
\end{aligned}$$

(18)

where $t_R$ are the relaxation times that correspond to the exchange times between selected species with corresponding absolute amplitudes of the sACF $AC_{x,x}^{(R)}$ and the relative normalized amplitudes of the sCCF $CC_{x,x}^{(R)}$. $T^{(x)}$ is the triplet amplitude, however, triplet states dynamic was not found in the measured samples. $N_{br}$ is the number of bright molecules in the sACF's in the focus and $N_{CC}$ of the sCCF's corresponds to the inverse of the initial amplitude $G_{i,m}(0)$. $G_B^{(x)}(t_c)$ is defined for bleaching term:

$$G_B^{(x)}(t_c) = 1 - B^{(x)} + B^{(x)} \cdot \exp\left(-t_c/t_B\right), \tag{19}$$

$G_{diff}^{(x)}(t_c)$ is the diffusion term of species $x$:

$$G_{diff}^{(x)}(t_c) = \left( 1 + \frac{t_c}{t_{diff}^{(x)}} \right)^{-1} \cdot \left( 1 + \left( \frac{\omega_0}{z_0} \right)^2 \cdot \frac{t_c}{t_{diff}^{(x)}} \right)^{-\frac{1}{2}} . \tag{20}$$

A 3-dimensional Gaussian shaped volume element parameters $\omega_0$ and $z_0$ is considered. We assume that $G_{diff}(t_c) = G_{diff}^{(i)}(t_c) = G_{diff}^{(m)}(t_c)$ take the form of Eq. (20).In fFCS the amplitudes are highly dependent on the brightness of the individual sub-states and the exchange rate constants. Thus a direct interpretation is not straightforward (Felekyan et al., 2012a).

**S1.5. Accessible volume (AV) model and inter-fluorophore distances**

To accurately compare FRET-derived distances with structural information provided by crystallography data it is imperative to consider the dimensions of the fluorophores. To do so, we compute the accessible volume of the dyes by considering them as hard sphere models connected to the protein via flexible linkers (modeled as a flexible cylindrical pipe) (Sindbert et al., 2011). The overall dimension (width and length) of the linker is based on their chemical structures. For Alexa 488 maleimide the five carbon linker length was set to 20 Å, the width of the linker is 4.5 Å and a three sphere model was used to model the dye $R_1$= 5.0 Å, $R_2$= 4.5 Å and $R_3$= 1.5 Å. For Alexa 647 maleimide the dimensions used were: length = 22 Å, width = 4.5 Å and the dye radii $R_1$= 11.0 Å, $R_2$= 4.7 Å and $R_3$= 1.5 Å.

To account for dye linker mobility we generated a series of AV's for donor and acceptor dyes attached to p27 placing the dyes at multiple separation distances. For each pair of AV's, we calculated the distance between dye mean positions ($R_{\mathrm{mp}}$)

$$R_{mp} = \left| \left\langle \vec{R}_{D(i)} \right\rangle - \left\langle \vec{R}_{A(j)} \right\rangle \right| = \left| \frac{1}{n} \sum_{i=1}^{n} \vec{R}_{D(i)} - \frac{1}{m} \sum_{j=1}^{m} \vec{R}_{A(j)} \right|, \quad (21)$$

where $\vec{R}_{D(i)}$ and $\vec{R}_{A(i)}$ are all the possible positions that the donor fluorophore and the acceptor fluorophore can take. However, in single-molecule FRET experiment where a ratiometric FRET is calculated the distances is weighted by the average fluorescence thus the mean donor-acceptor distance observed is

$$\left\langle R_{DA} \right\rangle_E = R_0 \left( \left\langle E \right\rangle^{-1} - 1 \right)^{1/6} \quad (22)$$

where the average efficiency is defined as $\left\langle E \right\rangle = \frac{1}{nm} \sum_{i=1}^{n} \sum_{j=1}^{m} \left( \frac{R_0}{R_0 + \left| \overrightarrow{R_D^{(i)}} - \overrightarrow{R_A^{(j)}} \right|^6} \right)$.

**S2. Supplemental Tables**

**Supplemental Table 1. Data collection and refinement statistics for determination of the structure of the Cdk2/cyclin A/p27-KID-ΔC complex using X-ray crystallography.**

The structure of the Cdk2/cyclin A/p27-KID-ΔC complex has been deposited and validated bythe PDB with the file name 6ATH for release upon publication.

| | **Cdk2/cyclin A/p27-KID-ΔC** |
|---|---|
| **Data Collection** [a] | |
| Space Group | $P2_12_12_1$ |
| Cell dimensions | |
| *a*,*b*,*c* (Å) | 74.2, 77.6, 137.5 |
| $\alpha$, $\beta$, $\gamma$ (°) | 90.0, 90.0, 90.0 |
| No. of crystals | 2 |
| Wavelength (Å) | 1.0 |
| Resolution (Å) | 50.0-1.82 (1.89-1.82) |
| No. unique reflections | 71,452 (6,848) |
| $R_{merge}$ [b] | 0.085 (0.522) |
| Completeness (%) | 99.2 (96.5) |
| Redundancy | 10.7 (4.2) |
| $I/\sigma$ | 26.6 (2.1) |
| **Refinement** | |
| Resolution (Å) | 29.6-1.82 |
| No. of reflections | 71,378 |
| $R_{work}/R_{free}$ [c] | 0.171/0.187 |
| No. atoms | |
| Protein | 4,774 |
| Ion | 5 |
| Water | 399 |
| Average B-factor (Å$^2$) | 28.0 |
| R.m.s. deviations | |
| Bond lengths (Å) | 0.005 |
| Bond angles (°) | 0.94 |
| Ramachandran plot | |
| Favored (%) | 98.1 |
| Allowed (%) | 1.4 |
| Outliers (%) | 0.5 |

[a] Values in parenthesis are for highest-resolution shell.

[b] $R_{merge} = \sum |I - \langle I \rangle| / \sum I$, where *I* is the observed intensity.

[c] $R_{free}$ is the *R* value obtained for a test set of reflections consisting of randomly selected 5% subset of the data set excluded from refinement.

**Supplemental Table 2. Fit parameters of smFA experiments obtained by PDA** (Sections S1.2, S1.3)**.**

| Residue number | Anisotropy (Fraction (%)) **No phosphorylation** | | $\chi_r^2$ | Anisotropy (Fraction (%)) **pY88** | | $\chi_r^2$ | Anisotropy (Fraction (%)) **pY74/pY88** | | $\chi_r^2$ |
|---|---|---|---|---|---|---|---|---|---|
| | Low $r_D$ | High $r_D$ | | Low $r_D$ | High $r_D$ | | Low $r_D$ | High $r_D$ | |
| 29 | 0.08±0.01 (12.9±1.9) | 0.26±0.01 (87.1±13.1) | 1.26 | 0.07±0.01 (25.6±3.1) | 0.26±0.01 (74.4±8.9) | 1.37 | 0.06±0.01 (26.7±3.5) | 0.25±0.01 (73.3±9.7) | 1.39 |
| 40 | 0.13±0.01 (57.7±7.6) | 0.25±0.01 (42.3±5.5) | 1.54 | 0.12±0.01 (62.0±5.8) | 0.24±0.01 (38.0±3.6) | 2.06 | 0.13±0.01 (61.8±7.0) | 0.25±0.01 (38.2±4.3) | 1.82 |
| 54 | 0.04±0.01 (30.8±5.7) | 0.24±0.01 (69.2±12.9) | 1.45 | 0.08±0.01 (29.9±4.9) | 0.24±0.01 (70.1±11.6) | 1.29 | 0.07±0.01 (71.7±6.8) | 0.23±0.01 (28.3±2.7) | 1.14 |
| 75 | 0.07±0.01 (27.2±4.8) | 0.20±0.01 (72.8±13) | 1.26 | 0.09±0.01 (25.4±3.5) | 0.26±0.01 (74.6±10.1) | 1.67 | 0.11±0.01 (40.5±4.6) | 0.26±0.01 (59.5±6.5) | 1.97 |
| 93 | 0.08±0.01 (33.3±5.1) | 0.28±0.01 (66.7±10.1) | 1.18 | 0.11±0.02 (65.9±12.3) | 0.27±0.02 (34.1±6.3) | 1.32 | 0.09±0.01 (65.1±10.4) | 0.25±0.01 (34.9±5.6) | 1.28 |

**Supplemental Table 3. Brightness correction for smFA experiments.**

Normalized to the maximum lifetime: $Q^{(\text{short lifetime})} = \tau_D^{(\text{short})}/\tau_D^{(\text{long})}$.

| Sample in complex | $\tau_D^{(\text{High rD})}$ [ns] | $\tau_D^{(\text{Low rD})}$ [ns] | $Q^{(\text{High rD}}$ | $Q^{(\text{Low rD})}$ |
|---|---|---|---|---|
| p27 C29 No phosphorylation | 4.6 | 5.4 | 0.85 | 1 |
| pY88-p27 C29 | 4.6 | 5.4 | 0.85 | 1 |
| pY74/pY88-p27 C29 | 4.6 | 5.4 | 0.85 | 1 |
| p27 C40 No phosphorylation | 5.9 | 4.6 | 1 | 0.78 |
| pY88-p27 C40 | 5.9 | 4.6 | 1 | 0.78 |
| pY74/pY88-p27 C40 | 5.9 | 4.6 | 1 | 0.78 |
| p27 C54 No phosphorylation | 5.9 | 4.2 | 1 | 0.71 |
| pY88-p27 C54 | 5.9 | 4.2 | 1 | 0.71 |
| pY74/pY88-p27 C54 | 5.9 | 4.2 | 1 | 0.71 |
| p27 C75 No phosphorylation | 5.9 | 4.2 | 1 | 0.71 |
| pY88-p27 C75 | 5.9 | 4.2 | 1 | 0.71 |
| pY74/pY88-p27 C75 | 5.9 | 4.2 | 1 | 0.71 |
| p27 C93 No phosphorylation | 5.9 | 4.6 | 1 | 0.78 |
| pY88-p27 C93 | 5.9 | 4.6 | 1 | 0.78 |
| pY74/pY88-p27 C93 | 5.9 | 4.6 | 1 | 0.78 |

**Supplemental Table 4. Rotational correlation time for smFA experiments.**
Values were obtained using Table 2, Table 3 and Perrin's Equation [Eq. (9)] for Cdk2/cyclin A/p27 samples.

| Rotational Correlation (ns) | **No phosphorylation** | | **pY88** | | **pY74/pY88** | |
|---|---|---|---|---|---|---|
| Residue No. | Low $r_D$ | High $r_D$ | Low $r_D$ | High $r_D$ | Low $r_D$ | High $r_D$ |
| 29 | 1.5 | 10.5 | 1.2 | 10.5 | 1.0 | 9.3 |
| 40 | 2.5 | 11.9 | 2.2 | 10.6 | 2.5 | 11.9 |
| 54 | 0.5 | 10.6 | 1.1 | 10.6 | 1.0 | 9.4 |
| 75 | 1.0 | 6.8 | 1.3 | 13.5 | 1.8 | 13.5 |
| 93 | 1.3 | 17.6 | 1.9 | 15.3 | 1.5 | 11.9 |

**Supplemental Table 5. Fluorophore properties of dyes used in smFRET experiments and generated FRET lines according to S1.1 equations (3) and (7).**

| Sample in complex | $\Phi_{D(0)}$ | $\Phi_A$ | *Static FRET Line* | *Dynamic FRET Line* |
|---|---|---|---|---|
| p27 C29/54 No phosphorylation | 0.70 | 0.39 | $(0.6988/0.39)/((3.7494/((-0.0519*\langle\tau_{D(A)}\rangle_f^3)+(0.3131*\langle\tau_{D(A)}\rangle_f^2)+0.5690*\langle\tau_{D(A)}\rangle_f+-0.0529))-1)$ | $0.1913/(0.39*((1/2.044+1/1.264-(1.4769*\langle\tau_{D(A)}\rangle_f+-1.0373)/(1.2640*2.0440))-1/3.7494))$ |
| pY88-p27 C29/54 | 0.72 | 0.36 | $(0.7220/0.36)/((3.9015/((-0.0470*\langle\tau_{D(A)}\rangle_f^3)+(0.3002*\langle\tau_{D(A)}\rangle_f^2)+0.5507*\langle\tau_{D(A)}\rangle_f+-0.0505))-1)$ | $0.1957/(0.36*((1/2.122+1/1.1340-(1.5937*\langle\tau_{D(A)}\rangle_f+-1.3355)/(1.1340*2.122))-1/3.9015))$ |
| pY74/pY88-p27 C29/54 | 0.72 | 0.37 | $(0.7239/0.37)/((4.0239/((-0.0483*\langle\tau_{D(A)}\rangle_f^3)+(0.2978*\langle\tau_{D(A)}\rangle_f^2)+0.5875*\langle\tau_{D(A)}\rangle_f+-0.0564))-1)$ | $0.1872/(0.37*((1/2.04+1/1.06-(1.5244*\langle\tau_{D(A)}\rangle_f+-1.1138)/(1.06*2.04))-1/4.0239))$ |
| E88-p27 C29/54 | 0.75 | 0.39 | $(0.6036/0.37)/((3.511/((-0.0574*\langle\tau_{D(A)}\rangle_f^3)+(0.2965*\langle\tau_{D(A)}\rangle_f^2)+0.6838*x+-0.0638))-1)$ | |
| E74/E88-p27 C29/54 | 0.60 | 0.37 | $(0.5948/0.394)/((3.5383/((-0.0556*\langle\tau_{D(A)}\rangle_f^3)+(0.282*\langle\tau_{D(A)}\rangle_f^2)+0.7161*\langle\tau_{D(A)}\rangle_f+-0.067))-1)$ | |
| p27 C54/93 No phosphorylation | 0.72 | 0.42 | $(0.7194/0.42)/((3.8301/((-0.0501*\langle\tau_{D(A)}\rangle_f^3)+(0.3112*\langle\tau_{D(A)}\rangle_f^2)+0.5553*\langle\tau_{D(A)}\rangle_f+-0.0513))-1)$ | $0.1884/(0.42*((1/2.05+1/1.46-(1.3896*\langle\tau_{D(A)}\rangle_f +-0.8588)/(1.46*2.05))-1/3.8301))$ |
| pY88-p27 C54/93 | 0.75 | 0.37 | $(0.7519/0.37)/((3.9997/((-0.0443*\langle\tau_{D(A)}\rangle_f^3)+(0.2947*\langle\tau_{D(A)}\rangle_f^2)+0.5367*\langle\tau_{D(A)}\rangle_f+-0.0490))-1)$ | $0.1925/(0.37*((1/2.14+1/1.49-(1.4414*\langle\tau_{D(A)}\rangle_f+-1.029)/(1.49*2.14))-1/3.9997))$ |
| pY74/pY88-p27 C54/93 | 0.75 | 0.36 | $(0.7541/0.36)/((3.9983/((-0.0468*\langle\tau_{D(A)}\rangle_f^3)+(0.3068*\langle\tau_{D(A)}\rangle_f^2)+0.5289*\langle\tau_{D(A)}\rangle_f+-0.0476))-1)$ | $0.1943/(0.36*((1/2.14+1/1.49-(1.4602*\langle\tau_{D(A)}\rangle_f+-1.0788)/(1.49*2.14))-1/3.9983))$ |
| E88-p27 C54/93 | 0.63 | 0.33 | $(0.6036/0.363)/((3.511/((-0.0574*\langle\tau_{D(A)}\rangle_f^3)+(0.2965*\langle\tau_{D(A)}\rangle_f^2)+0.6838*\langle\tau_{D(A)}\rangle_f+-0.0638))-1)$ | |
| E74/E88-p27 C54/93 | 0.67 | 0.36 | $(0.683/0.33)/((3.7855/((-0.0475*\langle\tau_{D(A)}\rangle_f^3)+(0.2846*\langle\tau_{D(A)}\rangle_f^2)+0.6179*\langle\tau_{D(A)}\rangle_f+-0.0575))-1)$ | |
| p27 C75/110 No phosphorylation | 0.63 | 0.39 | $(0.6324/0.39)/((3.7203/((-0.0510*\langle\tau_{D(A)}\rangle_f^3)+(0.2782*\langle\tau_{D(A)}\rangle_f^2)+0.6777*\langle\tau_{D(A)}\rangle_f+-0.0646))-1)$ | $0.175/(0.39*((1/2.13+1/1.02-(1.3731*\langle\tau_{D(A)}\rangle_f +-0.7835)/(1.02*2.13))-1/3.7203))$ |
| pY88-p27 C75/110 | 0.63 | 0.36 | $(0.6332/0.36)/((3.7901/((-0.0477*\langle\tau_{D(A)}\rangle_f^3)+(0.2612*\langle\tau_{D(A)}\rangle_f^2)+0.7019*\langle\tau_{D(A)}\rangle_f +-0.0666))-1)$ | $0.1681/(0.36*((1/2.11+1/1.56-(1.1605*\langle\tau_{D(A)}\rangle_f +-0.2981)/(1.56*2.11))-1/3.5651))$ |

| | | | |
|---|---|---|---|
| pY74/pY88-p27 C75/110 | 0.60 | 0.37 | $\frac{(0.6025/0.37)/((3.7301/((-0.0486*\langle\tau_{D(A)}\rangle_f^3)+(0.2502*\langle\tau_{D(A)}\rangle_f^2)+0.7510*\langle\tau_{D(A)}\rangle_f +-0.0711))-1)}{0.1624/(0.37*((1/1.77+1/1.32-(1.1961*\langle\tau_{D(A)}\rangle_f +-0.3101)/(1.3200*1.77))-1/3.7301))}$ |
| E88-p27 C75/110 | 0.67 | 0.38 | $(0.7129/0.385)/((3.8577/((-0.0456*\langle\tau_{D(A)}\rangle_f^3)+(0.2868*\langle\tau_{D(A)}\rangle_f^2)+0.5847*\langle\tau_{D(A)}\rangle_f +-0.0539))-1)$ |
| E74/E88-p27 C75/110 | 0.71 | 0.39 | $(0.6482/0.375)/((3.7123/((-0.0497*\langle\tau_{D(A)}\rangle_f^3)+(0.2802*\langle\tau_{D(A)}\rangle_f^2)+0.6602*\langle\tau_{D(A)}\rangle_f +-0.0616))-1)$ |

**Supplemental Table 6. Fit parameters of PDA analysis for smFRET experiments.**

Global fitting $\Delta T$=1, 2 and 3 ms. $\chi^2_r$ is shown for $\Delta T$=3 ms. A) All fractions B) Renormalized fractions for only FRET species. Half-width distribution ($hw_{DA}$) of the PDA distribution should not be confused with $\sigma_{DA}$ in Eq. (5). Data was corrected as described in S1.2 and using values from Table 3.

| **A)**<br>Samples<br>Cdk2/cyclin A/p27 | $\langle R_{DA}\rangle_E$ [Å]<br>(Fraction (%))<br>$hw_{DA}$ [Å]<br>**No P** | | | | $\chi_r^2$ | $\langle R_{DA}\rangle_E$ [Å]<br>(Fraction (%))<br>$hw_{DA}$ [Å]<br>**pY88** | | | | $\chi_r^2$ | $\langle R_{DA}\rangle_E$ [Å]<br>(Fraction (%))<br>$hw_{DA}$ [Å]<br>**pY74/pY88** | | | | $\chi_r^2$ |
|---|---|---|---|---|---|---|---|---|---|---|---|---|---|---|---|
| | Low $\langle R_{DA}\rangle_E$ | High $\langle R_{DA}\rangle_E$ | Dirt | $D_{only}$ | | Low $\langle R_{DA}\rangle_E$ | High $\langle R_{DA}\rangle_E$ | Dirt | $D_{only}$ | | Low $\langle R_{DA}\rangle_E$ | High $\langle R_{DA}\rangle_E$ | Dirt | $D_{only}$ | |
| C29-54 | 43.1±0.1<br>(11.1±0.5)<br>4.6 | 52.3±0.2<br>(59.9±2.7)<br>2.7 | 74.9<br>(22.8)<br>17.7 | $10^6$<br>(6.2)<br>-- | 4.8 | 41.6±0.1<br>(9.2±0.2)<br>3.3 | 52.5±0.1<br>(30.0±0.6)<br>2.8 | 83.6<br>(48.6)<br>18.8 | $10^6$<br>(12.2)<br>-- | 37.4 | 40.9±0.3<br>(14.4±1.4)<br>2.8 | 52.0±0.3<br>(22.2±2.1)<br>2.5 | 79.5<br>(44.3)<br>17.6 | $10^6$<br>(19.1)<br>-- | 5.2 |
| C54-93 | 45.2±0.3<br>(55.7±2.8)<br>3.0 | 52.3±0.3<br>(11.6±3.2)<br>3.5 | 76.4<br>(24.7)<br>16 | $10^6$<br>(8.0)<br>-- | 5.3 | 45.2±0.4<br>(20.2±1.7)<br>3.7 | 52.3±0.4<br>(16.9 ±2.2)<br>3.0 | 82.9<br>(52.9)<br>15.9 | $10^6$<br>(10.0)<br>-- | 2.1 | 45.2±0.2<br>(18.9±1.4)<br>3.6 | 52.3±0.2<br>(7.4±0.5)<br>2.0 | 82.9<br>(44.7)<br>21.2 | $10^6$<br>(29.0)<br>-- | 10.1 |
| C75-110 | 40.7±0.2<br>(6.4±0.5)<br>3.4 | 54.6±0.3<br>(37.1±2.7)<br>3.8 | 81.5<br>(43.1)<br>16.6 | $10^6$<br>(13.5)<br>-- | 5.1 | 46.8±0.1<br>(17.8±0.3)<br>4.4 | 54.4±0.1<br>(13.2±0.2)<br>5.9 | 83.5<br>(63.7)<br>16.1 | $10^6$<br>(5.2)<br>-- | 71.0 | 44.3±0.3<br>(23.3±3.0)<br>2.6 | 49.9±0.4<br>(15.4±2.0)<br>6.2 | 88.2<br>(37.5)<br>18.1 | $10^6$<br>(23.9)<br>-- | 3.1 |

| **B)**<br>Samples<br>Cdk2/cyclin A/p27 | **No P**<br>Fraction (%) | | **pY88**<br>Fraction (%) | | **pY74/pY88**<br>Fraction (%) | |
|---|---|---|---|---|---|---|
| | Low $\langle R_{DA}\rangle_E$ | High $\langle R_{DA}\rangle_E$ | Low $\langle R_{DA}\rangle_E$ | High $\langle R_{DA}\rangle_E$ | Low $\langle R_{DA}\rangle_E$ | High $\langle R_{DA}\rangle_E$ |
| C29-54 | 15.6 | 84.4 | 23.5 | 76.5 | 39.3 | 60.7 |
| C54-93 | 82.8 | 17.2 | 54.4 | 45.6 | 71.9 | 28.1 |
| C75-110 | 14.7 | 85.3 | 57.4 | 42.6 | 60.2 | 39.8 |

**Supplemental Table 7. $F_D/F_A$ levels for High and Low $\langle R_{DA} \rangle_E$.**
$\langle R_{DA} \rangle_E$ were converted to FD/FA with Eq. (6) and (3) (Table 6).

| Sample in ternary complex | $F_D/F_A$ ($\langle R_{DA} \rangle_E^{(Low)}$) | $F_D/F_A$ ($\langle R_{DA} \rangle_E^{(High)}$) |
|---|---|---|
| p27 C29/54 No P | 0.7 | 2.02 |
| pY88-p27 C29/54 | 0.61 | 2.33 |
| pY74/pY88-p27 C29/54 | 0.52 | 2.04 |
| E88-p27 C29/54 | | |
| E74/E88-p27 C29/54 | | |
| p27 C54/93 No P | 0.84 | 1.85 |
| pY88-p27 C54/93 | 0.96 | 2.18 |
| pY74/pY88-p27 C54/93 | 0.99 | 2.27 |
| E88-p27 C54/93 | | |
| E74/E88-p27 C54/93 | | |
| p27 C75/110 No P | 0.47 | 2.2 |
| pY88-p27 C75/110 | 1.1 | 2.24 |
| pY74/pY88-p27 C75/110 | 1.1 | 1.42 |
| E88-p27 C75/110 | 0.78 | 1.42 |
| E74/E88-p27 C75/110 | | |

**Supplemental Table 8. Inter-fluorophore distances of p27 FRET variants in complex with Cdk2/cyclin A based on accessible volume calculations (Section S1.5) and comparison with the "major" state in the non-phosphorylated state.**

| Sample | **1JSU** [Å] | | **Cdk2/cyclin A/p27 Fig 1**. pdb:1JSU [Å] | | **Cdk2/cyclin A/p27-ΔC** [Å] | | **Experiment** [Å] |
|---|---|---|---|---|---|---|---|
| | $\langle R_{DA} \rangle_E$ | Cα-Cα | $\langle R_{DA} \rangle_E$ | Cα-Cα | $\langle R_{DA} \rangle_E$ | Cα-Cα | $\langle R_{DA} \rangle_E$ |
| C29-54 | 53.7 | 39.3 | 52.2 | 39.6 | 54.3 | 40.4 | 52.3 |
| C54-93 | 49.5 | 29.8 | 48.4 | 30.3 | - | - | 45.2 |
| C75-110 | - | - | 55.3 | 54.1 | - | - | 54.6 |

**Supplemental Table 9. Relaxation rate constants obtained by fFCS for Figure 4E in main text (Section S1.4, Equations (18)-(20)).**

| Correlation | | $b$ | $N_{CC}/N_{Br}$ | $t_{diff}$ [ms] | $\omega 0/z0$ | $CC$ | $t_{R1}$ [µs] | $X_1/AC_1$ | $t_{R2}$ [µs] | $X_2/AC_2$ | $t_{R3}$ [µs] | $X_3/AC_3$ | $t_{R4}$ [µs] | $X_4/AC_4$ | $B$ | $t_B$ [ms] |
|---|---|---|---|---|---|---|---|---|---|---|---|---|---|---|---|---|
| **p27 C29 C54 in complex with Cdk2/cyclin A** | | | | | | | | | | | | | | | | |
| No P | LF-HF | 1.01 | 0.37 | 1.76 | 3.70 | 1.24 | 0.058 | 0.52 | 1.96 | 0.17 | 22.7 | 0.24 | 207 | 0.07 | 0.10 | 4.03 |
| | HF-LF | 1.01 | 0.29 | | | 1.27 | | | | | | | | | 0.30 | |
| | LF-LF | 1.01 | 0.049 | | | | | 0.35 | | 0.078 | | 0.066 | | 0.047 | 0.10 | |
| | HF-HF | 1.00 | 0.060 | | | | | 0.37 | | 0.062 | | 0.10 | | 0.023 | 0.00 | |
| pY88 | LF-HF | 1.00 | 0.77 | 1.69 | 4.87 | 2.05 | 0.058 | 0.49 | 1.77 | 0.16 | 21.6 | 0.22 | 349 | 0.13 | 0.16 | 2.98 |
| | HF-LF | 1.00 | 0.58 | | | 2.21 | | | | | | | | | 0.38 | |
| | LF-LF | 1.00 | 0.066 | | | | | 0.16 | | 0.066 | | 0.087 | | 0.18 | 0.29 | |
| | HF-HF | 1.00 | 0.051 | | | | | 0.33 | | 0.083 | | 0.10 | | 0.049 | 0.15 | |
| pY88-pY74 | LF-HF | 1.00 | 0.81 | 1.86 | 3.95 | 2.82 | 0.058 | 0.50 | 1.77 | 0.14 | 20.5 | 0.23 | 355 | 0.13 | 0.25 | 2.43 |
| | HF-LF | 1.01 | 0.57 | | | 2.75 | | | | | | | | | 0.47 | |
| | LF-LF | 1.00 | 0.040 | | | | | 0.19 | | 0.060 | | 0.079 | | 0.19 | 0.34 | |
| | HF-HF | 0.99 | 0.043 | | | | | 0.30 | | 0.097 | | 0.089 | | 0.067 | 0.24 | |
| **p27 C54 C93 in complex with Cdk2/cyclin A** | | | | | | | | | | | | | | | | |
| No P | LF-HF | 1.00 | 0.34 | 1.77 | 3.42 | 0.82 | 0.058 | 0.50 | 1.39 | 0.12 | 18.3 | 0.29 | 247 | 0.09 | 0.01 | 5.76 |
| | HF-LF | 1.02 | 0.20 | | | 0.89 | | | | | | | | | 0.44 | |
| | LF-LF | 1.01 | 0.039 | | | | | 0.30 | | 0.075 | | 0.086 | | 0.084 | 0.10 | |
| | HF-HF | 1.01 | 0.041 | | | | | 0.35 | | 0.065 | | 0.096 | | 0.033 | 0.05 | |
| pY88 | LF-HF | 1.00 | 0.78 | 1.48 | 4.83 | 1.45 | 0.058 | 0.66 | 1.27 | 0.15 | 25.1 | 0.14 | 345 | 0.05 | 0.00 | 3.94 |
| | HF-LF | 1.00 | 0.41 | | | 1.55 | | | | | | | | | 0.50 | |
| | LF-LF | 1.00 | 0.061 | | | | | 0.17 | | 0.091 | | 0.074 | | 0.12 | 0.17 | |
| | HF-HF | 0.99 | 0.052 | | | | | 0.38 | | 0.093 | | 0.088 | | 0.022 | 0.12 | |
| pY88-pY74 | LF-HF | 1.00 | 1.12 | 1.38 | 4.27 | 3.16 | 0.058 | 0.62 | 2.22 | 0.16 | 27.7 | 0.13 | 351 | 0.09 | 0.00 | 3.70 |
| | HF-LF | 1.01 | 0.44 | | | 3.64 | | | | | | | | | 0.66 | |
| | LF-LF | 1.00 | 0.034 | | | | | 0.20 | | 0.078 | | 0.074 | | 0.16 | 0.19 | |
| | HF-HF | 1.00 | 0.048 | | | | | 0.37 | | 0.087 | | 0.080 | | 0.035 | 0.11 | |
| **p27 C75 C110 in complex with Cdk2/cyclin A** | | | | | | | | | | | | | | | | |
| No P | LF-HF | 1.00 | 0.69 | 1.52 | 4.05 | 2.30 | 0.043 | 0.77 | 1.06 | 0.14 | 15.1 | 0.05 | 188 | 0.04 | 0.00 | 2.53 |
| | HF-LF | 1.01 | 0.35 | | | 0.96 | | | | | | | | | 0.35 | |
| | LF-LF | 1.01 | 0.076 | | | | | 0.17 | | 0.072 | | 0.11 | | 0.10 | 0.03 | |
| | HF-HF | 1.00 | 0.053 | | | | | 0.39 | | 0.12 | | 0.088 | | 0.049 | 0.10 | |
| pY88 | LF-HF | 1.00 | 1.42 | 1.31 | 3.99 | 2.74 | 0.032 | 0.72 | 1.78 | 0.17 | 18.2 | 0.08 | 195 | 0.03 | 0.01 | 2.76 |
| | HF-LF | 1.00 | 0.47 | | | 1.85 | | | | | | | | | 0.56 | |
| | LF-LF | 1.01 | 0.090 | | | | | 0.13 | | 0.079 | | 0.092 | | 0.086 | 0.00 | |
| | HF-HF | 1.00 | 0.062 | | | | | 0.23 | | 0.095 | | 0.088 | | 0.025 | 0.13 | |
| pY88-pY74 | LF-HF | 1.00 | 1.01 | 1.45 | 5.57 | 1.91 | 0.094 | 0.63 | 1.35 | 0.19 | 27.2 | 0.15 | 286 | 0.03 | 0.00 | 3.60 |
| | HF-LF | 1.01 | 0.34 | | | 1.47 | | | | | | | | | 0.55 | |
| | LF-LF | 1.00 | 0.049 | | | | | 0.23 | | 0.028 | | 0.065 | | 0.080 | 0.13 | |
| | HF-HF | 0.99 | 0.050 | | | | | 0.16 | | 0.21 | | 0.093 | | 0.025 | 0.087 | |

**Supplemental Table 10. Result of the analysis of triplicate measurements of p27-KID, pY88-p27-KID, and pY74/Y88-p27-KID binding to Cdk2/cyclin A, and separately to cyclin A or Cdk2, using surface plasmon resonance (SPR).**

The data were fit to a 1:1 Langmuir interaction model.

| Interacting species | $k_a$ ($M^{-1}s^{-1}$) | $k_d$ ($s^{-1}$) | $K_D$ (nM) | Rmax (RU) |
|---|---|---|---|---|
| p27-KID + Cdk2/cyclin A | $1.52 \pm 0.01 \times 10^6$ | $5.0 \pm 0.2 \times 10^{-4}$ | 0.33 ± 0.01 | 32.2 ± 0.1 |
| pY88-p27-KID + Cdk2/cyclin A | $1.63 \pm 0.01 \times 10^6$ | $9.8 \pm 0.1 \times 10^{-4}$ | 0.60 ± 0.01 | 17.4 ± 0.1 |
| pY74/Y88-p27-KID + Cdk2/cyclin A | $1.64 \pm 0.01 \times 10^6$ | $4.80 \pm 0.03 \times 10^{-3}$ | 2.93 ± 0.02 | 34.8 ± 0.1 |
| p27-KID + cyclin A | $4.8 \pm 0.1 \times 10^6$ | $1.36 \pm 0.03 \times 10^{-1}$ | 28 ± 1 | 21.7 ± 0.2 |
| pY88-p27-KID + cyclin A | $3.93 \pm 0.08 \times 10^6$ | $1.46 \pm 0.03 \times 10^{-1}$ | 37 ± 1 | 11.0 ± 0.1 |
| pY74/Y88-p27-KID + cyclin A | $2.91 \pm 0.05 \times 10^6$ | $1.51 \pm 0.02 \times 10^{-1}$ | 52 ± 1 | 22.9 ± 0.2 |
| p27-KID + Cdk2 | $1.17 \pm 0.01 \times 10^4$ | $4.4 \pm 0.1 \times 10^{-4}$ | 37.4 ± 0.09 | 26.9 ± 0.1 |
| pY88-p27-KID + Cdk2 | $1.18 \pm 0.01 \times 10^4$ | $8.6 \pm 0.1 \times 10^{-4}$ | 73 ± 1 | 14.3 ± 0.1 |
| pY74/Y88-p27-KID + Cdk2 | $9.28 \pm 0.02 \times 10^3$ | $3.37 \pm 0.01 \times 10^{-3}$ | 363 ± 1 | 29.0 ± 0.1 |

**S3. Supplemental Figures**

**Supplemental Figure 1. Incremental phosphorylation of tyrosine residues in p27 exerts rheostat-like control over Cdk2/cyclin A activity.**

(A) Representative results of kinase activity assays for Cdk2/cyclin A in the presence of increasing concentrations of unphosphorylated and mono and dual Y phosphorylated p27-KID and p27. Autoradiography was used to monitor incorporation of $^{32}$P-labeled phosphate into the substrate, Histone H1, which was resolved using SDS-PAGE. The regions of the autoradiograms corresponding to the Histone H1 protein in the gels are shown. The experiments were performed in triplicate. The results for p27-KID, pY88-p27-KID and pY74/pY88-p27-KID are quantified in Fig. 2A as average kinase activity (± standard deviation of the mean) relative to that in the absence of p27-KID. (B) Quantification of the results presented in (A) for mono and dual Y phosphorylated p27. The results are quantified as in Fig. 2A.

**A**

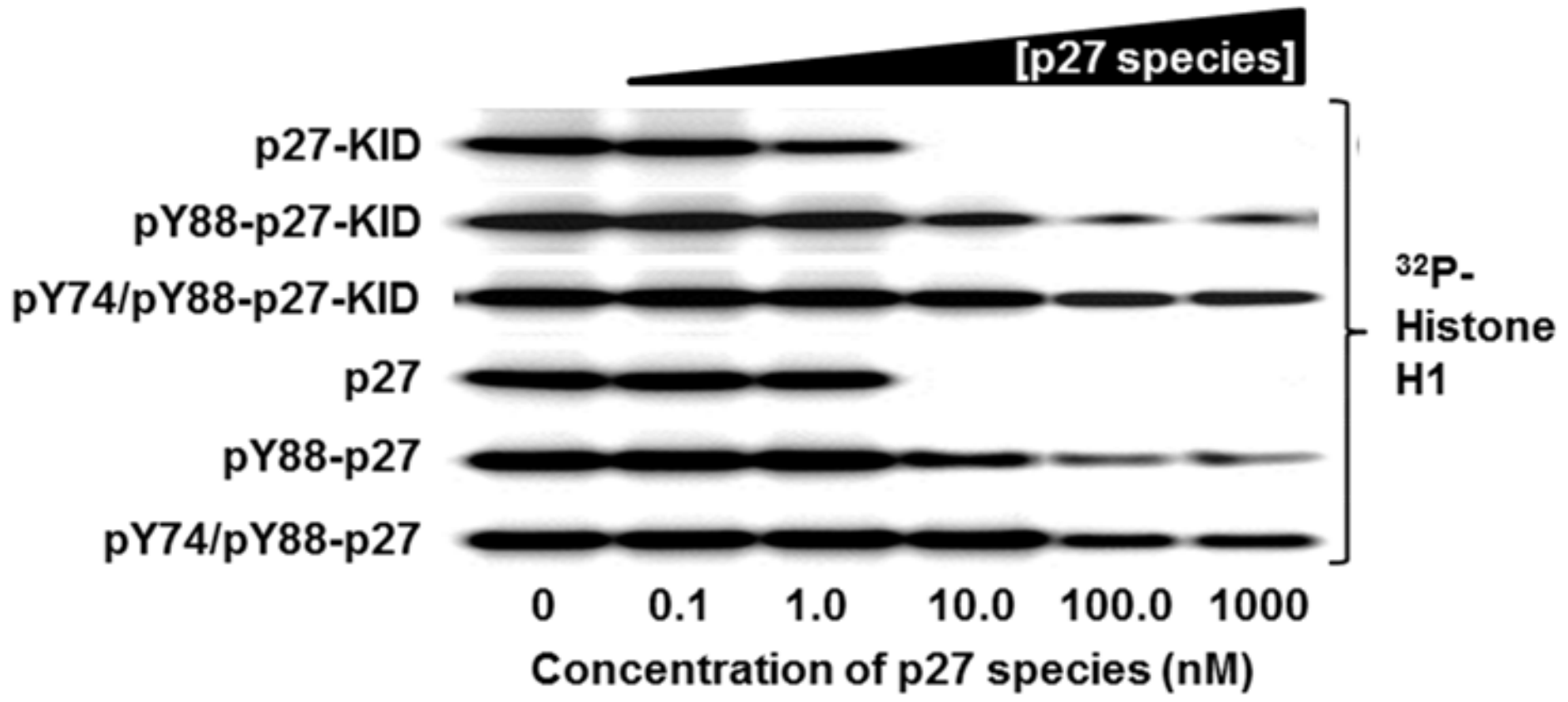

**B**

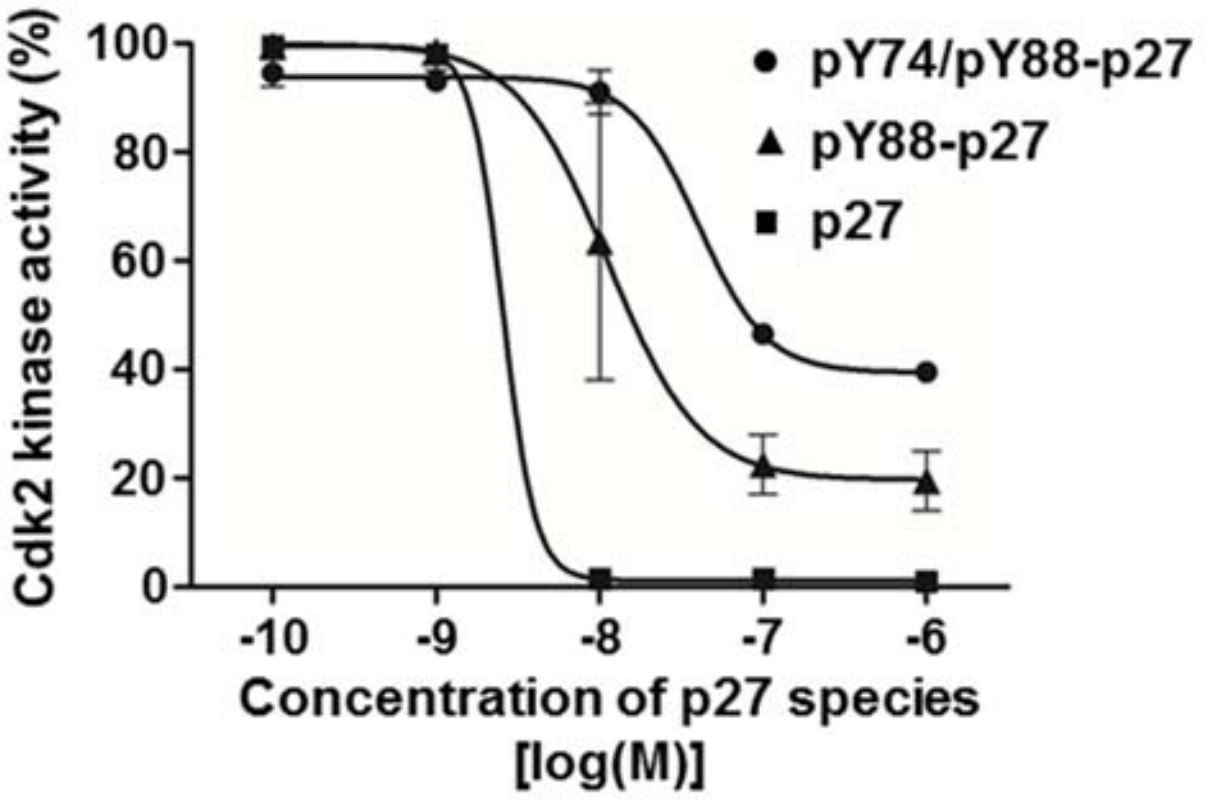

**Supplemental Figure 2 (next page). Incremental phosphorylation of tyrosine residues in p27 exerts rheostat-like control over Cdk2/cyclin A and promotes phosphorylation of p27 on T187.**

(A) Representative results of kinase activity assays for different concentrations of 1:1:1 complexes of Cdk2/cyclin A with p27, pY88-p27 or pY74/pY88-p27 with T187 (of p27) as the substrate. Autoradiography was used to monitor incorporation of $^{32}$P-labeled phosphate into T187 of p27, which was resolved using SDS-PAGE. The regions of the autoradiograms corresponding to the p27 protein in the gels are shown. The experiments were performed in triplicate. These results are quantified in Fig. 2C as average kinase activity (± standard deviation of the mean) relative to that for the highest concentration of pY74/pY88-p27/Cdk2/cyclin A. (B) Representative results of kinase activity assays for 1.0 and 2.0 µM 1:1:1 complexes of Cdk2/cyclin A with p27, pY88-p27 or pY74/pY88-p27. In addition, equimolar amounts of either Nus-tagged Rb C-terminus (Nus-Rb) or GST/His-tagged p107 C-terminus (p107) were included as "in trans" Cdk2 substrates. Autoradiography was used to monitor incorporation of $^{32}$P-labeled phosphate into T187 of p27 and Nus-Rb or p107, which were resolved using SDS-PAGE. The regions of the autoradiograms corresponding to the p27 and Nus-Rb or p107 proteins in the gels are shown. The experiments were performed in triplicate. The results for the 2.0 µM 1:1:1 complexes and Nus-Rb or p107 are quantified in Fig. 2D as average kinase activity (± standard deviation of the mean) relative to that for $^{32}$P-labeled Nus-Rb.

A

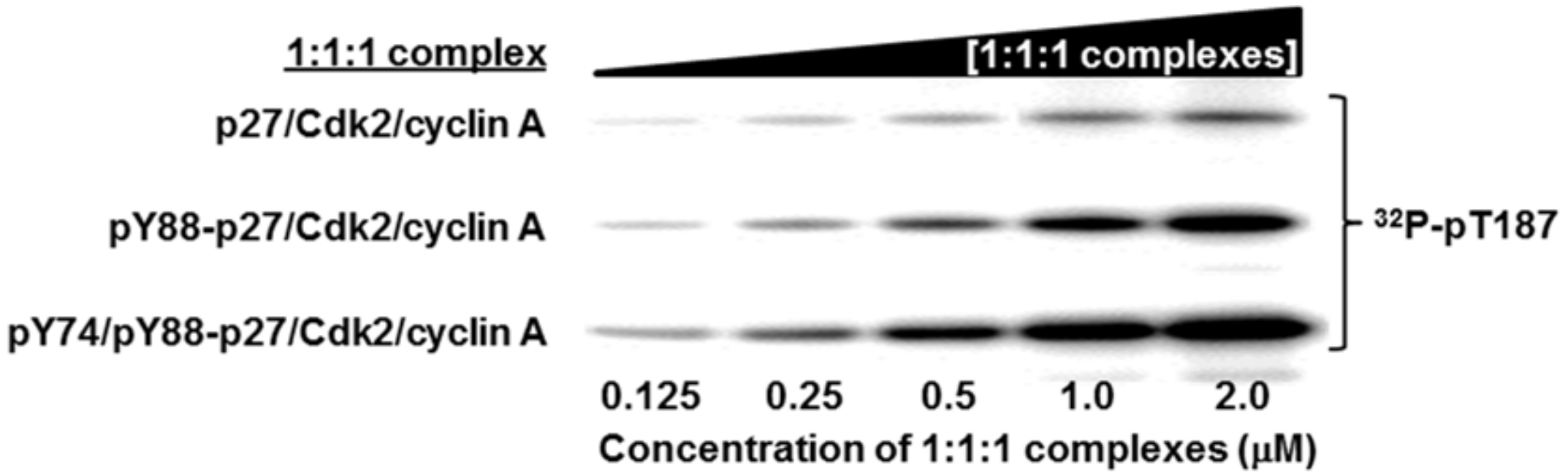

B

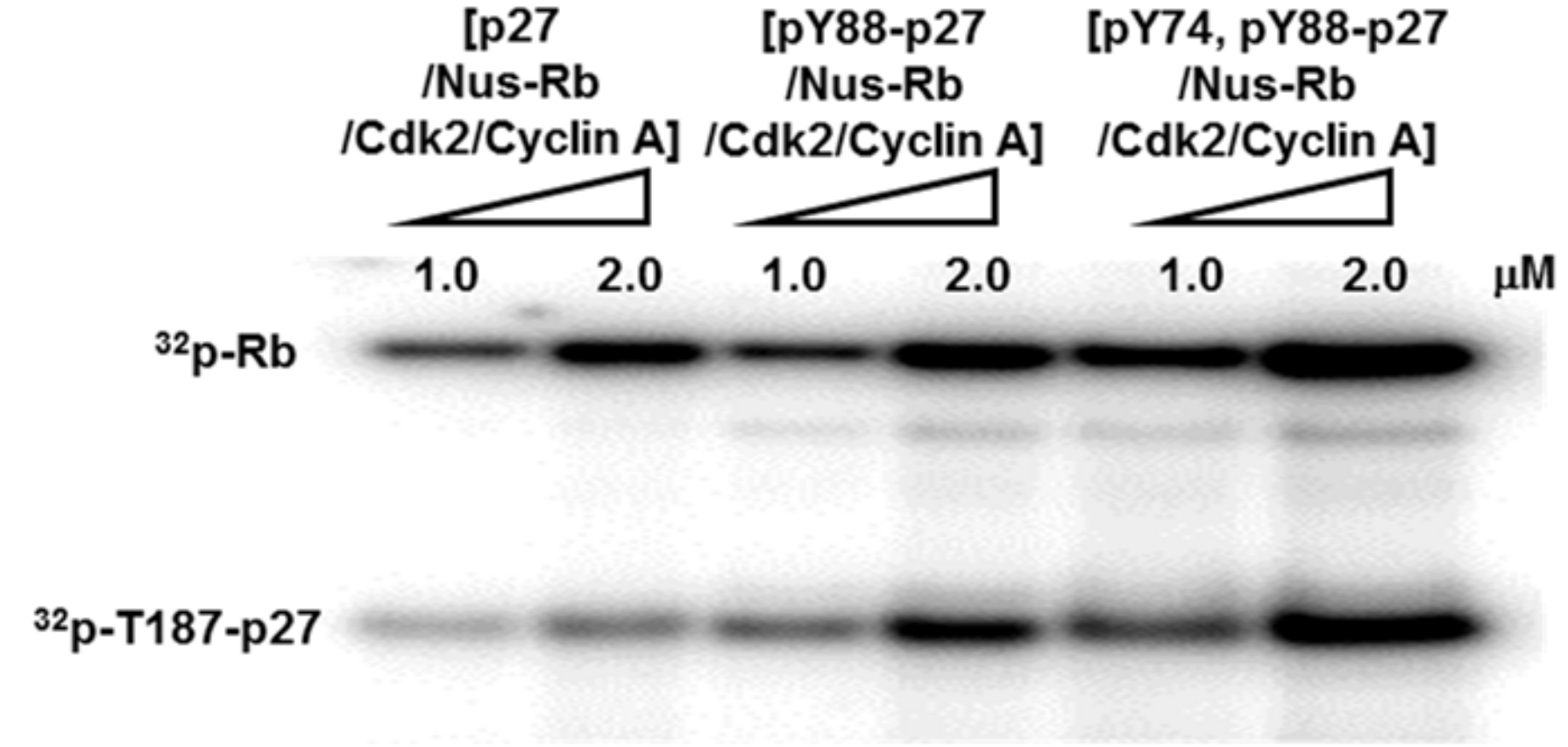

[p27 /Nus-Rb /Cdk2/Cyclin A]
[pY88-p27 /Nus-Rb /Cdk2/Cyclin A]
[pY74, pY88-p27 /Nus-Rb /Cdk2/Cyclin A]
1.0
2.0
1.0
2.0
1.0
2.0
μM
32p-Rb
32p-T187-p27

C

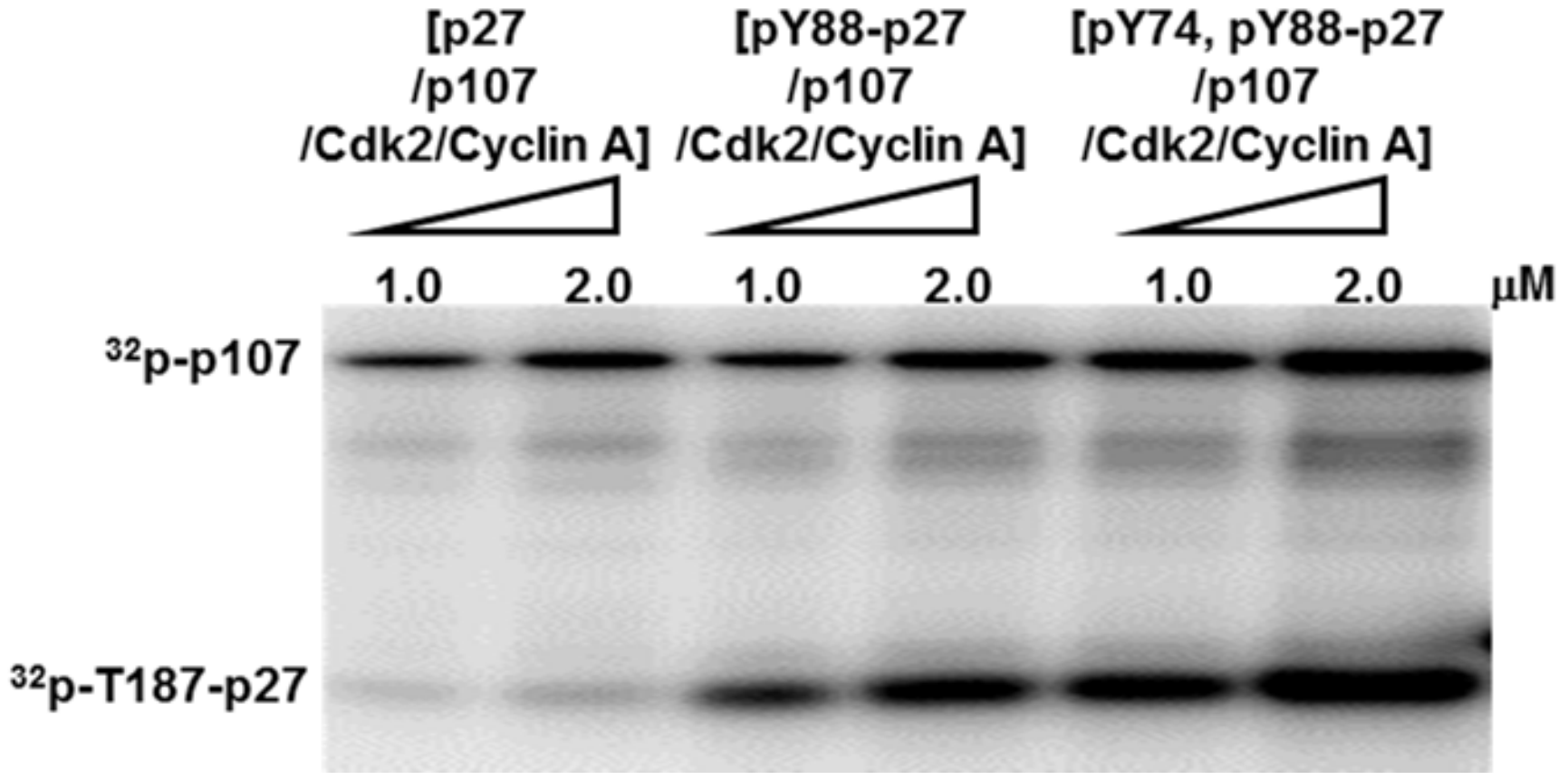

[p27 /p107 /Cdk2/Cyclin A]
[pY88-p27 /p107 /Cdk2/Cyclin A]
[pY74, pY88-p27 /p107 /Cdk2/Cyclin A]
1.0
2.0
1.0
2.0
1.0
2.0
μM
32p-p107
32p-T187-p27

**Supplemental Figure 3. Ejection of pY88 and the residue 83-89 region from the Cdk2 active site is mimicked by truncation of p27-KID at residue 79.**

(A) Left, representative results of kinase activity assays for Cdk2/cyclin A in the presence of increasing concentrations of p27-KID, pY88-p27-KID or p27-KID-ΔC (p27-KID with residues 80-94 deleted). Autoradiography was used to monitor incorporation of $^{32}$P-labeled phosphate into the substrate, Histone H1, which was resolved using SDS-PAGE. The regions of the autoradiograms corresponding to the Histone H1 protein in the gels are shown. The experiments were performed in triplicate. Right, quantification of the results on the left showing average kinase activity (± standard deviation of the mean) relative to that for 4 pM p27-KID. (B) Superposition of the structures of Cdk2/cyclin A bound to ATP (PDB:1JST; Cdk2/cyclin A:ATP), p27-KID bound to Cdk2/cyclin A (PDB:1JSU; p27-KID/Cdk2/cyclin A) and p27-KID-ΔC bound to Cdk2/cyclin A (p27-KID- ΔC/Cdk2/cyclin A, determined in this study). The structures are superimposed on backbone heavy atoms of cyclin A. The color code is indicated in the illustration and the boxed region is illustrated in Fig. 3B. The PyMOL Molecular Graphics System (Schrödinger, LLC) was used to prepare the illustration.

**A**

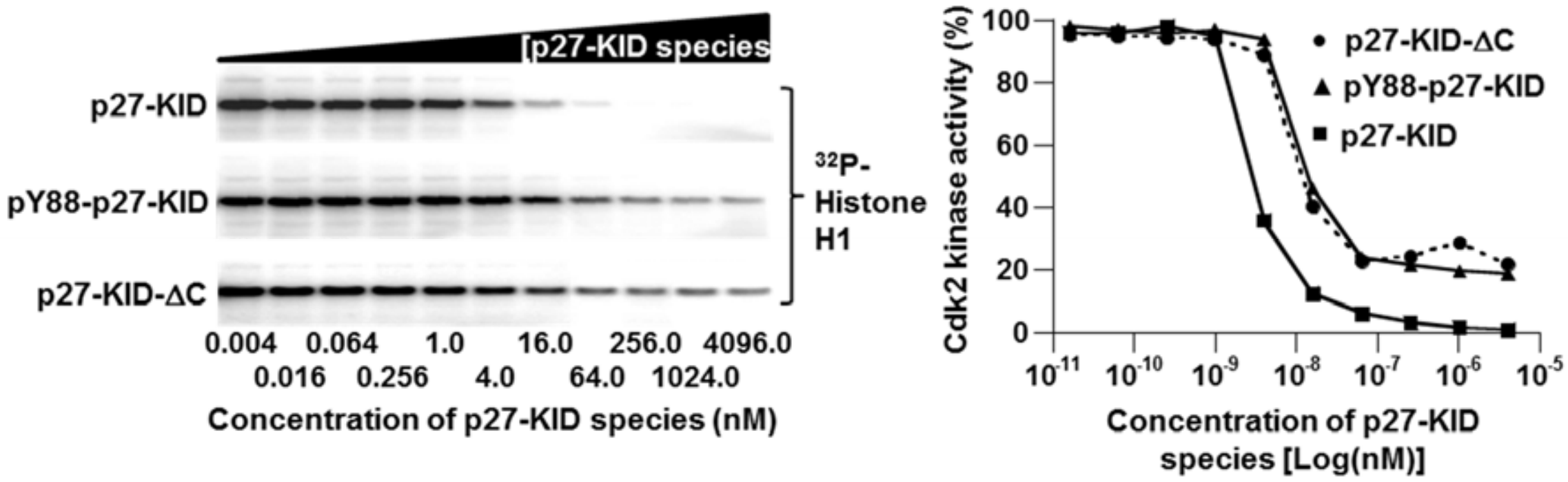

**B**

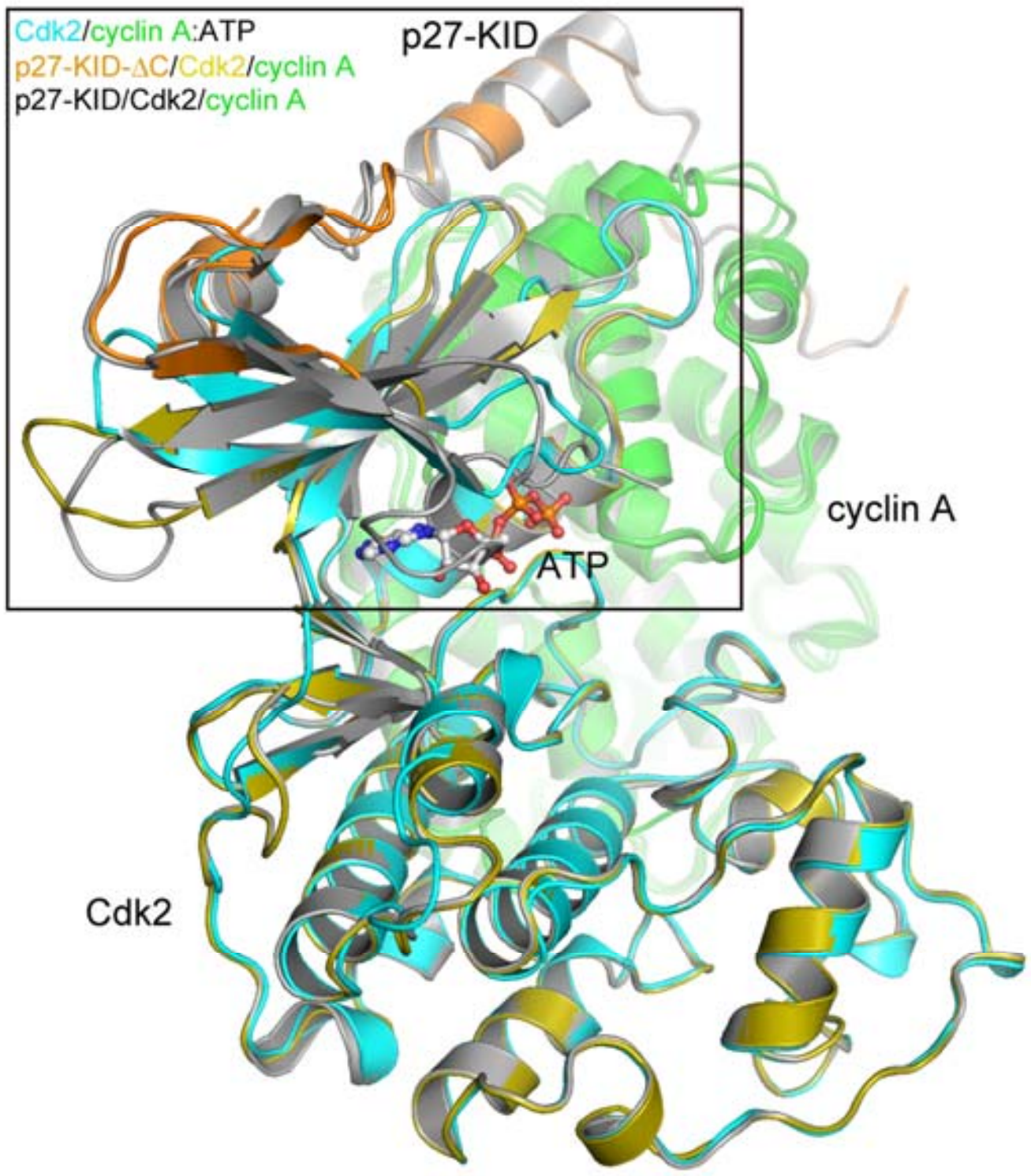

**Supplemental Figure 4. Multidimensional smFA histograms of p27/Cdk2/cyclin A at various phosphorylation states.**

Two-dimensional histogram of scatter anisotropy ($r_D$) vs. $\langle\tau_{D(A)}\rangle_f$ for Bodipy labeled single Cys p27 variants in complex with Cdk2/Cyclin A. "Burstwise" analysis of A) C29, B) C40, C) C54, D) C75 and E) C93 variants. For all cases one dimensional projections for $\langle\tau_{D(A)}\rangle_f$ and anisotropy are also shown. Pure donor fluorescence ($F_D$) is corrected for background ($\langle B_G\rangle$ = 1.07 kHz,). Perrin's equation for $r_{low}$ (blue) and $r_{high}$ (purple) using $\tau_D^{(\text{High r})}$ and $\tau_D^{(\text{Lowr})}$ are shown (Suppl. Table 2 and Suppl. Table 4). Rotational correlation times $\rho^{(Low\ rD)}$ and $\rho^{(High\ rD)}$ are given in the respective plot.

**A** **Cdk2/cyclin A/p27-C29**

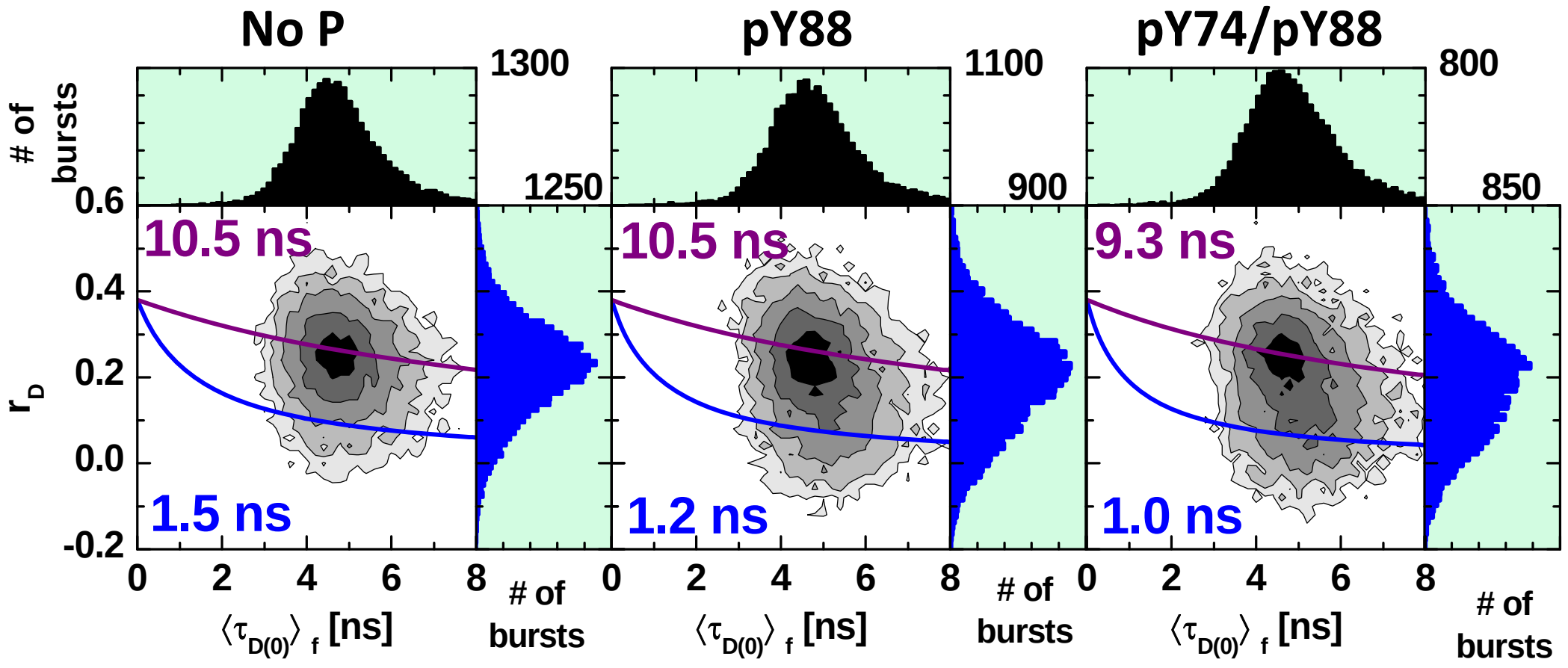

**B** **Cdk2/cyclin A/p27-C40**

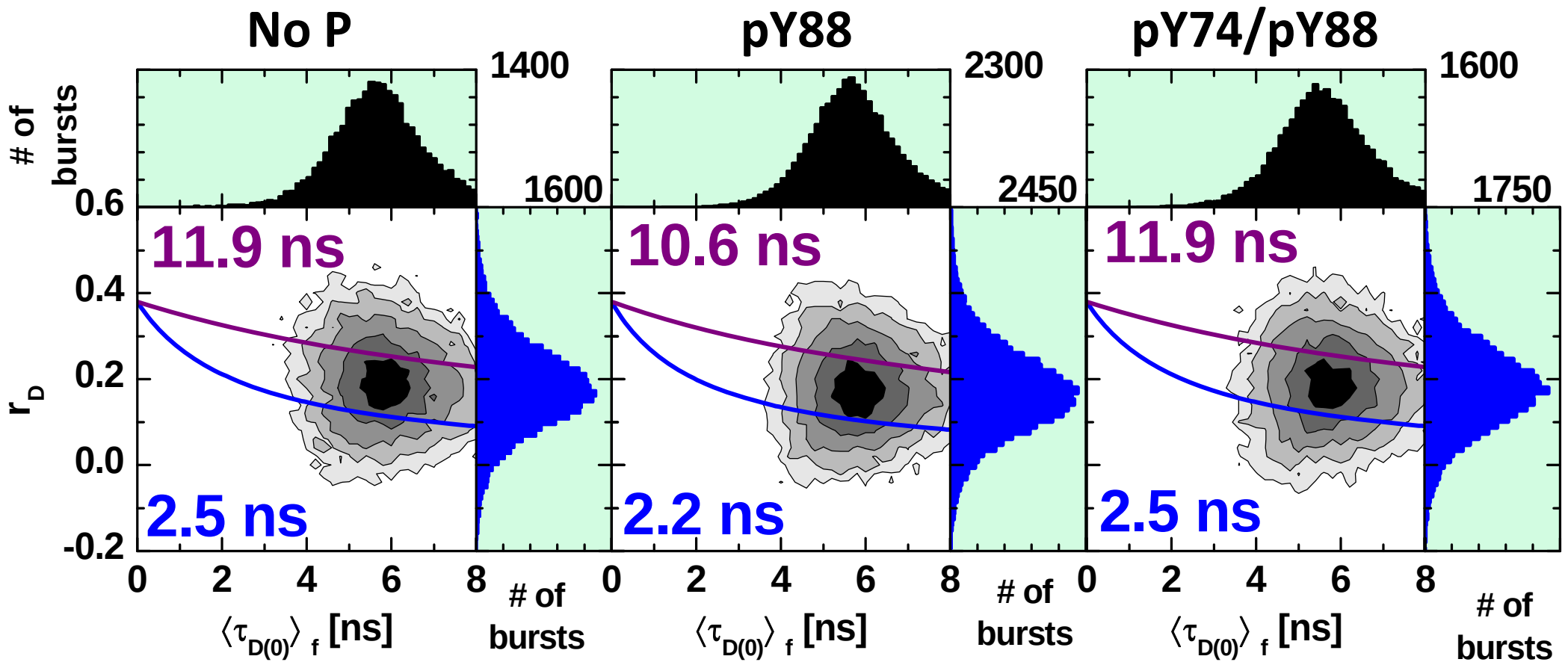

C

Cdk2/cyclin A/p27-C54

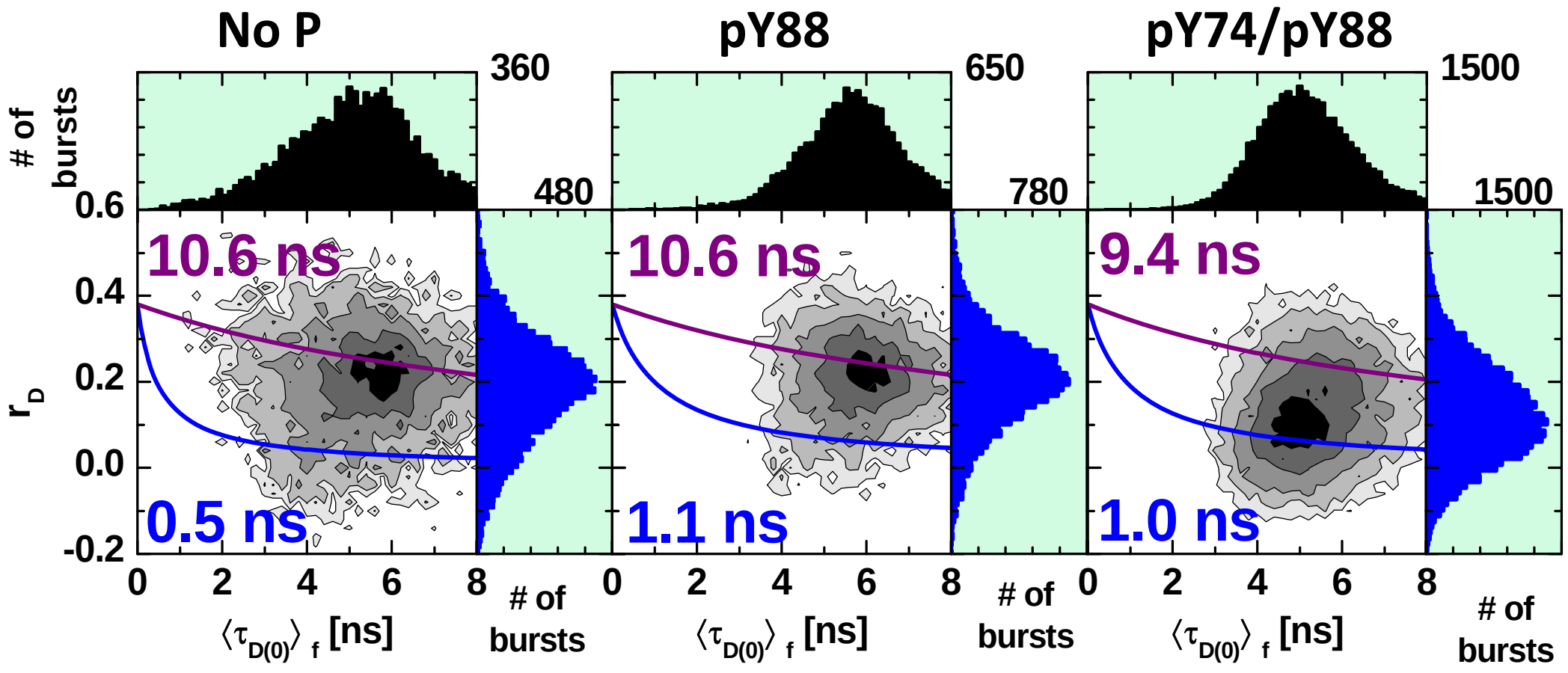

D

Cdk2/cyclin A/p27-C75

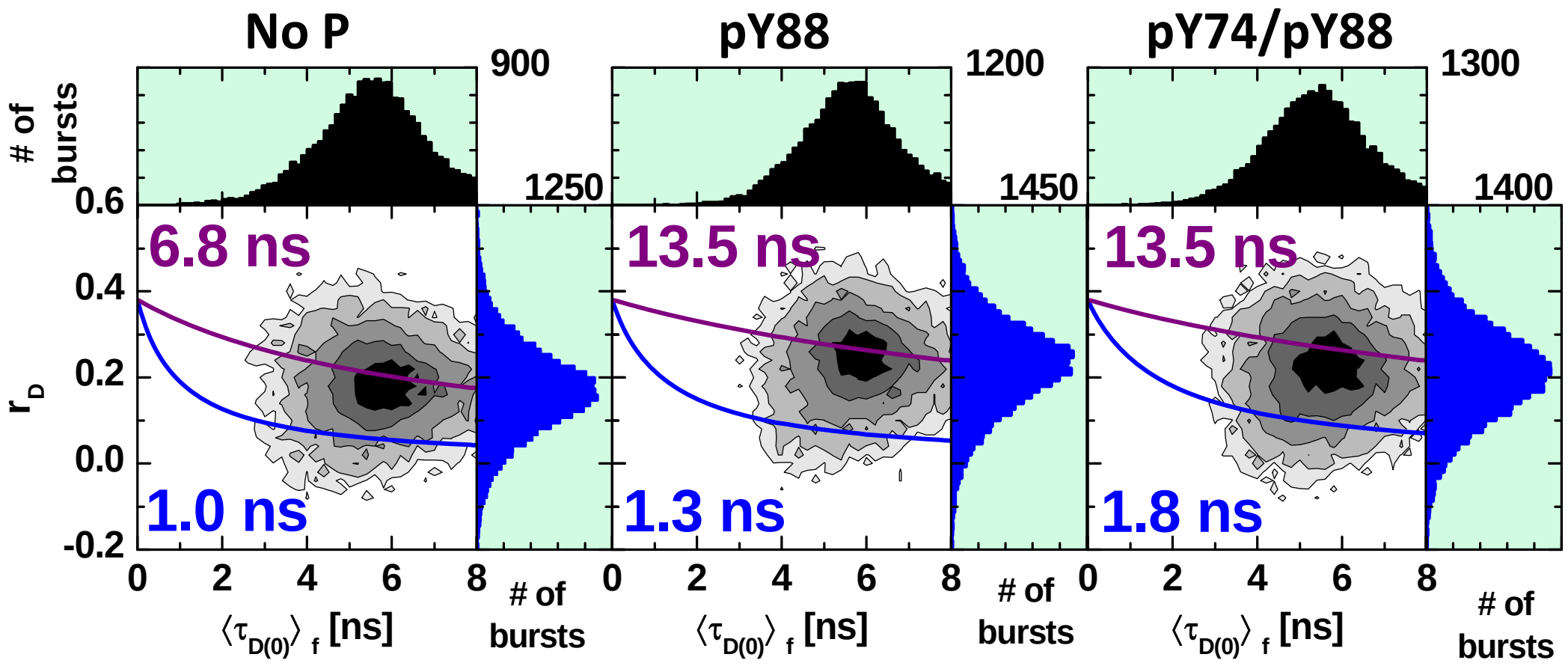

E

Cdk2/cyclin A/p27-C93

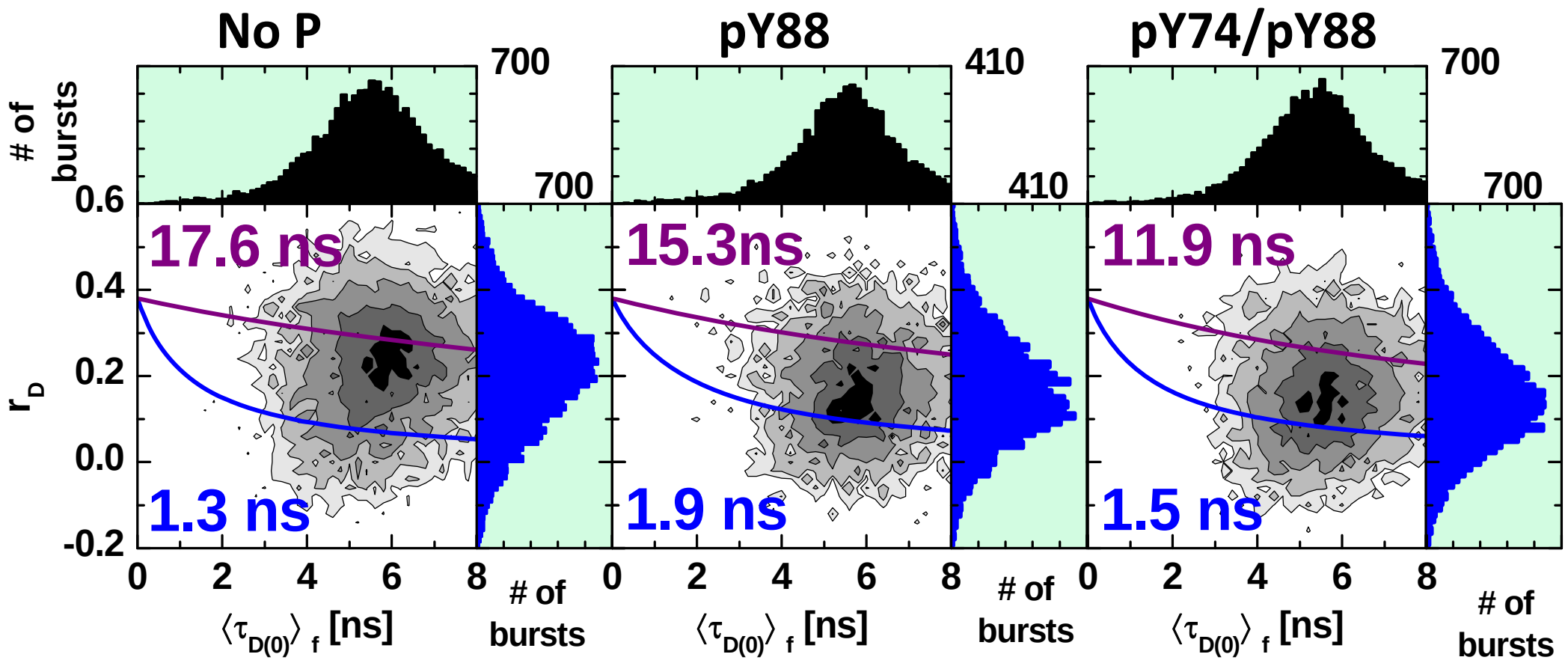

**Supplemental Figure 5. Multidimensional smFRET histograms of p27/Cdk2/cyclin A at various phosphorylated states.**

Two-dimensional histogram $F_D/F_A$ vs. lifetime of donor in the presence of acceptor $\langle\tau_{D(A)}\rangle_f$, and scatter corrected donor anisotropy ($r_D$) vs. $\langle\tau_{D(A)}\rangle_f$ for Cdk2/cyclin A/p27 with donor and acceptor dyes at various positions in "burstwise" mode. A) C29-54, B) C54-93 and C) C75-110. For all cases one dimensional projections for $F_D/F_A$, $\langle\tau_{D(A)}\rangle_f$ and anisotropy are also shown. Pure donor and acceptor fluorescence ($F_D$ and $F_A$) are corrected for background ($\langle B_G\rangle$ = 1.57 kHz A, B) or 0.65 kHz for C), $\langle B_R\rangle$ = 0.94 kHz A,B) or 0.42 kHz C), spectral cross-talk ($\alpha$ = 1.7 %) and detection efficiency ratio ($g_G/g_R$ = 0.8). Static FRET lines [Eq. (3)] are shown in blue. Dynamic FRET lines [Eq. (7)] between the Low and High $\langle R_{DA}\rangle_E$ states (Suppl. Tables 5-7) are shown in magenta. Light and dark horizontal lines mark the $F_D/F_A$ ratio corresponding to the Low and High $\langle R_{DA}\rangle_E$ states. Perrin's equation with rotational correlation time $\rho$ indicated in the 2D plot is shown as blue line.

**A** **Cdk2/cyclin A/p27-C29-54**

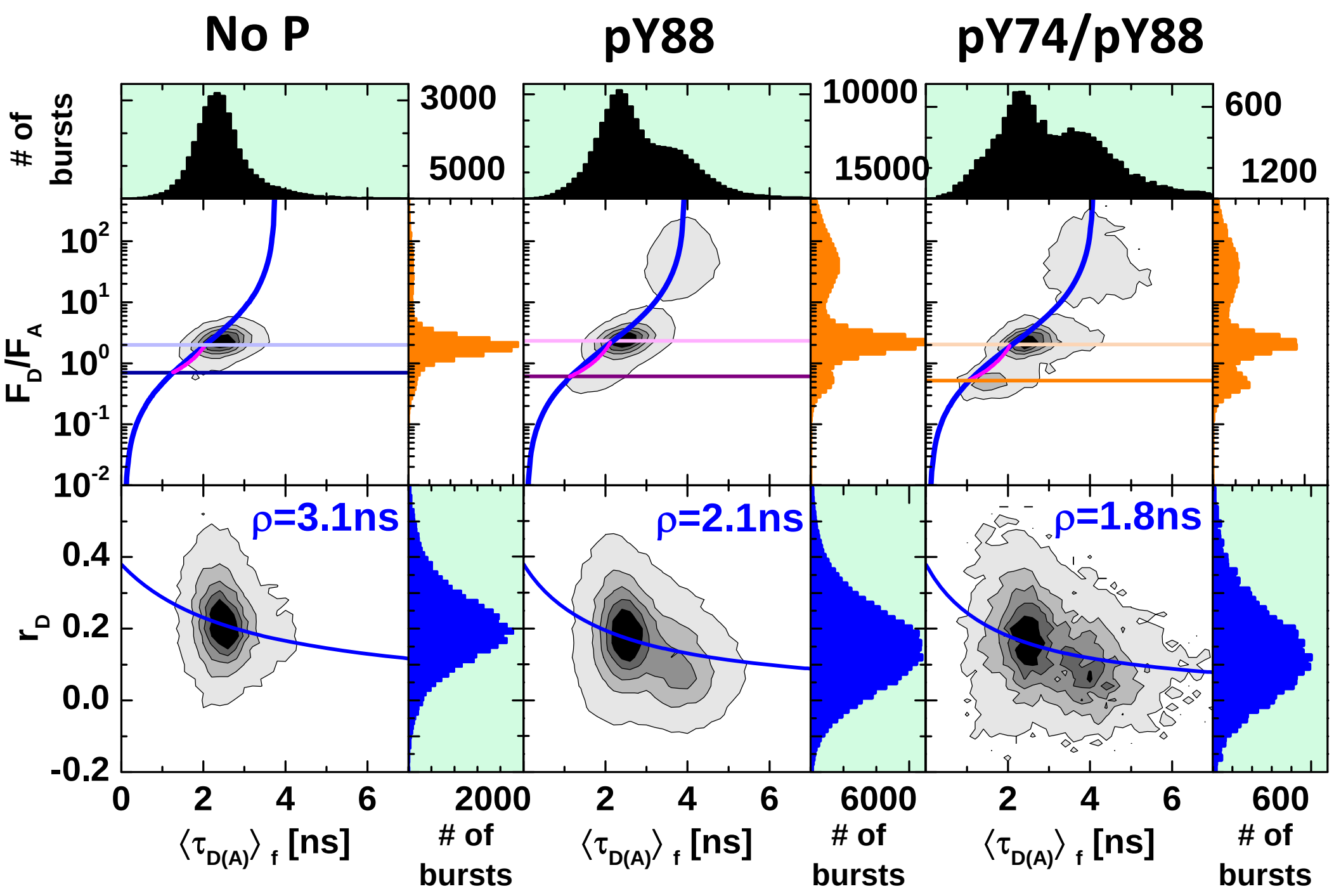

B

Cdk2/cyclin A/p27-C54-93

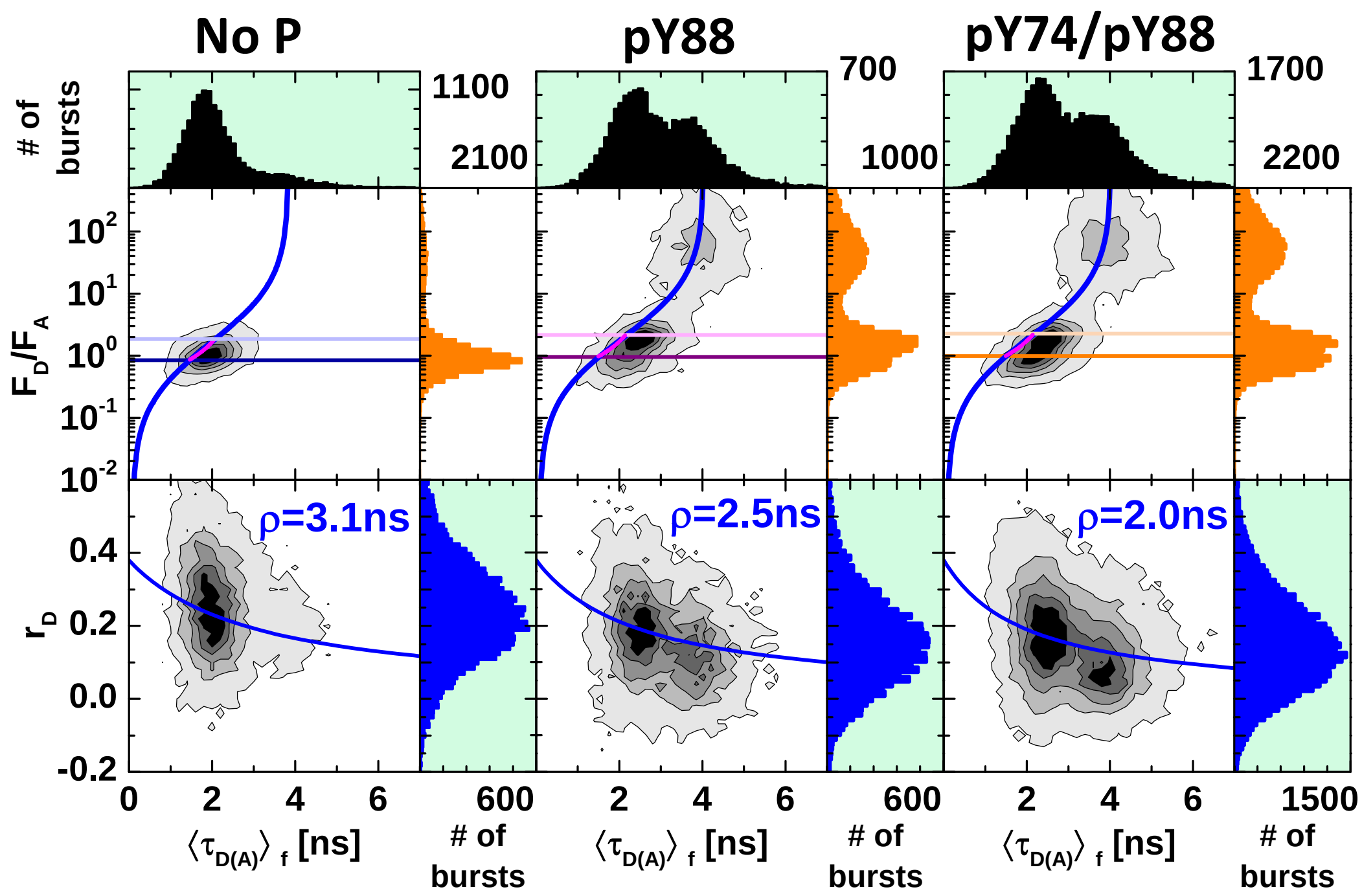

C

Cdk2/cyclin A/p27-C75-110

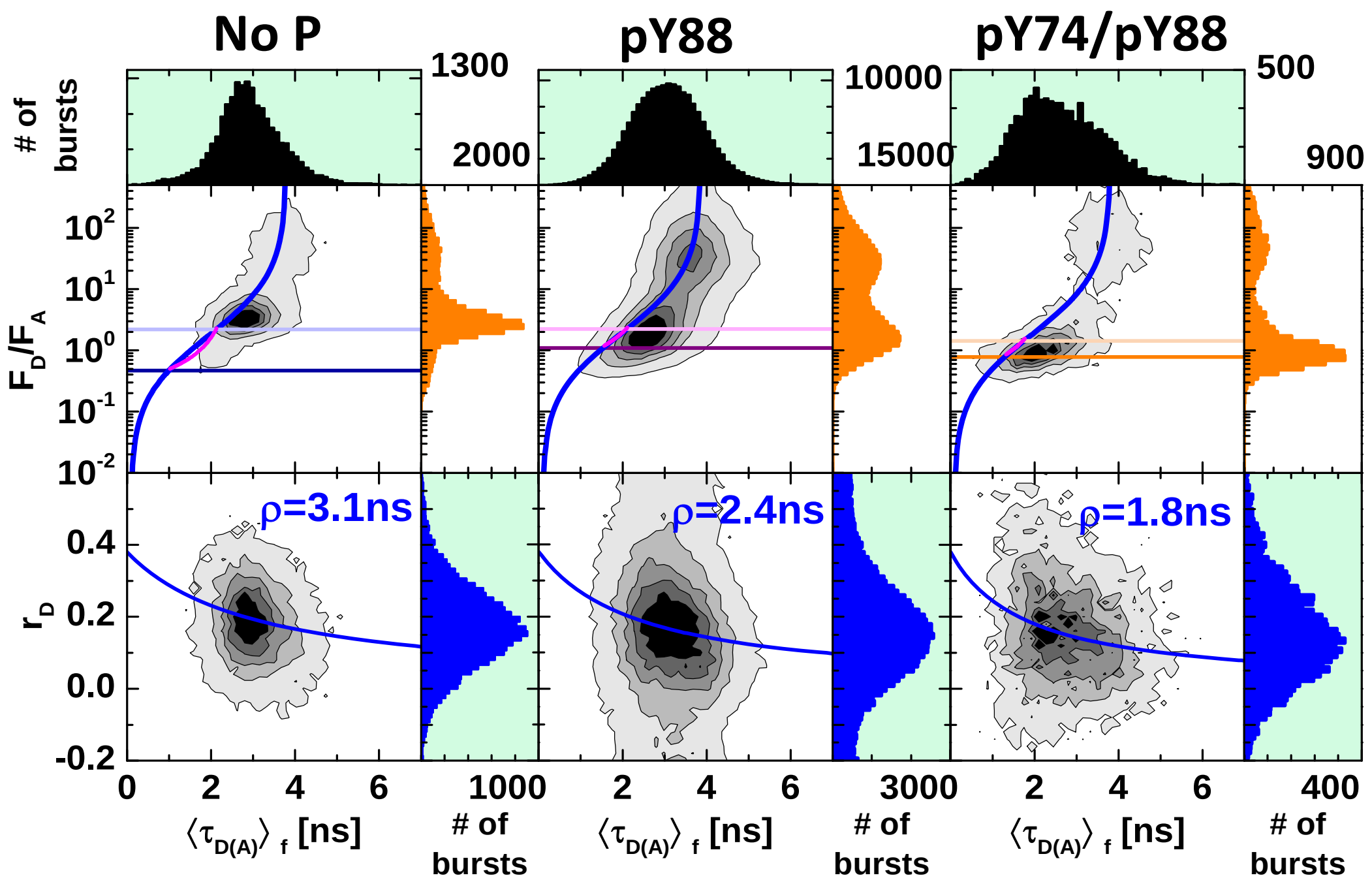

**Supplemental Figure 6. Filtered FCS Species auto and cross-correlation (sACF and sCCF) function of smFRET experiments for Cdk2/cyclin A/p27 samples.**

Filtered Fluorescence Auto and Cross-Correlation, left sACF and right sCCF, respectively. Filters were selected by "burstwise" selection based on FD/FA arbitrary cutoffs to select low-FRET (LF) or high-FRET (HF) populations. The integrated fluorescence of these burst corresponds to two independent species. The two sCCF (right) (HF to LF and LF to HF) and the sACF where globally fit using Eq. (18) to determine the number of relaxation times and their amplitudes (Suppl. Table 9). Residuals of the fit for the sACF and sCCF are shown on top of each correlation curve. Similar treatment was used for the pY88, data not shown.

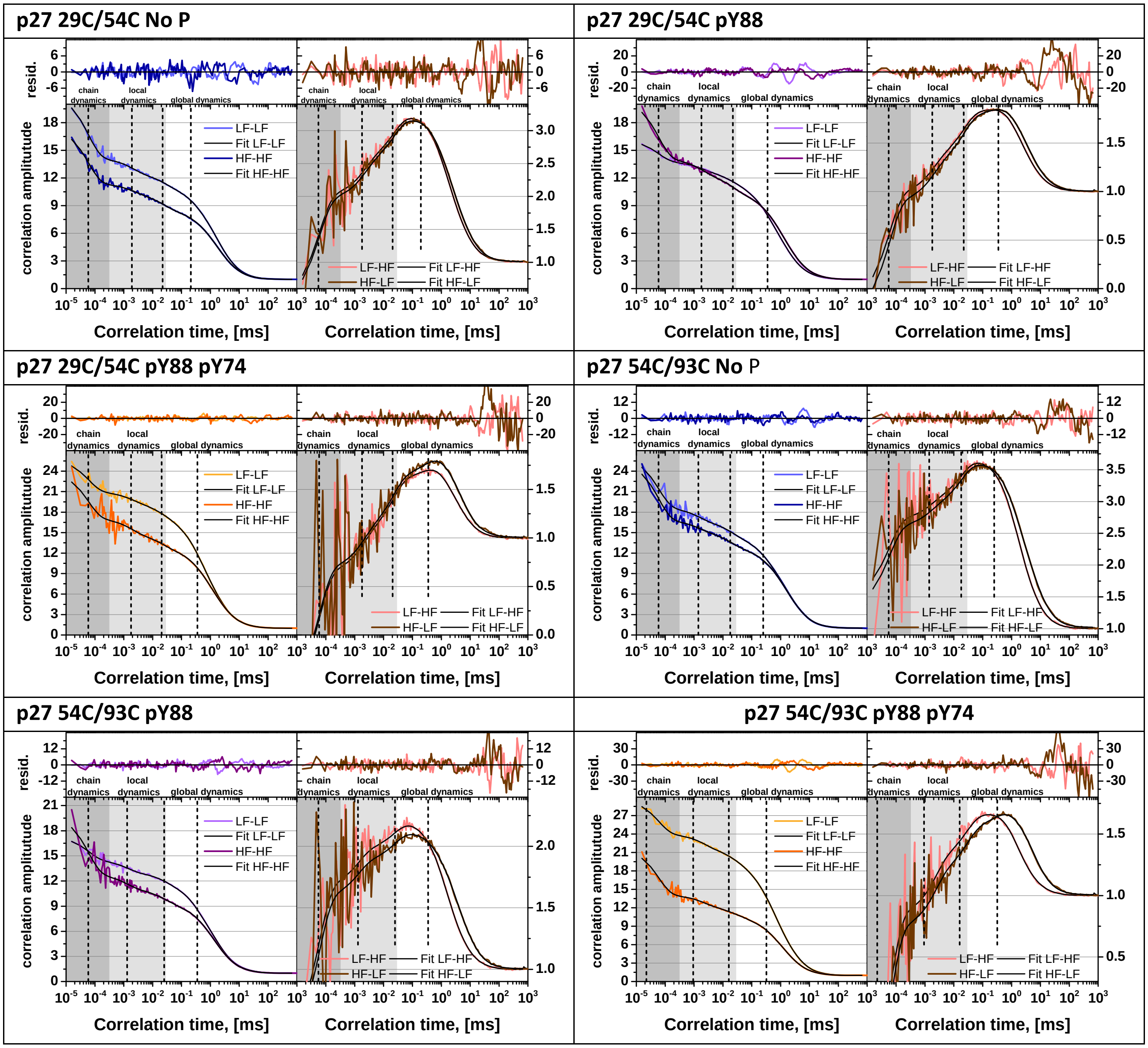

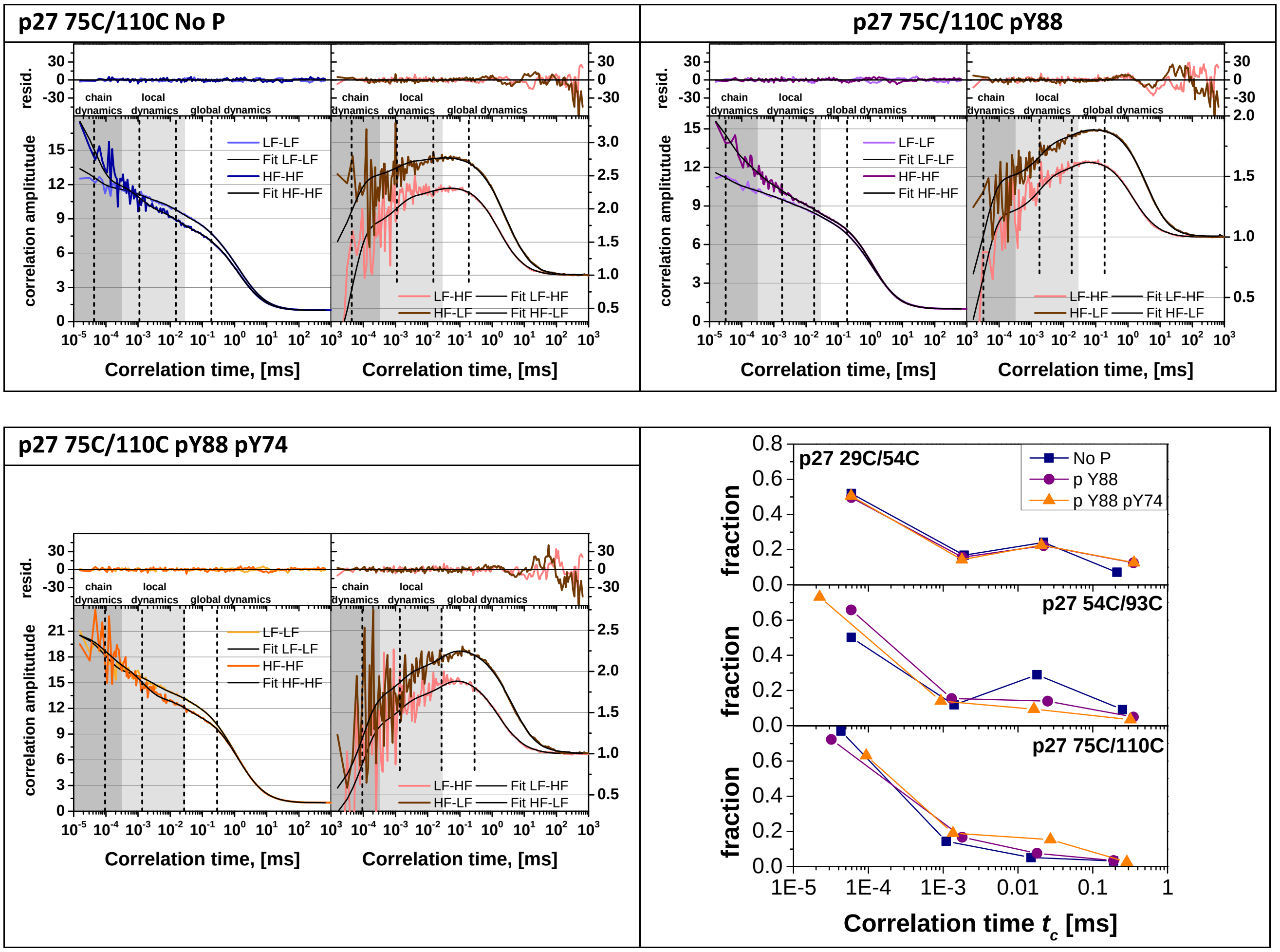

**Supplemental Figure 7. Representative results for varied concentrations of Cdk2/cyclin A, cyclin A, or Cdk2 binding to p27-KID, pY88-p27-KID, or pY74/Y88-p27-KID separately immobilized on the sensor surface.**

The results of triplicate injections are shown with fits of a 1:1 Langmuir interaction model shown as solid orange curves. The kinetic constants and $K_D$ values derived from these analyses are provided in Supplemental Table 10.

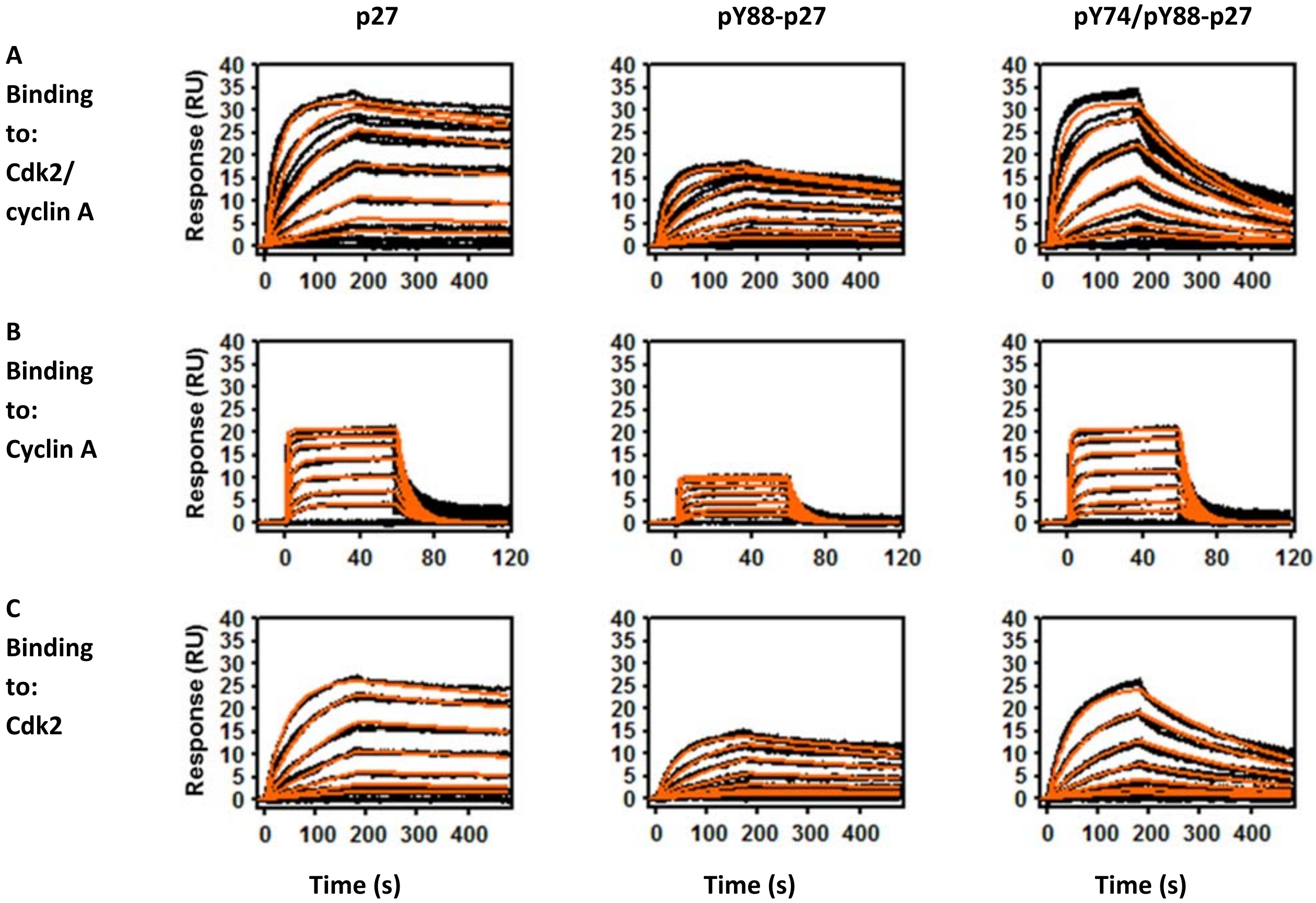

**Supplemental Figure 8 (next page). Representative binding isotherms for injected p27 (A), Y88E-p27 (B), or Y74E/Y88E-p27 (C) binding to Cdk2/cyclin A at temperatures from 5 °C to 25 °C recorded using isothermal titration calorimetry (ITC).**

For each experiment, the upper panel shows the power and the lower panel the heat associated with each injection. The enthalpy of binding (ΔH) values derived from analysis of triplicate measurements using a 1:1 binding model are provided in Table 1. Heat capacity change for binding ($\Delta C_p$) values were determined from the slope of ΔH *versus* temperature plots (D); $\Delta C_p$ values are provided in Table 1.

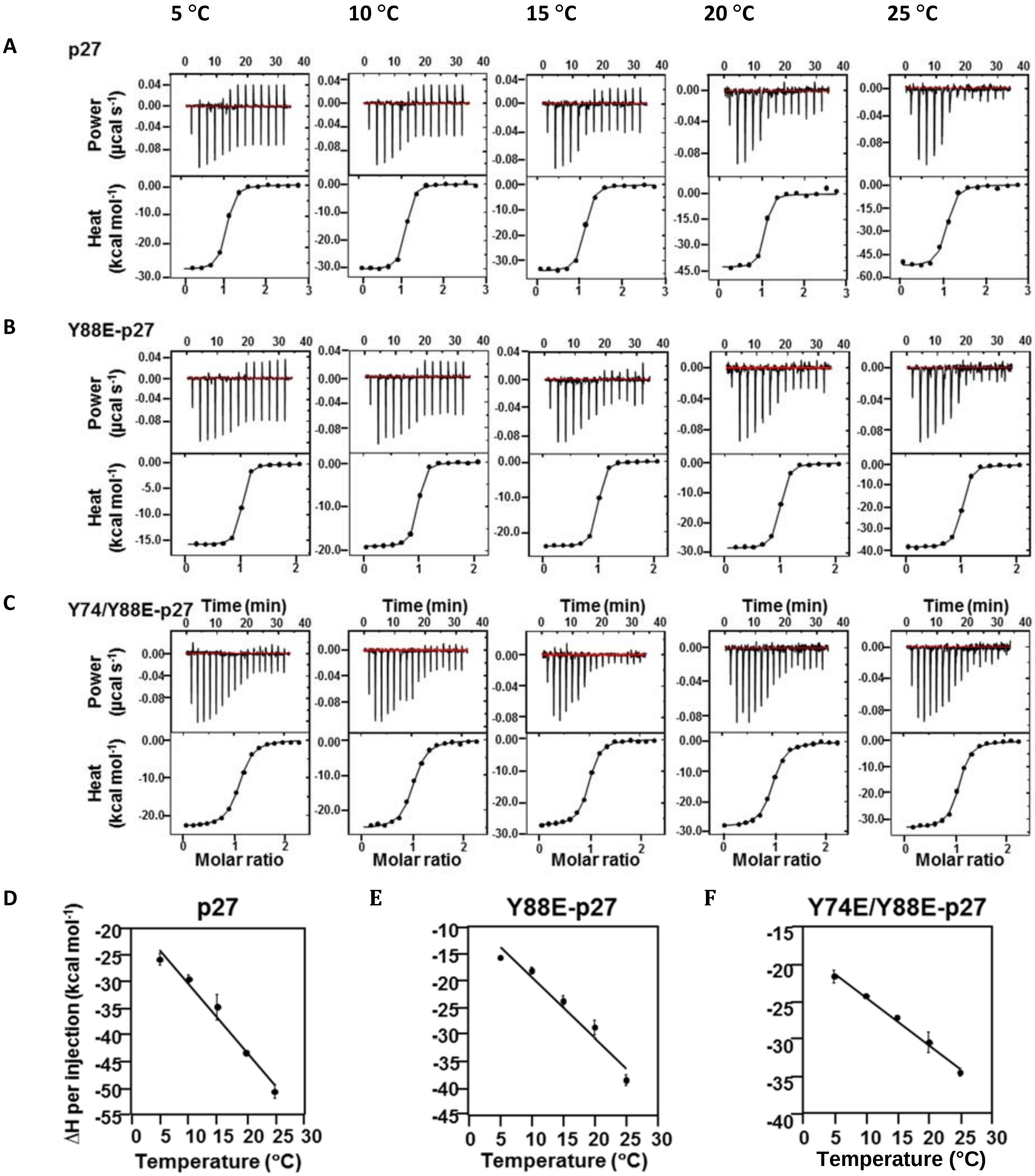

**Supplemental Figure 9. Multidimensional smFRET histograms of p27/Cdk2/Cyclin A with phosphomimetic variants.**

Two-dimensional histogram $F_D/F_A$ vs. lifetime of donor in the presence of acceptor $\langle\tau_{D(A)}\rangle_f$, and scatter corrected donor anisotropy ($r_D$) vs. $\langle\tau_{D(A)}\rangle_f$ for Cdk2/Cyclin A/p27 with donor and acceptor dyes at various positions with phosphomimetic mutations at positions E88 and E74/88. "Burstwise" mode of A) C29-54, B) C54-93 and C) C75-110 samples. For all cases one dimensional projections for $F_D/F_A$, $\langle\tau_{D(A)}\rangle_f$ and anisotropy are also shown. Pure donor and acceptor fluorescence ($F_D$ and $F_A$) are corrected for background ($\langle B_G\rangle$ = 1.80 kHz, $\langle B_R\rangle$ = 0.27 kHz), spectral cross-talk ($\alpha$ = 1.7%), direct acceptor excitation ($\beta$ = 1.3 %) and detection efficiency ratio ($g_G/g_R$ = 0.8). Values for "No P" samples are given in the legend to Figure S5. Static FRET lines (Eq. (3)) are shown in blue. Dynamic FRET lines (Eq. (7)) between the Low and High $\langle R_{DA}\rangle_E$ states (Table 5,7) are shown in magenta. Perrin's equation with rotational correlation time $\rho$ indicated in the 2D plot is shown as blue line.

**A** **Cdk2/cyclin A/p27-C29-54**

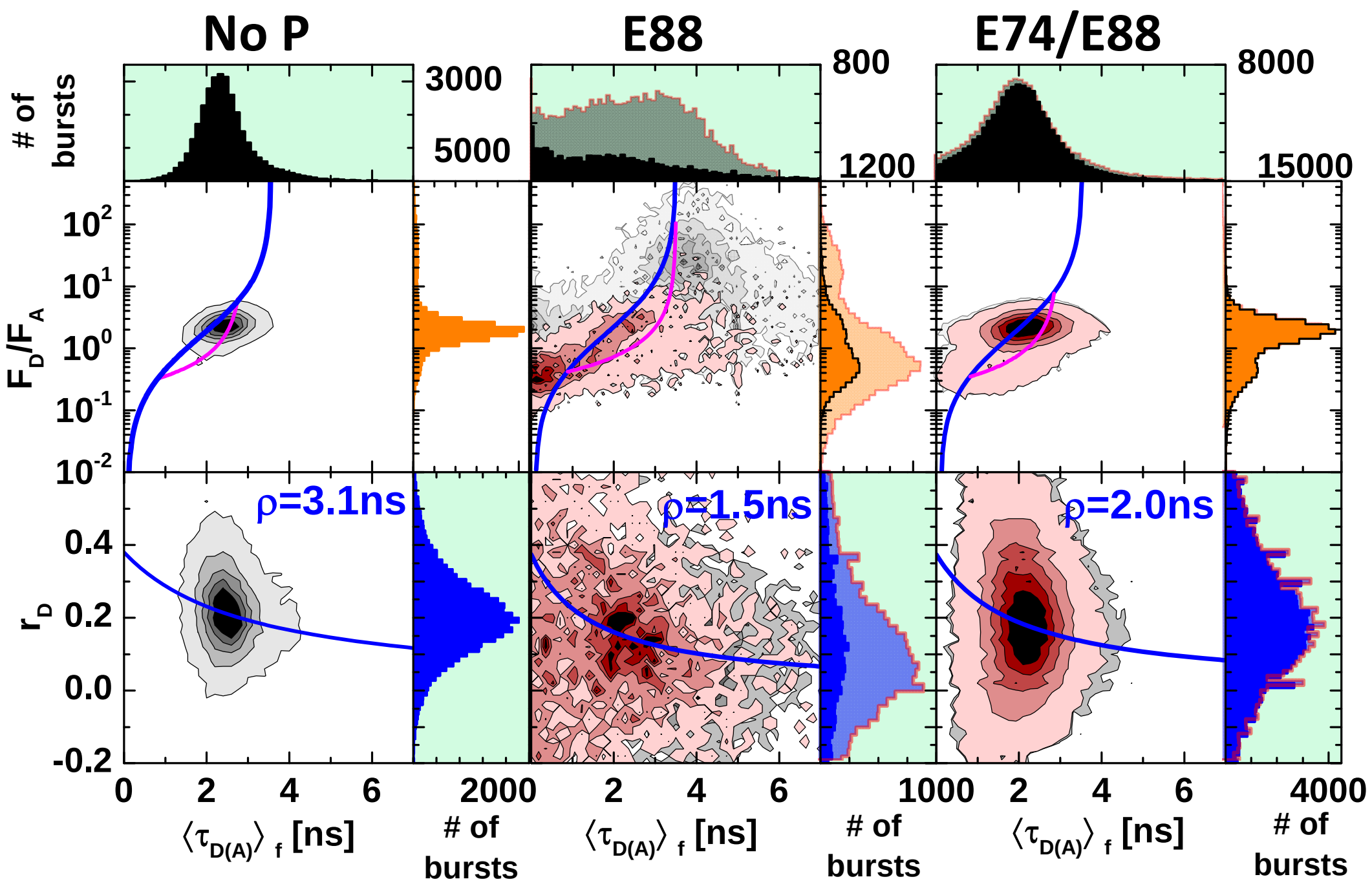

**B**

**Cdk2/cyclin A/p27-C54-93**

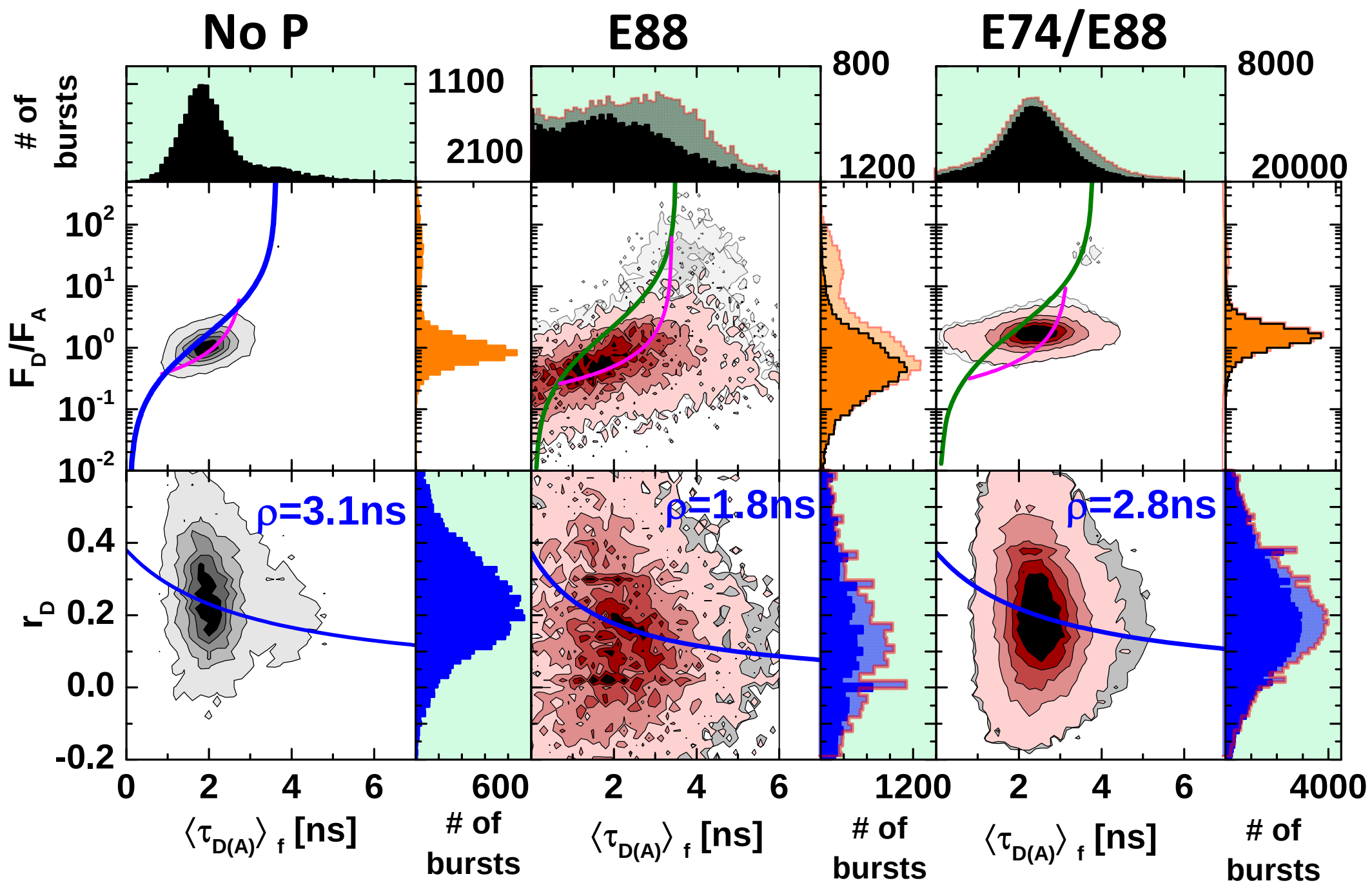

**C**

**Cdk2/cyclin A/p27-C75-110**

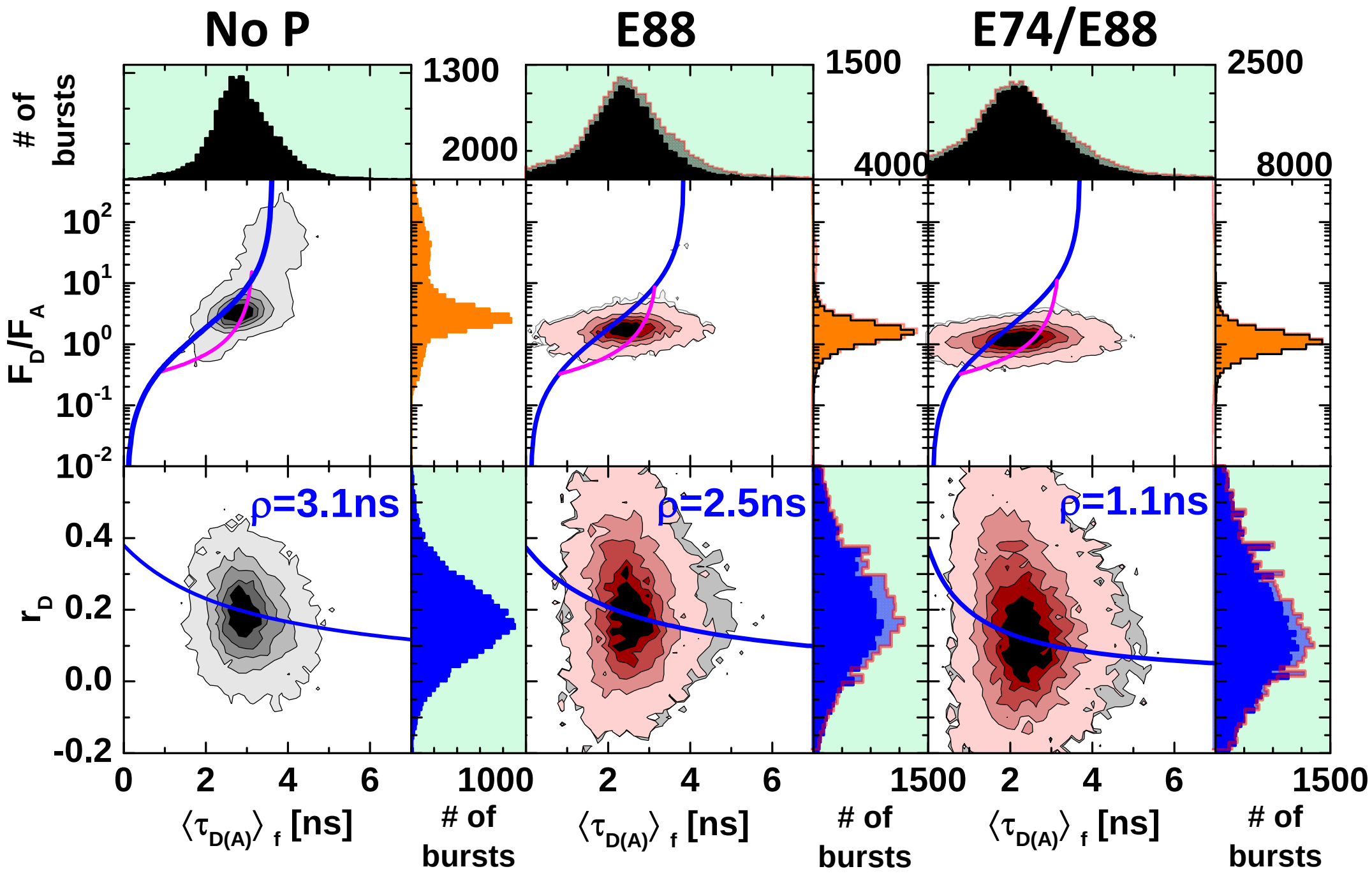

**Supplemental Figure 10. The effects of mono and dual tyrosine phosphorylation on regulation of Cdk2 by p27 can be mimicked by mutation of Y88, and Y74 and Y88, to glutamate (E).**
(A) Results of kinase activity assays for Cdk2/cyclin A in the presence of increasing concentrations of p27, Y88E-p27 or Y74E/Y88E-p27. Autoradiography was used to quantify incorporation of $^{32}$P-labeled phosphate into the substrate, Histone H1, which was resolved using SDS-PAGE (not shown). The experiments were performed in triplicate and the results quantified as average kinase activity (± standard deviation of the mean) relative to that in the presence of the lowest concentration of p27 expressed as percentage activity values. (B) NMR analysis of the influence of tyrosine to glutamate mutagenesis on interactions between p27-KID and Cdk2/cyclin A. Chemical shift differences for residues in unmutated (top) and tyrosine to glutamate mutated (Y88E-p27-KID, middle; and Y74E/Y88E-p27-KID, bottom) p27-KID bound to Cdk2/cyclin A. Residues near Y88 and within the entire D2 subdomain adopt free state-like conformations in Y88E-p27-KID and Y74E/Y88E-p27-KID, respectively. $\Delta\delta$ values were calculated using the equation: $\Delta\delta = [(\Delta\delta\ ^1H_N)^2 + 0.0289 \times (\Delta\delta\ ^{15}N_H)^2]^{1/2}$.

**A**

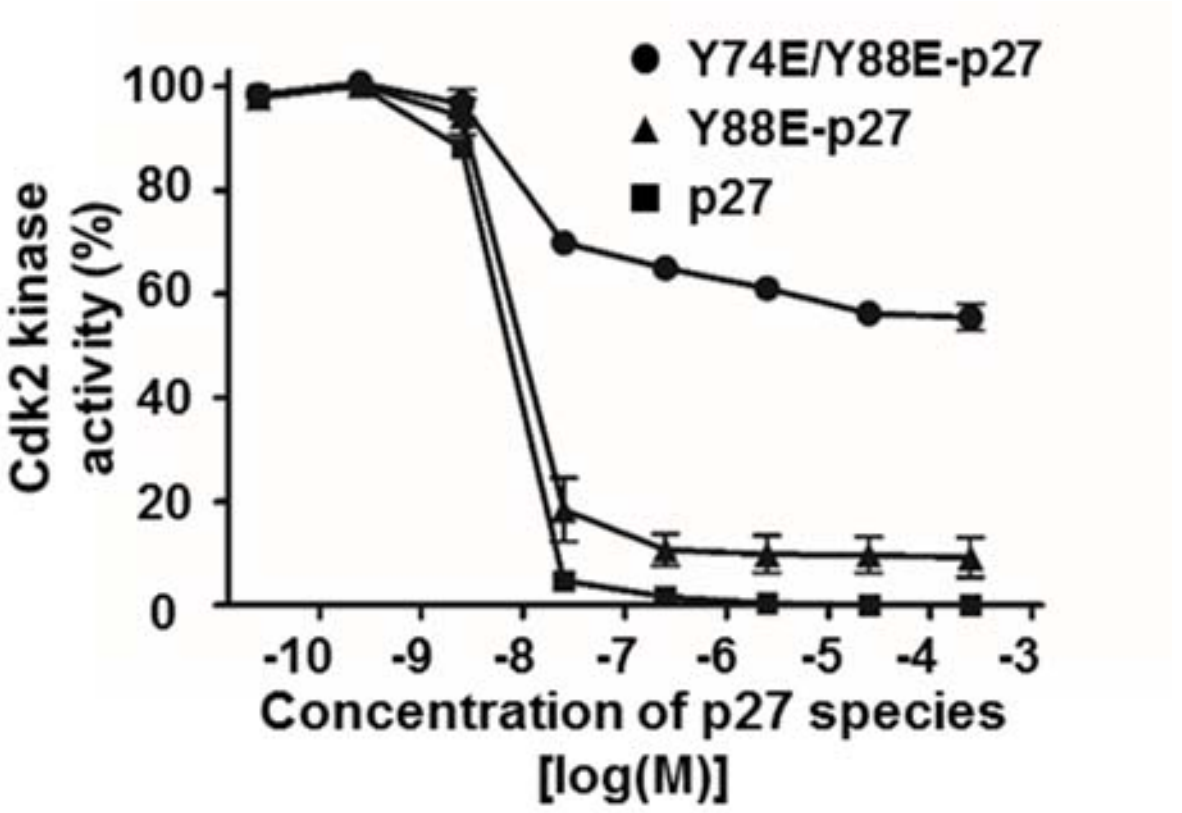

**B**

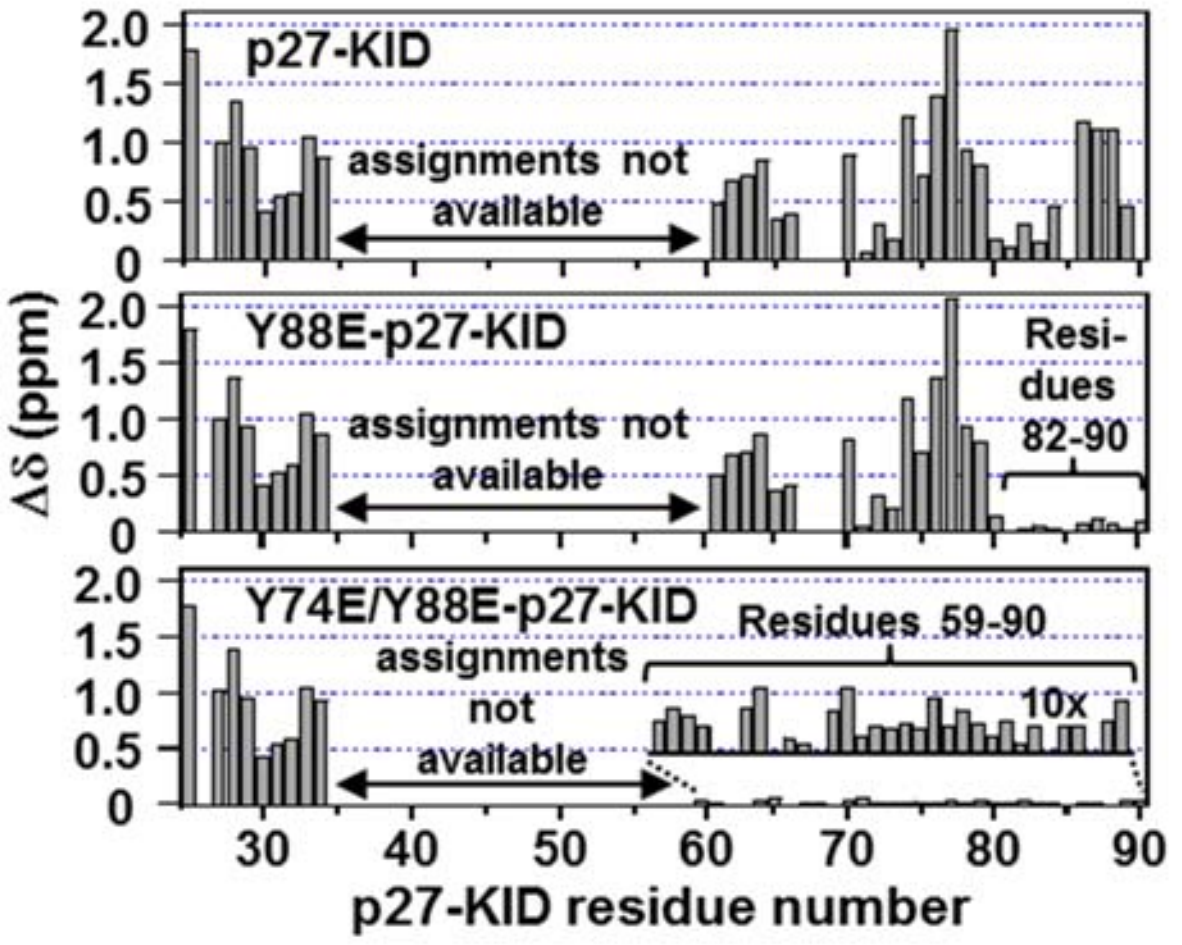

**Supplemental references**

Antonik, M., Felekyan, S., Gaiduk, A., and Seidel, C.A. (2006). Separating structural heterogeneities from stochastic variations in fluorescence resonance energy transfer distributions via photon distribution analysis. J Phys Chem B *110*, 6970-6978.

Böhmer, M., Wahl, M., Rahn, H.J., Erdmann, R., and Enderlein, J. (2002). Time-resolved fluorescence correlation spectroscopy. Chem Phys Lett *353*, 439-445.

Elson, E.L., and Magde, D. (1974). Fluorescence Correlation Spectroscopy. I. Conceptual Basis and Theory. Biopolymers *13*, 1-27.

Felekyan, S., Kalinin, S., Sanabria, H., Valeri, A., and Seidel, C.A. (2012a). Filtered FCS: species auto- and cross-correlation functions highlight binding and dynamics in biomolecules. Chemphyschem *13*, 1036-1053.

Felekyan, S., Kalinin, S., Sanabria, H., Valeri, A., and Seidel, C.A.M. (2012b). Filtered FCS: species auto- and cross-correlation functions highlight binding and dynamics in biomolecules. ChemPhysChem *13*, 1036-1053.

Felekyan, S., Sanabria, H., Kalinin, S., Kühnemuth, R., and Seidel, C.A.M. (2013). Analyzing Förster resonance energy transfer (FRET) with fluctuation algorithms. Methods Enzymol *519*, 39-85.

Kalinin, S., Felekyan, S., Antonik, M., and Seidel, C.A.M. (2007). Probability distribution analysis of single-molecule fluorescence anisotropy and resonance energy transfer. Journal of Physical Chemistry B *111*, 10253-10262.

Kalinin, S., Felekyan, S., Valeri, A., and Seidel, C.A. (2008). Characterizing multiple molecular States in single-molecule multiparameter fluorescence detection by probability distribution analysis. J Phys Chem B *112*, 8361-8374.

Kalinin, S., Sisamakis, E., Magennis, S.W., Felekyan, S., and Seidel, C.A. (2010). On the origin of broadening of single-molecule FRET efficiency distributions beyond shot noise limits. J Phys Chem B *114*, 6197-6206.

Koshioka, M., Sasaki, K., and Masuhara, H. (1995). Time-Dependent Fluorescence Depolarization Analysis in 3-Dimensional Microspectroscopy. Appl Spectrosc *49*, 224-228.

Maus, M., Cotlet, M., Hofkens, J., Gensch, T., De Schryver, F.C., Schaffer, J., and Seidel, C.A. (2001). An experimental comparison of the maximum likelihood estimation and nonlinear least-squares fluorescence lifetime analysis of single molecules. Anal Chem *73*, 2078-2086.

Schaffer, J., Volkmer, A., Eggeling, C., Subramaniam, V., Striker, G., and Seidel, C.A.M. (1999). Identification of single molecules in aqueous solution by time-resolved fluorescence anisotropy. J Phys Chem A *103*, 331-336.

Sindbert, S., Kalinin, S., Nguyen, H., Kienzler, A., Clima, L., Bannwarth, W., Appel, B., Muller, S., and Seidel, C.A. (2011). Accurate distance determination of nucleic acids via Forster resonance energy transfer: implications of dye linker length and rigidity. J Am Chem Soc *133*, 2463-2480.

Soong, T.T. (2004). Fundamentals of Probability and Statistics for Engineers (Wiley VCH).